\documentclass[iop]{emulateapj}

\usepackage{amsmath}
\usepackage{calrsfs}

\def\jcomp{Journal of Computational Physics}

\slugcomment{Not to appear in Nonlearned J., 45.}

\shorttitle{Effects of Resistivity on Magnetized Core-Collapses}
\shortauthors{Sawai et al.}

\begin{document}

\title{Effects of Resistivity on Magnetized Core-Collapse Supernovae}

\author{H. Sawai\altaffilmark{1}, S. Yamada\altaffilmark{2},
  K. Kotake\altaffilmark{3}, and H. Suzuki\altaffilmark{1}} 

\altaffiltext{1}{Tokyo University of Science, Chiba 278-8510, Japan}
\altaffiltext{2}{Waseda University, Shinjuku, Tokyo 169-8555, Japan}
\altaffiltext{3}{National Astronomical Observatory of Japan, Mitaka, Tokyo
  181-8588, Japan} 

\begin{abstract}
We studied roles of a turbulent resistivity in the core-collapse of a strongly
magnetized massive star, carrying out 2D-resistive-MHD simulations. The three
cases with different initial strengths of magnetic field and rotation
are investigated; 1.~strongly magnetized rotating core;
2.~moderately magnetized rotating core; 3.~very strongly magnetized
non-rotating core. In each case, both an ideal-MHD model and resistive-MHD
models are computed. As a result of computations, each model shows
a matter eruption helped by a magnetic acceleration (and also
by a centrifugal acceleration in the rotating cases). We found that a
resistivity attenuates 
the explosion in case~1 and 2, while it enhances the explosion in
case~3. We also found that in the rotating cases, main
mechanisms for the amplification of a magnetic field in the post-bounce
phase are an outward advection of magnetic field and a winding of
poloidal magnetic field-lines by differential rotation, which
are somewhat dampened down with the presence of a
resistivity. Although the magnetorotational instability seems to occur
in the rotating models, it will play only a minor role in a magnetic
field amplification. Another
impact of resistivity is that on the aspect ratio. In the rotating
cases, a large aspect ratio of the ejected matters,
$> 2.5$, attained in a ideal-MHD model is reduced to some extent in a
resistive model. These results indicate
that a resistivity possibly plays an important role in the dynamics of
strongly magnetized supernovae. 
\end{abstract}

\keywords{magnetohydrodynamics (MHD) --- methods: numerical  ---
  stars: magnetars --- supernovae: general}

\section{Introduction}\label{sec.intro}
Studies on magnetized core-collapse supernovae (CCSNe) has
gathered stream for past several years. Numerical simulations so far
have shown that the presence of strong magnetic field together with
rapid rotation results in a vigorous matter eruption accompanied by
bipolar jets with an explosion energy
of $\sim 10^{51}$~erg \citep[magnetorotational
explosion:][]{leb70,sym84,yam04}. One of the main 
driving forces is a toroidal magnetic 
pressure amplified by differential rotation that becomes intense
after the
collapse in the vicinity of proto-neutron star surface. This mechanism
requires a magnetic field strength of $\gtrsim 10^{15}$~G after collapse,
which is comparable to an inferred surface magnetic field of magnetar
candidates, soft-gamma repeaters (SGR) and anomalous X-ray 
pulsars (AXP). Magnetically-driven explosions may be related to such
supernovae that produce magnetars.

Most numerical simulations of magnetized core-collapse so far have been
done in the regime of ideal magnetohydrodynamics (MHD)
\citep[e.g.][]{kot04,saw05,moi06,bur07,saw08,tak09,obe11,end12}.
The only exception is the numerical study by \citet{gui11}, in which a
resistivity is introduced to 1D-MHD simulations investigating the
dynamics of an Alfv\'en surface in the context of core-collapse supernovae.
Although a numerical computation with a finite
differential scheme inevitably involves numerical diffusion, effects of
electric resistivity on the dynamics have not been investigated
systematically. The reason why a resistivity has been neglected is that
it would be infinitesimally small 
assuming that the source is the Coulomb scattering (Spitzer
resistivity; \citet{spi56}). The magnetic Reynolds number, the ratio of the
resistive timescale to dynamical timescale, estimated with a typical
parameters of proto-neutron star surface is 
\begin{eqnarray}
R_{\textrm{m}}&\sim& 8\times 10^{15}
\left(\displaystyle\frac{Z}{26}\right)^{-1} 
\left(\displaystyle\frac{T}{5\times10^{10} \textrm{K}}\right)^{3/2}\nonumber\\
&&\times\left(\displaystyle\frac{L}{4\times10^{5} \textrm{cm}}\right)
\left(\displaystyle\frac{v}{2\times10^{8} \textrm{cm s}^{-1}}\right),\nonumber
\end{eqnarray}
where $Z$, $T$, $L$, and $v$ are, respectively, an atomic charge number,
temperature, scale length of magnetic field, and flow velocity.
Since the magnetic Reynolds number is quite larger than unity, a
resistivity apparently seems not important for the dynamics.

However, it is uncertain whether the Coulomb scattering is the unique
origin of resistivity in a supernova core, and there may exist
other sources that give rise to a dynamically important value of
resistivity. One of such candidates is a turbulence. In the
collapsed core of a massive 
star, a convection, which occurs due to negative gradient of entropy
and/or electron fraction, may play a key role to produce a turbulent state
and a turbulent resistivity (and viscosity) along with
that. \citet{thot05} roughly estimated the amplitude of a
turbulent \textit{viscosity} arising from convective motions in a
supernova core by 
the product of the correlation length and convective velocity,
$\xi_{\rm{con}}\sim l v_{\rm{con}}/3$. They found that $\xi_{\rm{con}}$
is around $10^{13}-10^{14}$~cm$^2$~s$^{-1}$. Insofar as this level of 
estimation, the amplitude of a turbulent resistivity may be
evaluated by the same formula and may be comparable to a
turbulent viscosity, since magnetic field would have a similar
timescale and length-scale with those of velocity in the present
situation. \citet{yos90} showed in the frame work of 
so-called two-scale direct-interaction approximation that the magnitude
of a turbulent viscosity and turbulent resistivity are same order,
albeit in the context of an incompressible MHD turbulence. With the
above amplitude of the turbulent resistivity, the magnetic Reynolds
number becomes $\sim 1-10$ around the surface of a proto-neutron star, and
then a resistivity is possibly important for the dynamics. 

In this paper, we investigate how a (turbulent) resistivity alters the
dynamics of a magnetized core-collapse, paying particular attention to
the explosion energy, the magnetic field amplification, and the aspect
ratio of ejected matters. To this end 
we carried out axisymmetric 2D-resistive-MHD simulations of the
core-collapse of a massive star, assuming a strong magnetic field and
a large resistivity. A constant resistivity of $10^{13}$ and
$10^{14}$~cm$^2$~s$^{-1}$ are taken according to the above
discussion. Both rapidly rotating and non-rotating cores are studied.  
In the computations, we omitted any treatments of neutrinos, and
adopted a nuclear equation of state (EOS) produced by \cite{she98a, she98b}.

Before proceeding to the next section, we go into a little more detail
on a convection as a source of a turbulent resistivity, and also mention
our position in choosing 
the initial strength of a magnetic field and a rotation.
 
In a collapsed stellar core, there are mainly two convectively unstable
regions (see e.g. \citet{her94}). One is a region behind the shock
surface, where a negative entropy gradient is created 
as the shock surface propagates with its amplitude decreasing, and is
maintained by a neutrino heating. The other is a region around the
proto-neutron star surface, where a negative gradient of lepton
fraction is created because a neutrino is easier to escape from the core
for a lager radius. Since we do not deal with neutrinos, convections
related to them are not captured in the computations. Only convection
that seems to appear in our computation is one due to the shock
propagation with a decreasing amplitude\footnote{In each computation,
  we found that the 
  square of Brunt-B\"ais\"al\"a frequency is negative due to a negative
  entropy gradient in some locations behind the shock surface, and
  that a relatively large vorticity develops around there.}. 
Nevertheless, we assume all of the above convections as sources of a
turbulent resistivity adopted in this study, since they will occur in
the nature. What we have done here is to effectively introduce an
impact of these convections to the simulations. In so doing, it does
not seem quite important whether they are properly captured in the
simulations. Note that we do not consider that the standing accretion
shock instability (SASI), which do not present in our computations,
is one of the origins of a turbulent resistivity, because all of our
models explode before this instability can develop (several~100 ms after
bounce), owing to strong magnetic fields initially assumed.

Although a turbulent resistivity will appear only around the
convectively unstable regions, in our computations a constant
resistivity is assumed everywhere except in the vicinity of the center
(see \S~\ref{sec.setup}). Also, we should note that the estimation
made by \citet{thot05} is very uncertain. Moreover, strong magnetic
fields initially assumed may decrease the strength of a
turbulent resistivity. Therefore, the strengths of an adopted turbulent
resistivity are perhaps too large to be realistic. However, at
present a probable value of a turbulent resistivity in a
collapsed-stellar core is very unclear. Under such a circumstance, it
is meaningful to parametrically study its effect with some possible
values, and to grasp dynamical trends. We consider that the adopted
strengths of a turbulent resistivity may be maximum possible values. 

A strongly magnetized core prior to collapse assumed in the present
study is based on so-called "{\it  fossil-field hypothesis},'' which
supposes that the progenitor of a magnetar already has a magnetar-class
magnetic flux during the main sequence 
stage. Assuming this hypothesis, \citet{fer06} have done
population synthesis calculations from main sequence stars to neutron
stars, to fit observational data of radio pulsars. Their calculation 
produces consistent number of magnetars with those observed in the
Galaxy, where both the age and rotational period of magnetars are taken
into account. The fossil field hypothesis is also supported by
observations. There are several O-type stars whose surface
magnetic flux is inferred to be a magnetar-class; e.g. HD148937
\citep{wad12} and HD19612 \citep{don06}. \citet{aur10} measured surface
magnetic fields of Betelgeuse, a red supergiant star, to be $\sim 1$~G,
which indicates a magnetar-class magnetic flux. 
Note, however, that the origin of a strong magnetic fields in magnetars
is still controversial. Alternatively, they may be produced during the
core-collapse by a dynamo mechanism \citep{thoc93}. 

\citet{heg05} found that the so-called Tayler-Spruit dynamo
\citep{spr02} drastically slows down the rotation of a star especially during
an early phase of red super giant, where an angular momentum of the
rapidly rotating helium core is transported into the slowly rotating
hydrogen envelope. According to their computation, an inferred rotational
period of a pulser is $\sim 10$~ms for a $15~M_{\odot}$ progenitor, in
which the available rotational energy is insufficient for the explosion.
On the other hand, \citet{woo06} carried out stellar evolution
computations of inherently rapid 
rotators, and showed that the rotation of a pre-supernova core
could be fast. Due to the fast rotation, matters are almost
completely mixed, and instead of forming a red supergiant it becomes a
compact helium core star, where a magnetic torque works less
efficiently. One of their $16 M_{\odot}$ magnetic star models with
the solar metalicity results in the expected pulsar rotation period of
$2.3$~ms, sufficient for a magnetorotational explosion. Note that the
both works involve uncertainties about such as 
mass loss rate and multi-dimensional effects. At present, it is unclear
either of slow or fast rotation is appropriate for the progenitor of
a magnetar. Hence, in this study both rapidly rotating models and
non-rotating models are investigated, where the latter, in effect,
corresponds to a slow rotation case.

The rest of this paper is organized as follows. 
We describe the governing equations and essentials of our resistive-MHD
code in \S~\ref{sec.eq}, and computational setups in \S~\ref{sec.setup}. The
results are presented in \S~\ref{sec.result}. Discussion and
conclusion are given in \S~\ref{sec.conc}.


\section{Governing Equations and Numerical Schemes}\label{sec.eq}
In order to follow the dynamics of magnetized core-collapses with
resistivity, the resistive-MHD equations below are solved: 
{\allowdisplaybreaks
\begin{eqnarray}
&&\frac{\partial\rho}{\partial
  t}+\nabla\cdot(\rho\mbox{\boldmath$v$})=0\label{eq.mhd.mass},
\\ 
&&\frac{\partial}{\partial t} (\rho\mbox{\boldmath$v$})+
\nabla\cdot\left(\rho\mbox{\boldmath$v$}\mbox{\boldmath$v$}-
\frac{\mbox{\boldmath$B$}\mbox{\boldmath$B$}}{4\pi}\right)\nonumber\\
&&\hspace{1pc}=-\nabla\left(p+\frac{B^2}{8\pi}\right)-\rho\nabla\Phi 
\label{eq.mhd.mom},
\\  
&&\frac{\partial}{\partial t}\left(e+\frac{\rho
    v^2}{2}+\frac{B^2}{8\pi}\right)
\nonumber\\
&&\hspace{1pc}+\nabla\cdot\left[\left(e+p+\frac{\rho
      v^2}{2}+\frac{B^2}{4\pi}\right) 
\mbox{\boldmath$v$}\right.
\nonumber\\
&&\hspace{4pc}
\left.-\frac{(\mbox{\boldmath$v$}\cdot
\mbox{\boldmath$B$}) 
\mbox{\boldmath$B$}}{4\pi} 
+\frac{\eta}{c}\mbox{\boldmath$j$}\times\mbox{\boldmath$B$}\right]
\nonumber\\
&&\hspace{4pc}
=-\rho(\nabla\Phi)\cdot\mbox{\boldmath$v$}\label{eq.mhd.eng},
\\
&&\frac{\partial\mbox{\boldmath $B$}}{\partial t}
+c\nabla\times\mbox{\boldmath$E$}=0\label{eq.mhd.far},
\\
&&\mbox{\boldmath$E$}=-\frac{1}{c}\mbox{\boldmath$v$}\times\mbox{\boldmath$B$}
+\frac{4\pi\eta}{c^2}\mbox{\boldmath$j$}\label{eq.mhd.ohm},
\\
&&\mbox{\boldmath$j$}=\frac{c}{4\pi}\nabla\times\mbox{\boldmath$B$}
\label{eq.mhd.amp},
\end{eqnarray}}in which notations of the physical variables follow custom.

To solve above equations we have developed 2D-resistive-MHD
code, "\textit{Yamazakura}.'' This is a time explicit, Eulerian code
based on the high resolution central scheme formulated by \citet{kur00}
(Here after KT scheme). Below we briefly describe the features of
\textit{Yamazakura}. For sake of simplicity, we deal in the case
where the equations are written in Cartesian coordinate with plane
symmetry in $z$-direction.
 
The KT scheme adopts a finite volume method to solve conservation
equations. Although the induction equations~(\ref{eq.mhd.far})
apparently seem written in non-conservation forms, they are rewritten
into conservation forms \citep{zie04};
\begin{eqnarray}
\frac{\partial B^x}{\partial t}
+c\nabla\cdot\left(
\begin{array}{cc}
\hphantom{-}0\hphantom{^-},  &
\hphantom{-}E^z
\end{array}
\right)=0,
\nonumber\\
\frac{\partial B^y}{\partial t}
+c\nabla\cdot\left(
\begin{array}{cc}
           -E^z,& 
\hphantom{-}0\hphantom{^-}
\end{array}
\right)=0,
\label{eq.mhd.far.matrix}
\\
\frac{\partial B^z}{\partial t}
+c\nabla\cdot\left(
\begin{array}{cc}
\hphantom{-}E^y,&
           -E^x
\end{array}
\right)=0.\nonumber
\end{eqnarray}
Then evolutionary
Eqs.~(\ref{eq.mhd.mass})$-$(\ref{eq.mhd.eng}) and
(\ref{eq.mhd.far.matrix}) are all written in conservation forms with
source terms; 
{\allowdisplaybreaks
\begin{eqnarray}
&&\frac{\partial \mbox{\boldmath$u$}}{\partial t} 
+\nabla\cdot\left( 
\begin{array}{ccc}
\mbox{{\boldmath$f$}}^x\left(\mbox{{\boldmath$u$}} \right), &
\mbox{{\boldmath$f$}}^y\left(\mbox{{\boldmath$u$}} \right) &
\end{array}
\right)
\label{eq.kt.vector}\\
&&\hspace{1.4pc}
+\nabla\cdot \left( 
\begin{array}{ccc}
\mbox{{\boldmath$g$}}^x\left(\mbox{{\boldmath$u$}}, 
\mbox{{\boldmath$u$}}_{,x},\mbox{{\boldmath$u$}}_{,y}\right),&
\mbox{{\boldmath$g$}}^y\left(\mbox{{\boldmath$u$}},
\mbox{{\boldmath$u$}}_{,x},\mbox{{\boldmath$u$}}_{,y}\right) &
\end{array}
\right)
=\mbox{{\boldmath$s$}}(\mbox{{\boldmath$u$}}), 
\nonumber
\end{eqnarray}
where a expression such as $\nabla\cdot(\mbox{{\boldmath$f$}}^x,
\mbox{{\boldmath$f$}}^y)$ means $\partial\mbox{{\boldmath$f$}}^x/\partial x
  + \partial\mbox{{\boldmath$f$}}^y/\partial y$.
The vectors in Eq.~(\ref{eq.kt.vector}), each of which has eight
components, are given as follows:
\begin{eqnarray}
\mbox{\boldmath$u$}&=&
\left(
\begin{array}{c}
\rho\\
\rho v^x\\
\rho v^y\\
\rho v^z\\
e+\rho v^2/2+B^2/8\pi\\
B^x\\
B^y\\
B^z 
\end{array}
\right),\nonumber\\
\mbox{\boldmath$f$}^x&=&
\left(
\begin{array}{c}
\rho v^x\\
\rho v^x v^x - B^xB^x/4\pi + p + B^2/8\pi\\
\rho v^x v^y - B^xB^y/4\pi\\
\rho v^x v^z - B^xB^z/4\pi\\
(e+p+\rho
v^2/2+B^2/4\pi)v^x-(\mbox{\boldmath$v$}\cdot\mbox{\boldmath$B$})B^x/4\pi\\ 
0\\
 v^xB^y-v^yB^x\\
-v^zB^x+v^xB^z
\end{array}
\right),\nonumber\\
\mbox{\boldmath$f$}^y&=&
\left(
\begin{array}{c}
\rho v^y\\
\rho v^y v^x - B^y B^x/4\pi\\
\rho v^y v^y - B^y B^y/4\pi + p + B^2/8\pi\\
\rho v^y v^z - B^y B^z/4\pi\\
(e+p+\rho
v^2/2+B^2/4\pi)v^y-(\mbox{\boldmath$v$}\cdot\mbox{\boldmath$B$})B^y/4\pi\\ 
-v^xB^y+v^yB^x\\
0\\
 v^yB^z-v^zB^y
\end{array}
\right),
\nonumber\\
\mbox{\boldmath$g$}^x&=&
\left(
\begin{array}{c}
0\\
0\\
0\\
0\\
\eta(j^y B^z-j^z B^y)/c\\
0\\
-4\pi\eta j_z/c\\
 4\pi\eta j_y/c
\end{array}
\right),
\nonumber\\
\mbox{\boldmath$g$}^y&=&
\left(
\begin{array}{c}
0\\
0\\
0\\
0\\
\eta(j^z B^x-j^x B^z)/c\\
 4\pi\eta j_z/c\\
0\\
-4\pi\eta j_x/c
\end{array}
\right),
\nonumber\\
\mbox{\boldmath$s$}&=&
\left(
\begin{array}{c}
0\\
-\rho \partial \Phi /\partial x\\
-\rho \partial \Phi /\partial y\\
-\rho \partial \Phi /\partial z\\
-\rho (\nabla\Phi)\cdot\mbox{\boldmath$v$}\\
0\\
0\\
0
\nonumber 
\end{array}
\right).
\end{eqnarray}

The KT scheme is written as follows: 
{\allowdisplaybreaks
\begin{eqnarray}
\frac{d \mbox{{\boldmath$u$}}_{i,j}(t)}{d t} &=& 
-\frac{\mbox{{\boldmath$F$}}^x_{i+1/2,j}(t)-\mbox{{\boldmath$F$}}^x_{i-1/2,j}(t)}
{\Delta x_i}
\nonumber\\
&&
-\frac{\mbox{{\boldmath$F$}}^y_{i,j+1/2}(t)-\mbox{{\boldmath$F$}}^y_{i,j-1/2}(t)}
{\Delta y_j}
\nonumber\\
&&
-\frac{\mbox{{\boldmath$G$}}^x_{i+1/2,j}(t)-\mbox{{\boldmath$G$}}^y_{i-1/2,j}(t)}
{\Delta x_i}
\nonumber\\
&&
-\frac{\mbox{{\boldmath$G$}}^y_{i,j+1/2}(t)-\mbox{{\boldmath$G$}}^y_{i,j-1/2}(t)}
{\Delta y_j}
+ \mbox{{\boldmath$s$}}_{i,j}(t),\label{eq.kt.semidis}
\end{eqnarray}}where $\Delta x_i = x_{i+1/2}-x_{i-1/2}$, $\Delta y_j = 
y_{j+1/2}-y_{j-1/2}$, and an integer and half-integer subscript
respectively means that a variable is evaluated at the numerical cell center
and interface. The numerical fluxes $\mbox{{\boldmath$F$}}$ and
$\mbox{{\boldmath$G$}}$ are for the non-resistive terms and resistive
terms, respectively.
The numerical fluxes of non-resistive terms are given by
{\allowdisplaybreaks
\begin{eqnarray}
\mbox{{\boldmath$F$}}^x_{i+1/2,j}(t)&\equiv&
\frac{\mbox{{\boldmath$f$}}^x\left(\mbox{{\boldmath$u$}}^{+}_{i+1/2,j}(t)\right)
     +\mbox{{\boldmath$f$}}^x\left(\mbox{{\boldmath$u$}}^{-}_{i+1/2,j}(t)\right)}
     {2}
\nonumber\\  
&&-\frac{a_{i+1/2,j}(t)}{2} 
\left[\mbox{{\boldmath$u$}}^+_{i+1/2,j}(t)-\mbox{{\boldmath$u$}}^-_{i+1/2,j}(y)\right],
\nonumber\\
\label{eq.kt.numfluxf}\\
\mbox{{\boldmath$F$}}^y_{i,j+1/2}(t)&\equiv&
\frac{\mbox{{\boldmath$f$}}^y\left(\mbox{{\boldmath$u$}}^{+}_{i,j+1/2}(t)\right)
     +\mbox{{\boldmath$f$}}^y\left(\mbox{{\boldmath$u$}}^{-}_{i,j+1/2}(t)\right)}
     {2}\nonumber\\
&&-\frac{a_{i,j+1/2}(t)}{2}
\left[\mbox{{\boldmath$u$}}^+_{i,j+1/2}(t)-\mbox{{\boldmath$u$}}^-_{i,j+1/2}(t)\right].
\nonumber
\end{eqnarray}}The numerical fluxes of the resistive terms,
$\mbox{{\boldmath$G$}}^x$ and $\mbox{{\boldmath$G$}}^y$, are to
appear later. 

In the original KT scheme, the interface values in
Eq.~(\ref{eq.kt.numfluxf}), which have a superscript "$+$" or "$-$", are
evaluated by a interpolation of conservative variables
$\mbox{{\boldmath$u$}}$: 
\begin{eqnarray}
\mbox{{\boldmath$u$}}^{+}_{i+1/2,j}(t)&=&\mbox{{\boldmath$u$}}_{i+1,j}(t)
-\frac{\Delta x_{i+1}}{2}(\mbox{{\boldmath$u$}}_{,x})_{i+1,j}(t),
\nonumber\\
\mbox{{\boldmath$u$}}^{-}_{i+1/2,j}(t)&=&\mbox{{\boldmath$u$}}_{i,j}(t)
_{\hphantom{+1}}+\frac{\Delta x_{i\hphantom{+1}}}{2}
(\mbox{{\boldmath$u$}}_{,x})_{i,j}(t),
\nonumber\\
\mbox{{\boldmath$u$}}^{+}_{i,j+1/2}(t)&=&\mbox{{\boldmath$u$}}_{i,j+1}(t)
-\frac{\Delta y_{j+1}}{2}(\mbox{{\boldmath$u$}}_{,y})_{i,j+1}(t),
\nonumber\\
\mbox{{\boldmath$u$}}^{-}_{i,j+1/2}(t)&=&\mbox{{\boldmath$u$}}_{i,j}(t)
_{\hphantom{+1}}+\frac{\Delta y_{j\hphantom{+1}}}{2}
(\mbox{{\boldmath$u$}}_{,y})_{i,j}(t).
\label{eq.kt.intpol}
\end{eqnarray}
Alternatively, we may be able to use interpolated interface values of
the primitive variables
$\mbox{{\boldmath$q$}}=(\rho,v_x,v_y,v_z,e,B_x,B_y,B_z)$,
{\allowdisplaybreaks
\begin{eqnarray}
\mbox{{\boldmath$q$}}^{+}_{i+1/2,j}(t)&=&\mbox{{\boldmath$q$}}_{i+1,j}(t)
-\frac{\Delta x_{i+1}}{2}(\mbox{{\boldmath$q$}}_{,x})_{i+1,j}(t),
\nonumber\\
\mbox{{\boldmath$q$}}^{-}_{i+1/2,j}(t)&=&\mbox{{\boldmath$q$}}_{i,j}(t)
_{\hphantom{+1}}+\frac{\Delta x_{i\hphantom{+1}}}{2}
(\mbox{{\boldmath$q$}}_{,x})_{i,j}(t),
\nonumber\\
\mbox{{\boldmath$q$}}^{+}_{i,j+1/2}(t)&=&\mbox{{\boldmath$q$}}_{i,j+1}(t)
-\frac{\Delta y_{j+1}}{2}(\mbox{{\boldmath$q$}}_{,y})_{i,j+1}(t),
\nonumber\\
\mbox{{\boldmath$q$}}^{-}_{i,j+1/2}(t)&=&\mbox{{\boldmath$q$}}_{i,j}(t)
_{\hphantom{+1}}+\frac{\Delta y_{j\hphantom{+1}}}{2}
(\mbox{{\boldmath$q$}}_{,y})_{i,j}(t),
\label{eq.kt.qintpol}
\end{eqnarray}}to evaluate the interface values of the conservative
variables in Eqs.~(\ref{eq.kt.numfluxf}), e.g.,
\begin{eqnarray}
\mbox{{\boldmath$u$}}^+_{i+1/2,j}(t)&=&
\mbox{{\boldmath$u$}}\left(\mbox{{\boldmath$q$}}^+_{i+1/2,j}(t)\right).
\nonumber
\end{eqnarray}
In \textit{Yamazakura}, we adopt the latter prescription.
For the
calculation of the numerical derivatives, $\mbox{{\boldmath$q$}}_{,x}$
and $\mbox{{\boldmath$q$}}_{,y}$, in Eqs.~(\ref{eq.kt.qintpol}), we
employ a minmod-like limiter suggested 
by \citet{kur00}. Note that the interpolations~(\ref{eq.kt.qintpol})
are only used for the hydrodynamical quantities and the $z$-component
of magnetic field, which are placed at the center of a numerical
cell. Those for the  
$x$-component and $y$-component of magnetic field are given later.

In Eqs.~(\ref{eq.kt.numfluxf}), $a_{i+1/2,j}(t)$ and
$a_{i,j+1/2}(t)$ are the maximum characteristic speeds, i.e. the sum
of the fluid velocity and fast magnetosonic speed, which we evaluate
using the interpolated primitive variables in
Eqs.~(\ref{eq.kt.qintpol}): 
\begin{eqnarray}
a_{i+1/2,j}(t)&=&
\max\left\{
a\left(\mbox{{\boldmath$q$}}^+_{i+1/2,j}(t)\right),
a\left(\mbox{{\boldmath$q$}}^-_{i+1/2,j}(t)\right)
\right\},\nonumber\\
a_{i-1/2,j}(t)&=&
\max\left\{
a\left(\mbox{{\boldmath$q$}}^+_{i-1/2,j}(t)\right),
a\left(\mbox{{\boldmath$q$}}^-_{i-1/2,j}(t)\right)
\right\}.\nonumber\\
\end{eqnarray}
In KT scheme, we only need to know these maximum characteristic speeds
instead of carrying out a complicated characteristic decomposition
for wave propagations.

The original KT scheme is formulated for uniform spatial cells. We have
followed the procedure to deduce the KT scheme described in
\citet{kur00} with non-uniform cells, and found that the final
semi-discrete form~(\ref{eq.kt.semidis})$-$(\ref{eq.kt.intpol}) is
unchanged, except that the subscripts to $\Delta x$  and $\Delta y$
appear. 

In order to obtain the time evolution of conservative variables
$\mbox{{\boldmath$u$}}_{i,j}(t)$, the semi-discrete
equation~(\ref{eq.kt.semidis}) is time integrated 
utilizing a third order Runge-Kutta method according to \citet{kur00}.
With this and the spatial interpolations in Eqs.~(\ref{eq.kt.intpol}), the
original KT scheme is a third order in time and second order in
space. In Appendix ("Linear Wave Propagation"), it is shown that
Yamazakura performs, 
approximately, at least second order in time and second order in space,
even though we adopt a non-uniform cell distribution and the spatial
interpolations of the primitive variables (Eqs.~(\ref{eq.kt.qintpol})).

\begin{figure}
\epsscale{1}
\plotone{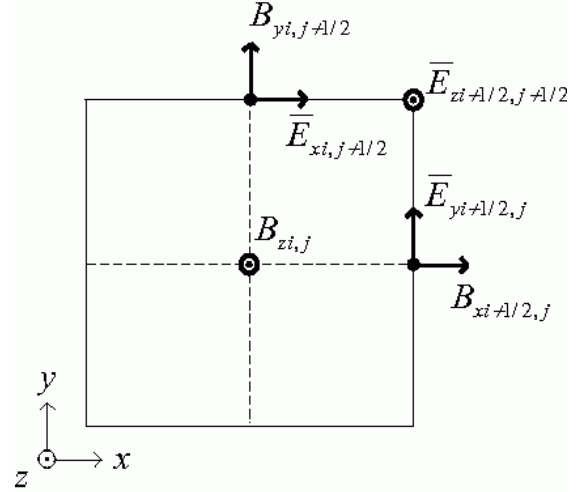}
  \caption{Positional relation of magnetic field vectors and
    numerical fluxes (or electric field vectors),
    $\mbox{{\boldmath$\bar{E}$}}$.} 
 \label{fig.ct}
\end{figure}

In solving MHD equations, it is necessary to satisfy the divergence-free
constraint of magnetic field. To accomplish this, we
apply a constraint transport (CT) method to KT scheme based on
\citet{zie04}, extending it into resistive-MHD case. In a 3D-CT method,
a magnetic field vector is placed at the center of a cubic cell interface
while a numerical flux (or an electric field vector) is at a cell edge
so that $\nabla\cdot\mbox{{\boldmath$B$}}$ does not evolve
\citep{eva88}. The positional relation between them in 2D case are shown in
Fig~\ref{fig.ct}. Due to this placement, the semi-discrete equations
and the numerical fluxes of the induction equations should be different from 
Eqs.~(\ref{eq.kt.semidis}) and (\ref{eq.kt.numfluxf}). With 
numerical fluxes,
$\mbox{{\boldmath$\bar{E}$}}$ = $\mbox{{\boldmath$\bar{F}$}} + 
\mbox{{\boldmath$\bar{G}$}}$, the semi-discrete form of the induction 
equations are written as
{\allowdisplaybreaks
\begin{eqnarray}
\frac{d}{dt}B^x_{i+1/2,j}&=&
-\frac{\Bar{E}^z_{i+1/2,j+1/2}-\Bar{E}^z_{i+1/2,j-1/2}}{\Delta y_{j}},
\nonumber\\
\frac{d}{dt}B^y_{i,j+1/2}&=&
\hphantom{-} 
\frac{\Bar{E}^z_{i+1/2,j+1/2}-\Bar{E}^z_{i-1/2,j+1/2}}{\Delta x_{i}},
\label{eq.kt.bsemidis}\\
\frac{d}{dt}B^z_{i,j}\hphantom{_{+1/2}}&=&
-\frac{\Bar{E}^y_{i+1/2,j}-\Bar{E}^y_{i-1/2,j}}{\Delta x_{i}}
+\frac{\Bar{E}^x_{i,j+1/2}-\Bar{E}^x_{i,j-1/2}}{\Delta y_{j}}.
\nonumber
\end{eqnarray}}
The numerical fluxes of the non-resistive terms are
{\allowdisplaybreaks
\begin{eqnarray}
\Bar{F}^{x}_{i,j+1/2\hphantom{+1/2}}&=&-F^{y(8)}_{i,j+1/2}
\nonumber\\
\Bar{F}^{y}_{i+1/2,j\hphantom{+1/2}}&=&\hphantom{-}F^{x(8)}_{i+1/2,j}
\label{eq.kt.fbar}\\
\Bar{F}^{z}_{i+1/2,j+1/2}&=&\frac{1}{4}
\left(-F^{x(7)}_{i+1/2,j}-F^{x(7)}_{i+1/2,j+1}\right.
\nonumber\\
&&
\hspace{1.2pc}\left.+F^{y(6)}_{i,j+1/2}+F^{y(6)}_{i+1,j+1/2}\right),
\nonumber
\end{eqnarray}}
where $F^{\{x,y\}(m)}$ denotes the m-th component of the vector
$\mbox{{\boldmath$F$}}^{\{x,y\}}$. The interpolations for the
$x$-component and $y$-component of a magnetic field
along the x-direction are given by
\begin{eqnarray}
B^{x+}_{i+1/2,j}&=&B^x_{i+1/2,j},
\nonumber\\
B^{x-}_{i+1/2,j}&=&B^x_{i+1/2,j},
\nonumber\\
B^{y+}_{i+1/2,j}&=&\frac{1}{2}\left[B^y_{i+1,j+1/2}+B^y_{i+1,j-1/2}\right.
\\
&&
\left.-\frac{\Delta x_{i+1}}{2}
\left\{(B^y_{,x})_{i+1,j+1/2}+(B^y_{,x})_{i+1,j-1/2}\right\}\right],
\nonumber\\
B^{y-}_{i+1/2,j}&=&
\frac{1}{2}\left[B^y_{i,j+1/2\hphantom{+1}}+B^y_{i,j-1/2\hphantom{+1}}\right.
\nonumber\\
&&
\left.+\frac{\Delta x_{i\hphantom{+1}}}{2}
\left\{(B^y_{,x})_{i,j+1/2\hphantom{+1}}+(B^y_{,x})_{i,j-1/2\hphantom{+1}}\right\}
\right].
\nonumber
\label{eq.kt.bintpol}
\end{eqnarray}
Those along the y-direction are given similar way to the above. Note again
that interpolations of $B^z$ are same as 
Eq.~(\ref{eq.kt.qintpol}), since in 2D it is defined at a cell center. 

In order to obtain the numerical fluxes of the resistive terms that
appear in the energy equation~(\ref{eq.mhd.eng}) and the induction
equations~(\ref{eq.mhd.far.matrix}), an evaluation of current 
density is required, which we simply give by
{\allowdisplaybreaks
\begin{eqnarray}
j^x_{i,j+1/2\hphantom{+1/2}}&=&
\hphantom{-} \frac{c}{4\pi}
\hphantom{\left[\right].}\frac{B^z_{i,j+1}-B^z_{i,j}}{\Delta 
  y_{j+1/2}},
\nonumber\\
j^y_{i+1/2,j\hphantom{+1/2}}&=&
-\frac{c}{4\pi}
\hphantom{\left[\right].}\frac{B^z_{i+1,j}-B^z_{i,j}}{\Delta
  x_{i+1/2}},
\label{eq.kt.j}\\ 
j^z_{i+1/2,j+1/2}&=&
\hphantom{-} \frac{c}{4\pi}\left[
 \frac{B^y_{i+1,j+1/2}-B^y_{i,j+1/2}}{\Delta x_{i+1/2}}\right.
\nonumber\\
&&\left.\hspace{2.5pc}
-\frac{B^x_{i+1/2,j+1}-B^x_{i+1/2,j}}{\Delta y_{j+1/2}}
  \right].
\nonumber
\end{eqnarray}}
By virtue of these definitions, the divergence-free condition of a current
density is automatically satisfied through a similar logic as the CT
scheme; 
\begin{eqnarray}
&&
 \frac{j^x_{i+1,j+1/2} - j^x_{i,j+1/2\hphantom{+1/2}}}{\Delta
   x_{i+1/2}}
\nonumber\\
&&
+\frac{j^y_{i+1/2,j+1} - j^y_{i+1/2,j\hphantom{+1/2}}}{\Delta y_{j+1/2}}=0.
\end{eqnarray}

The representations for the numerical fluxes of the resistive terms
in the induction equations are straightforward, since a current
density is defined at the same grid position as a numerical flux:
\begin{eqnarray}
\Bar{G}^x_{i,j+1/2\hphantom{+1/2}}&=&\frac{4\pi\eta}{c^2} j^x_{i,j+1/2},
\nonumber\\
\Bar{G}^y_{i+1/2,j\hphantom{+1/2}}&=&\frac{4\pi\eta}{c^2} j^y_{i+1/2,j},
\label{eq.kt.gbar}\\
\Bar{G}^z_{i+1/2,j+1/2}&=&\frac{4\pi\eta}{c^2} j^z_{i+1/2,j+1/2}.\nonumber
\end{eqnarray}
The numerical fluxes of the resistive terms in the energy equation are
given as 
\footnote{
  Numerical fluxes given here are a 
  little different from those found in the original KT scheme. In the
  original scheme the average of cell center values, $B^z_{i}$ and
  $B^z_{i+1}$, are used while we employ the average of left and
  right-interpolated values, 
  $B^{z-}_{i+1/2}$ and $B^{z+}_{i+1/2}$.} 
\begin{eqnarray}
&&G^{x(5)}_{i+1/2,j}(t)=\frac{\eta}{c}\left[
 j^y_{i+1/2,j}\frac{B^{z-}_{i+1/2,j}+B^{z+}_{i+1/2,j}}{2}\right.
\nonumber\\
&&
\hspace{1pc}\left.
-\frac{j^z_{i+1/2,j-1/2}+j^z_{i+1/2,j+1/2}}{2}
\frac{B^{y-}_{i+1/2,j}+B^{y+}_{i+1/2,j}}{2}\right],
\nonumber\\
\label{eq.kt.numfluxg}\\
&&G^{y(5)}_{i,j+1/2}(t)
\nonumber\\
&&\hspace{1pc}
=\frac{\eta}{c}\left[
\frac{j^z_{i-1/2,j+1/2}+j^z_{i+1/2,j+1/2}}{2}
\frac{B^{x-}_{i,j+1/2}+B^{x+}_{i,j+1/2}}{2}\right.
\nonumber\\
&&\left.\hspace{3pc}
-j^x_{i,j+1/2}
\frac{B^{z-}_{i,j+1/2}+B^{z+}_{i,j+1/2}}{2}\right].
\nonumber
\end{eqnarray}

In order to close the equation system (\ref{eq.mhd.mass})$-$(\ref{eq.mhd.amp}),
we further need to know a gravitational potential and the relation
between pressure and other thermodynamic quantities.
The former is done by solving the Poisson 
equation; $\triangle \Phi = 4 \pi G \rho$. In \textit{Yamazakura}, this is
numerically solved by Modified Incomplete Cholesky decomposition
Conjugate Gradient (MICCG) method \citep{gus83}. For the latter, we
adopt a tabulated 
nuclear equation of state produced by \cite{she98a, she98b},
which is commonly used in recent core-collapse simulations; 
e.g. \citet{sum05,mur08,mar09,iwa09}. To derive a pressure
from the EOS table, three thermodynamic quantities should be
specified: in our case, density, specific internal 
energy, and electron fraction are chosen. We do not solve the
evolution of an electron fraction. Instead they are 
assumed a function of density according to \citet{lie05}\footnote{In
  \citet{lie05}, prescriptions not only for an electron fraction, but 
  also for a neutrino stress and entropy change are suggested. In the
  present simulations, we only adopt the prescription for electron
  fraction.}. 
A neutrino transport that may be important in the dynamics of a
core-collapse is not dealt in the present simulations. Since a neutrino
cooling is not considered, only a photo-dissociation of heavy nuclei
takes energy away from shocked matters. Nevertheless, our
core-collapse simulation without magnetic field and rotation still indicate
that no explosion occurs (see \S~\ref{sec.result}). Although a 
neutrino heating is also omitted, it may not be very important since
the present computations are run until at most $\sim 100$~ms after
bounce, a several factor shorter than the heating time scale.

Although we adopt the Liebend{\"o}rfer's prescription for electron
fraction through a whole evolution, it is only valid
until bounce. For example, that prescription does not properly
reproduce a decrease in electron fraction around the neutrino sphere
due to the neutrino burst. A numerical
simulation done by \citet{sum07}, which deals with sophisticated
neutrino physics, shows that a electron fraction around the neutrino
sphere at 100~ms after bounce is $\sim 0.1$, while it is $\sim 0.3$ in
our simulations.  
We have tested in the simulation without magnetic field and rotation at
100~ms after bounce, how much a pressure around the neutrino sphere
varies when the electron fraction is replaced from the the
Liebend{\"o}rfer's value to 0.1, and found that the difference is at
most only 20~\%. Note that in the present simulations, an electron
fraction may influence dynamics only through a pressure and sound
speed, where the latter is just related to the strength of a numerical
diffusion. 

\textit{Yamazakura} has passed several numerical test problems, which are
demonstrated in Appendix.

\section{Computational Setups}\label{sec.setup}
We follow the collapse of the central 4000~km core of a 15~$M_{\odot}$
star provided by S. E. Woosley (1995, private communication).  
To construct the initial condition of the core, the density and temperature
distributions are taken from the stellar data. Note that the initial
profile of electron fraction is determined using the prescription by
\citet{lie05} as mentioned above. In order that the collapse
proceeds in the presence of strong magnetic field and 
rapid rotation, a temperature of the initial core is reduced as
\begin{eqnarray}\label{eq.model.temp}
T(r)=T_{\textrm{org}}(r)\left(1-\frac{r_{\textrm{T}}^2}{r_{\textrm{T}}^2+r^2}\right)
\end{eqnarray}
where $r$ is the distance from the center of the core,
$T_{\textrm{org}}(r)$ is an original temperature of the core, and 
$r_{\textrm{T}}$ is taken 1000~km 
\footnote{In what follows we denote a spatial point in the polar and
  cylindrical coordinate by $(r,\theta,\phi)$ and $(\varpi,\phi,z)$,
  respectively.}. 
The initial internal energy distribution is
obtained by the EOS table, using density, electron fraction,
and temperature as the three parameters. Radial velocities are initially
assumed to be zero. Magnetic field and rotation are initially put by
hand into the core. We assume that the initial magnetic field is purely
dipole-like, and the core is either rotating or non-rotating.  

The dipole-like magnetic field is produced by putting a toroidal
electric current of a 2D-Gaussian-like 
distribution centered at $(\varpi,z)=(\varpi_0,0)$,
\begin{eqnarray}\label{eq.model.dipole1}
j_\phi(\varpi,z)=j_0e^{-\tilde{r}^2/2\sigma^2}
\left(\frac{\varpi_0\varpi}{\varpi_0^2+\varpi^2}\right)
\end{eqnarray}
where $\tilde r = \sqrt{(\varpi-\varpi_0)^2+z^2}$.
The last factor is multiplied to impose $j_\phi=0$ along the pole. A width
$\sigma$ is a function of $\tilde \theta \equiv \arccos (z/\tilde r)$,
defined by 
\begin{eqnarray}\label{eq.model.dipole2}
\sigma(\tilde \theta)=
\frac{\tilde r_{\textrm{dec}}}{\sqrt{1-e^2\cos{\tilde \theta}}},
\end{eqnarray}
which traces a prolate ellipse centered at $(\varpi,z)=(\varpi_0,0)$ with
an eccentricity $e$ and major radius $\tilde r_{\textrm{dec}}$. Parameters in 
Eqs.~(\ref{eq.model.dipole1}) and (\ref{eq.model.dipole2}) are put as
$\varpi_0=1000$~km, $\tilde r_{\textrm{dec}}=710$~km and
$e=0.5$ in every computation. A parameter $j_0$, which determine
the field strength, is given later. From the above electric current
distribution, the vector potential is calculated by 
\begin{eqnarray}\label{eq.model.apot}
&&A_\phi(\varpi,z)
\\
&&
=\frac{1}{c}
\int_0^{\varpi_{\textrm{core}}}\int_0^{2\pi}\int_0^{z_{\textrm{core}}}
\frac{j_\phi(\varpi_c,z_c)\cos\phi_c}{R(\varpi,z,\varpi_c,\phi_c,z_c)}
\varpi_c d\varpi_c d\phi_c dz_c
\nonumber
\end{eqnarray}
where $R(\varpi,z,\varpi_c,\phi_c,z_c)$ is the distance between
$(\varpi,0,z)$ and $(\varpi_c,\phi_c,z_c)$, and
$\varpi_{\textrm{core}}=z_{\textrm{core}}=4000$~km. The magnetic
field is obtained via 
$\mbox{{\boldmath$B$}}=\nabla\times\mbox{{\boldmath$A$}}$. Note that,
evaluating $A_\phi$ at a cell corner, the initial magnetic
field automatically satisfies the divergence free condition as the
same way in the CT scheme. The initial magnetic field configuration
and the distribution of magnetic energy per unit mass are shown in
Fig~\ref{fig.binit}.

\begin{figure}
\epsscale{1}
\plotone{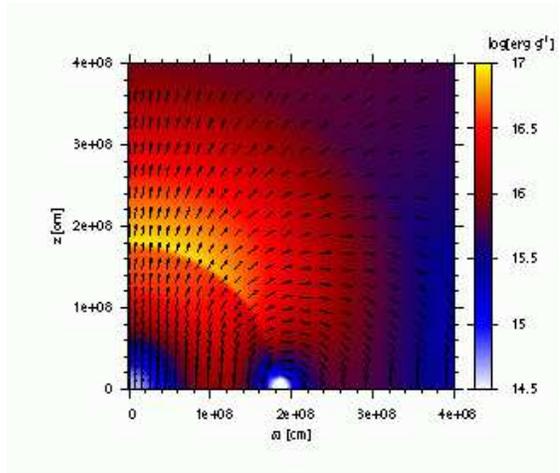}
  \caption{Initial magnetic field configuration (vector) and
    distribution of specific magnetic energy in logarithmic scale for
    a representative model (B\textit{s}-$\Omega$-$\eta_{-\infty}$; see
    text). Note that all models share the same initial
    field configuration.} 
 \label{fig.binit}
\end{figure}

In each rotating model, an initial angular velocity is given by 
\begin{eqnarray}
\Omega(r)=\Omega_{0}\frac{r_0^2}{r_0^2+r^2}.
\end{eqnarray}
where $r_0=1000$~km and $\Omega_{0}=3.9$~rad s$^{-1}$.

Employing the above magnetic field and rotation, we study three
different cases, namely, strong magnetic field and rapid
rotation (model-series B\textit{s}-$\Omega$), moderate 
magnetic field and rapid rotation (model-series
B\textit{m}-$\Omega$), and very strong magnetic field and no rotation
(model-series B\textit{ss}-$\mho$). Parameters for each model-series
are given in Table~1.  

\begin{table}
\label{tab.model}
\begin{center}
\caption{Parameters and some results for the computed models.}
\begin{tabular}{lccccc}
\tableline\tableline

Model name & $E_{\rm{m}}/W$\tablenotemark{a} [\%] &
$T/W$\tablenotemark{b} [\%] & $j_0$ [cgs-Gauss] & 
$B_{0,\rm{i}}\tablenotemark{c} [G]$ \\
\tableline
Bs-Rot     & 0.5   & 0.5 & $5.4\times10 ^{14}$ &
$9.7\times 10^{12}$\\
Bm-Rot     & 0.05  & 0.5 & $1.7\times10 ^{14}$ &
$3.1\times 10^{12}$\\
Bss-Nonrot & 5.0   & 0.0 & $1.7\times10 ^{15}$ &
$3.1\times 10^{13}$\\
\tableline
\end{tabular}
\tablenotetext{a}{Initial ratio of magnetic energy to gravitational energy.}
\tablenotetext{b}{Initial ratio of rotational energy to gravitational energy.}
\tablenotetext{c}{Initial magnetic field strength at the center.}
\end{center}
\end{table}

In order to study effects of resistivity on the dynamics, both an
ideal model and resistive models are run in each model-series. For the 
resistive models, we examine two different values of resistivity, say
$\eta=10^{13}$ and $10^{14}$~cm$^2$~s$^{-1}$, reminding the
discussion in~\S~\ref{sec.intro}. Resistivity is uniform in space
and time except that it is set zero inside the radius of 10~km to
save the computational time. There a magnetic field and thus a
resistivity are expected to be unimportant due to high density.
For a descriptive convenience, we use abbreviations $\eta_{-\infty}$,
$\eta_{13}$, and $\eta_{14}$, respectively, for models with $\eta=0$,
$10^{13}$ and $10^{14}$~cm$^2$~s$^{-1}$, attaching after name of a
model-series introduced above. For example, a 
model with strong magnetic field, rapid rotation, and
$\eta=10^{14}$~cm$^2$~s$^{-1}$ is referred to as model
B\textit{s}-$\Omega$-$\eta_{14}$.   

Each computation is done in cylindrical coordinate. Assuming
the axisymmetry and equatorial symmetry, we take the numerical domain as
$(\varpi,z)\in[0~\textrm{km},4000~\textrm{km}]
\times[0~\textrm{km},4000~\textrm{km}]$. 
Until the central density reaches $10^{12}$~g~cm$^{-3}$, the number
of numerical cells is $N_{\varpi}\times N_z=320\times 320$.
After that the number of cells is changed to $N_{\varpi}\times
N_z=720\times 720$. There, the spatial width of a cell increases outward
in each direction with a constant ratio of 1.0051 and 1.0056, before
and after the re-griding, respectively. Both the width $\Delta
\varpi$ of the inner-most cells of $\varpi$-coordinate and the width
$\Delta z$ of the inner-most cells of $z$-coordinate are 5~km and
400~m, before and after the re-griding, respectively. When the
numerical cells are redistributed, physical variables are linearly
interpolated from the coarse into the fine cells. In this procedure,
the divergence-free constraint of magnetic 
field is usually violated, which stems from the poloidal components. To avoid
this, we calculate $A_\phi$ from Eq.~(\ref{eq.model.apot}) using the
distribution of $j_\phi$ right after the cell redistribution, and  
then obtain a divergence-free poloidal magnetic field. 

Each simulation is run until the shock front reaches a radius of
2100, 3000, and 2300~km in model series B\textit{s}-$\Omega$,
B\textit{m}-$\Omega$,  B\textit{ss}-$\mho$, respectively.

\section{Results}\label{sec.result}
In this section, we will present results of our computations for each
model-series separately, seeing how a resistivity affect the dynamics
of a magnetized supernova. Particular attentions are paid to the
explosion energy, magnetic field amplification, and the aspect 
ratio of the ejecta.

Before proceeding to the main results, we here describe the dynamical
evolution in the simulation without magnetic field and rotation. Soon
after the start of computation, the core has a negative radial velocity
everywhere, and collapses towards the center. A bounce occurs at
$t=133$~ms due to nuclear force, and a shock wave is generated. The
shock wave propagating outward first stalls around $r\sim 200$~km, but
it starts gradual expansion around 165~ms. Afterward, the shock
surface alternatly expands and shrinks. We followed the evolution until
$t=350$~ms (217~ms after bounce) during which the maximum shock
position is $\sim 800$~km. In this way, our model without magnetic
field and rotation does not result in a stalled shock as seen
recent core-collapse simulations (see e.g. Fig.~2 of \citet{nor10}),
which may be because we only consider a photo-dissociation of heavy
nuclei but no neutrino cooling as cooling
processes. Fig.~\ref{fig.t-eng.sph} shows the evolutions of total,
internal, gravitational, positive kinetic, and negative kinetic
energies in the simulation without magnetic field\footnote{Here, the
  terms "positive kinetic energy" and "negative kinetic energy" mean the
  kinetic energy associated with fluid elements with a positive and negative
  radial velocity, respectively.}. This indicates that a part of fluid
has a positive radial velocity. Nontheless, we found that an estimated
explosion energy is very small; less than $\sim 10^{48}$~erg and
sometimes exactly zero. We assume that a fluid element is exploding if
the total fluid energy plus the gravitational 
potential energy at its position, and the radial velocity are both
positive, i.e. $e+\rho v^2/2+B^2/8\pi+\rho\Phi>0$ and $v_r>0$. Then
the explosion energy is obtained by the sum of the total energy of
fluid elements that fulfill the criterion plus the gravitational
potential energy for the exploding fluid. We calculate the latter by
$E_{\textrm{exp,grv}}=\int_{V_\textrm{exp}}[\rho\Phi_{\textrm{exp}}/2+\rho\tilde{\Phi}]dV$, 
where $\Phi_{\textrm{exp}}$ is the gravitational potential due to the
exploding fluid and $\tilde{\Phi}$ is that due to the other fluid, and 
$\Phi=\Phi_{\textrm{exp}}+\tilde{\Phi}$. Gravitational potentials
$\Phi_{\textrm{exp}}$ and $\tilde{\Phi}$ are obtained by solving a
Poisson equation 
with only the mass of the exploding fluid and non-exploding fluid,
respectively. The fact that the explosion
enrgy is quite small as mentioned above implies that most or all fluid
elements including those with a positive radial velocity do not
fulfill the criterion. It seems reasonable to assume that the
model without 
magnetic field and rotation \textit{does not} explode. 

The error in
the total energy conservation of the system is 51~\% at the end of the
simulation. We found that the error in the total energy
conservation is 24-33~\% at the end of the all simulations involving
magnetic field, except that it is 14~\% in a different resolution run
for model B\textit{s}-$\Omega$-$\eta_{14}$
described in \S~\ref{sec.res}. In \S~\ref{sec.res}, we will discuss
whether these errors are problematic for results presented in this
paper. 

\begin{figure}
\epsscale{1}
\plotone{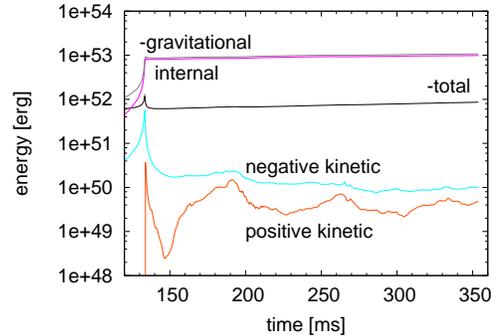}
  \caption{Evolutions of the total (black line), internal (magenta),
    gravitational (gray), positive kinetic (orange), and negative
    kinetic (cyan) energy integrated over the whole numerical domain
    for the simulation without magnetic field and rotation. The total
    and gravitational energy are multiplied by $-1$.} 
 \label{fig.t-eng.sph}
\end{figure}

\subsection{Strong Magnetic Field and Rapid Rotation --- Model Series
  B\textit{s}-$\Omega$}\label{result1}

\begin{figure*}
\epsscale{1}
\plotone{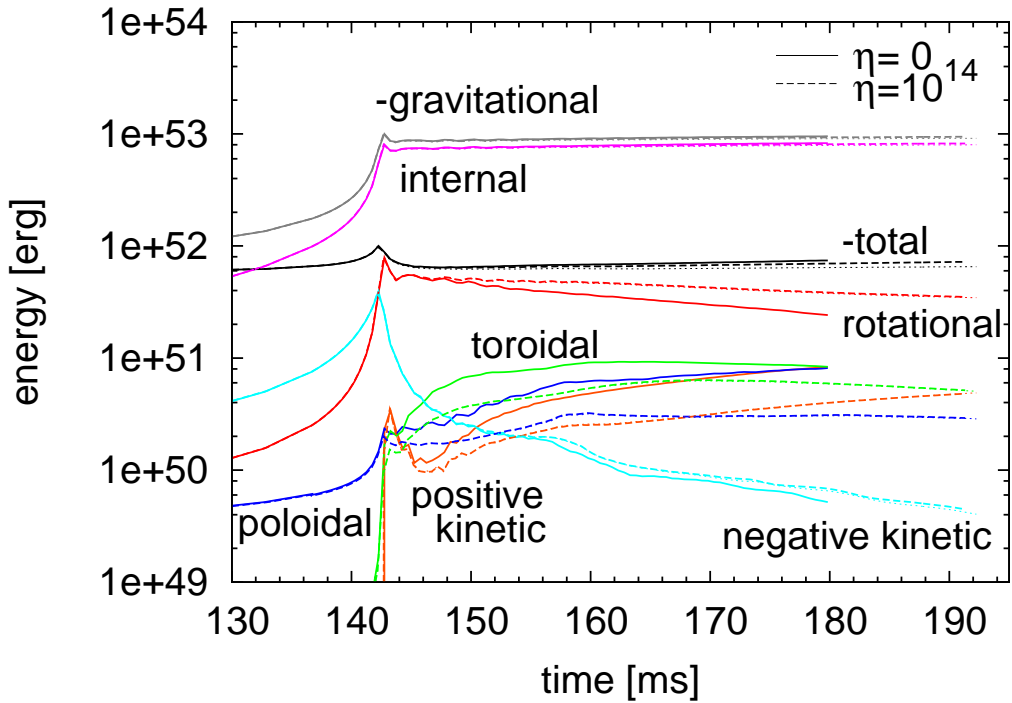}
  \caption{Evolutions of the total (black lines), internal (magenta),
    gravitational (gray), rotational (red), positive kinetic (orange),
    negative kinetic (cyan), poloidal magnetic (blue), and toroidal
    magnetic (green) energy integrated over the whole numerical
    domain. The total and gravitational energy are multiplied by
    $-1$. The solid and dashed  
  lines are drawn for model B\textit{s}-$\Omega$-$\eta_{-\infty}$ 
  and B\textit{s}-$\Omega$-$\eta_{14}$, respectively. The dotted lines,
  which are almost identical with the dashed lines, are for
  a different resolution run for model B\textit{s}-$\Omega$-$\eta_{14}$
  (see \S~\ref{sec.res})}
 \label{fig.t-eng.55}
\end{figure*}

We start from briefly describing the dynamical evolution in model
B\textit{s}-$\Omega$-$\eta_{-\infty}$. In this model, a rotation
hampers the collapse, and a bounce occurs at $t=143$~ms, 10~ms later
than in the case without magnetic field and rotation. During the
collapse, the core is largely spined up accompanied with an increase
in the degree of differential rotation: Just after bounce, a
rotational period reaches $\sim 1$~ms in a considerable part inside
the radius of 50~km, while it is initially at least
$\sim 1$~s.  A magnetic field, which initially plays little role, is 
greatly amplified by compression during collapse. In addition, the
differential rotation winds poloidal magnetic field-lines, and the
toroidal component of magnetic field is largely generated around the time of
bounce. An outward matter motion driven by bounce 
first decelerates, losing the energy due to a photo-dissociation of
heavy nuclei, but accelerates again helped by a  
magnetic pressure of toroidal field and centrifugal force. As a
result, a strong eruption of 
matter occurs preferentially along the pole. The above dynamical
sequence can be followed in Fig.~\ref{fig.t-eng.55} in terms of
energetics (see thick lines): i.e. a decrease of the gravitational
energy results in an increase of the rotational and magnetic energy;
the rotational energy is partially converted into the
toroidal magnetic energy; then the toroidal magnetic energy is
consumed to boost the positive kinetic 
energy. The top panels of Fig.~\ref{fig.vradbvec.55} shows the
distributions of velocity and magnetic field at 164~ms (21~ms after
bounce) for model B\textit{s}-$\Omega$-$\eta_{-\infty}$. A fast mass
eruption ($v_r \gtrsim 5\times 10^9$~cm~s$^{-1}$) is seen notably
around the pole, where the ratio of a magnetic pressure to matter
pressure is large. This also implies that magnetic force plays an
essential role for a fast mass eruption. Note that the dynamical
features described here is quite similar to those found in previous
works that employ similar strengths of magnetic field and rotation
(see e.g. \citet{yam04}, \citet{tak09}).

\begin{figure*}[t]
\begin{center}
\epsscale{1}
\plottwo{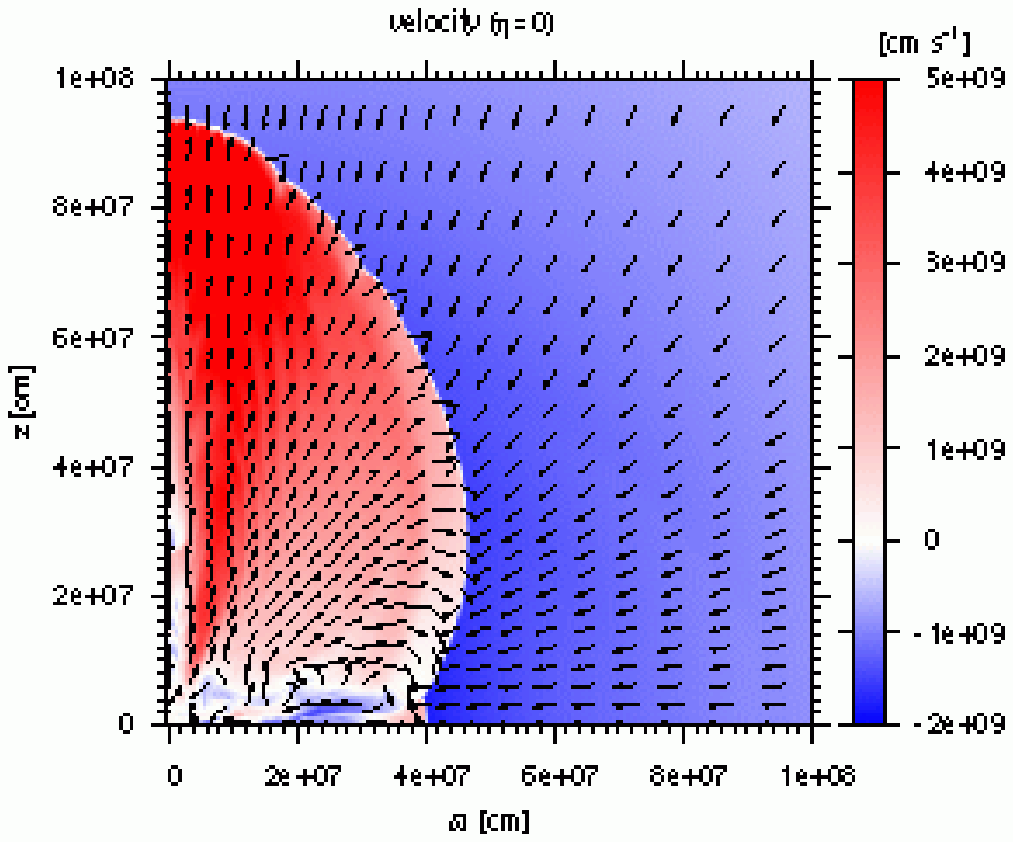}{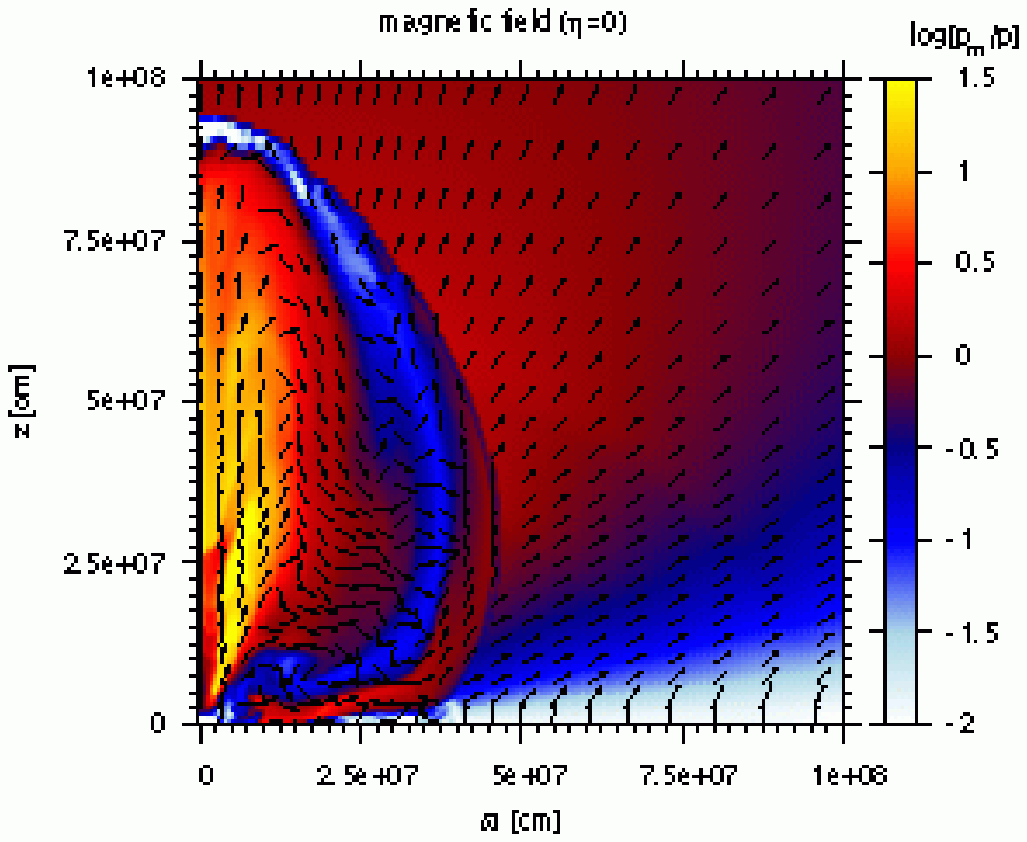}
\plottwo{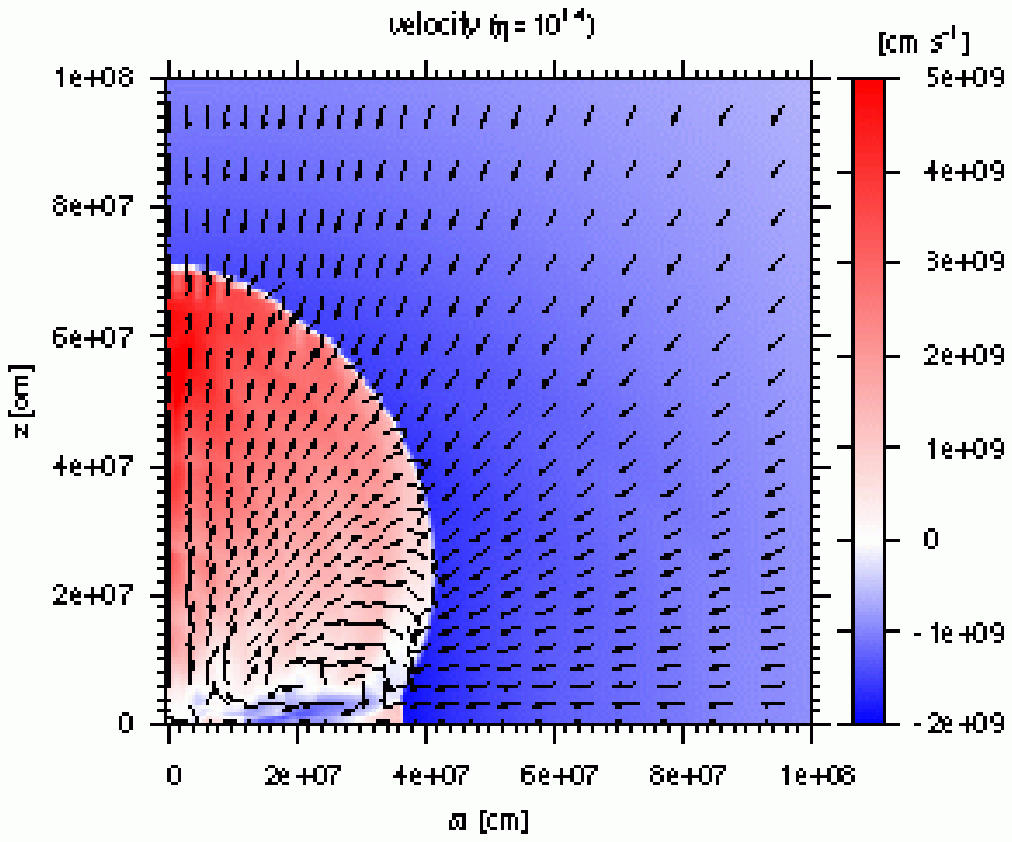}{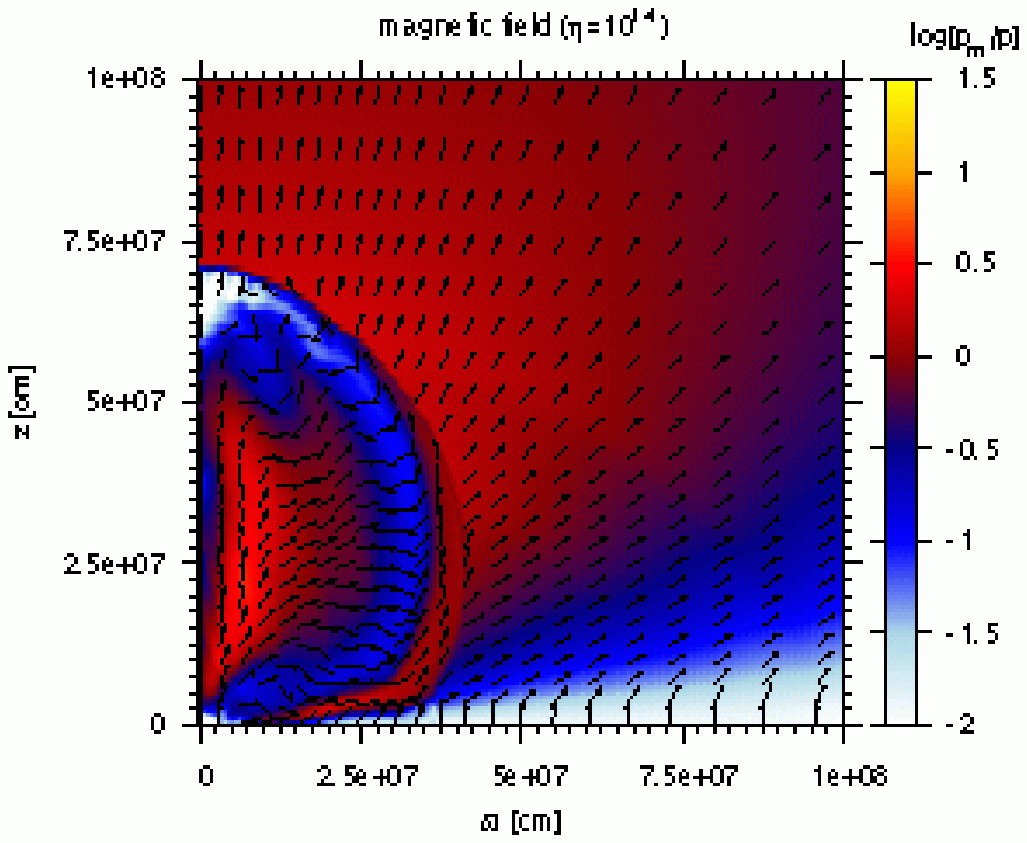}
  \caption{\textit{Left panels}: Distributions of radial
    velocity magnitude (color) and velocity direction
    (vectors). \textit{Right panels}: Distributions of ratio of
    magnetic pressure to matter pressure, $p_{\textrm{m}}/p$, in
    logalithmic scale (color), and magnetic 
    field direction (vectors). These figures are depicted at
    $t=164$~ms for model B\textit{s}-$\Omega$-$\eta_{-\infty}$ (upper
    panels) and B\textit{s}-$\Omega$-$\eta_{14}$ (lower panels).} 
 \label{fig.vradbvec.55}
\end{center}
\end{figure*}

In a resistive model B\textit{s}-$\Omega$-$\eta_{14}$, the evolution
proceeds in qualitatively similar way to the ideal model
B\textit{s}-$\Omega$-$\eta_{-\infty}$.
However, outgoing velocities are relatively slow compared with the
ideal model, which result in a smaller shock radius at a same physical
time (compare the left panels of Fig.~\ref{fig.vradbvec.55}). 
The right panels of Fig.~\ref{fig.vradbvec.55} implies that
this is due to a less strong magnetic pressure in model
$\eta_{14}$. As easily expected, a magnetic field amplification by
differential rotation is ineffective under the presence of
resistivity. This means that the rotational energy cannot be spent
efficiently as an 
energy source for the explosion. Indeed, it is observed in
Fig.~\ref{fig.t-eng.55} that the rotational energy in model
$\eta_{14}$ decreases slowly compared with that of model
$\eta_{-\infty}$ as well as both the positive kinetic and magnetic energy
increase slowly. 

\subsubsection{Explosion Energy}\label{sec.exp.55}
Below, we will see the effect of resistivity on the explosion energy
together with the detailed mechanism of the explosion. 
Fig.~\ref{fig.t-exp.55} shows the evolutions of the 
explosion energies, $E_{\textrm{exp}}$, in model-series
B\textit{s}-$\Omega$. It is found that a larger resistivity results
in a smaller explosion energy.  

\begin{figure}
\epsscale{1}
\plotone{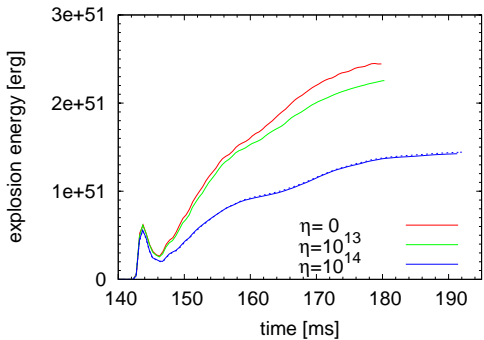}
\caption{Evolutions of the explosion energies in model-series
  B\textit{s}-$\Omega$. The red, green, and blue-solid lines are for
  model $\eta_{-\infty}, \eta_{13},$ and $\eta_{14}$, respectively. The
  blue-dotted line is for a different resolution run for model
  B\textit{s}-$\Omega$-$\eta_{14}$ (see \S~\ref{sec.res}). For the
  definition of an explosion energy, see text. }  
\label{fig.t-exp.55}
\end{figure}

As a preparation for detailed analyses, we consider dividing
the volume inside the shock surface into the two parts, say, the
\textit{eruption-region} and the \textit{infall-region}. The
definition of these parts are as follows. First, the volume inside the
shock surface is equally cut up with respect to 
$\theta$ into 30 volume segments with $\Delta\theta=3^\circ$ opening
angle. The eruption-region is defined by the sum of the segments whose
integrated radial momentum is positive, whereas the infall-region is
by the sum of those having the negative radial momentum. For example,
in each left panel of Fig.~\ref{fig.vradbvec.55}, the infall-region
appears in the vicinity of the equator ($\theta\gtrsim 70^\circ$) where
a negative momentum dominates over a positive one, while the other
part inside the shock surface corresponds to the eruption-region. 

\begin{figure*}
\epsscale{0.8}
\plotone{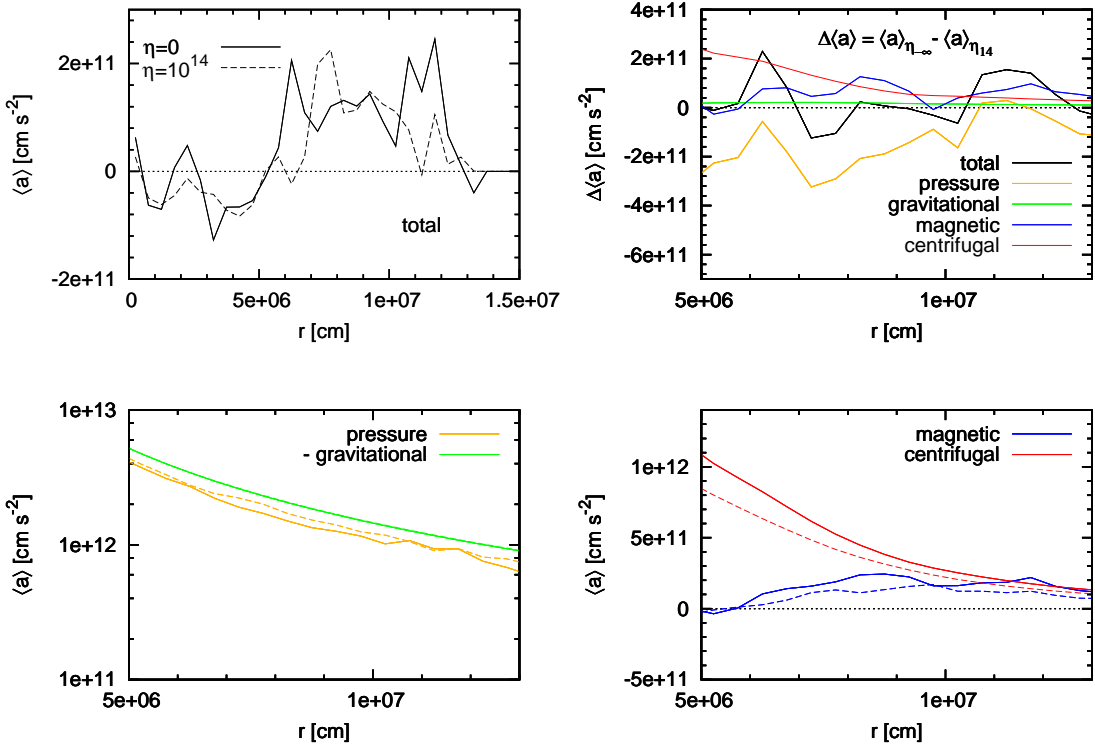}
\caption{Radial distributions of the radial accelerations,
  $\langle a \rangle$, 
  angularly averaged in the eruption-region and time averaged during
  $t=147-155$~ms, in model B\textit{s}-$\Omega$-$\eta_{-\infty}$ and
  B\textit{s}-$\Omega$-$\eta_{14}$. In each panel, the
  black, red, blue, and magenta lines represent the total,
  pressure, gravitational, magnetic, and centrifugal accelerations,
  respectively. Except for the right-top panel, the solid lines are
  for model $\eta_{-\infty}$, while the dashed lines are for model
  $\eta_{14}$. The right-top panel shows the  
  differences in the accelerations, $\Delta \langle a \rangle \equiv
  \langle a \rangle_{\eta_{-\infty}}-\langle a \rangle_{\eta_{14}}$,
  subtracting that of model~$\eta_{14}$ from that of model~$\eta_{-\infty}$. }  
\label{fig.force.55}
\end{figure*}

\begin{figure}
\begin{center}
\epsscale{1}
\plotone{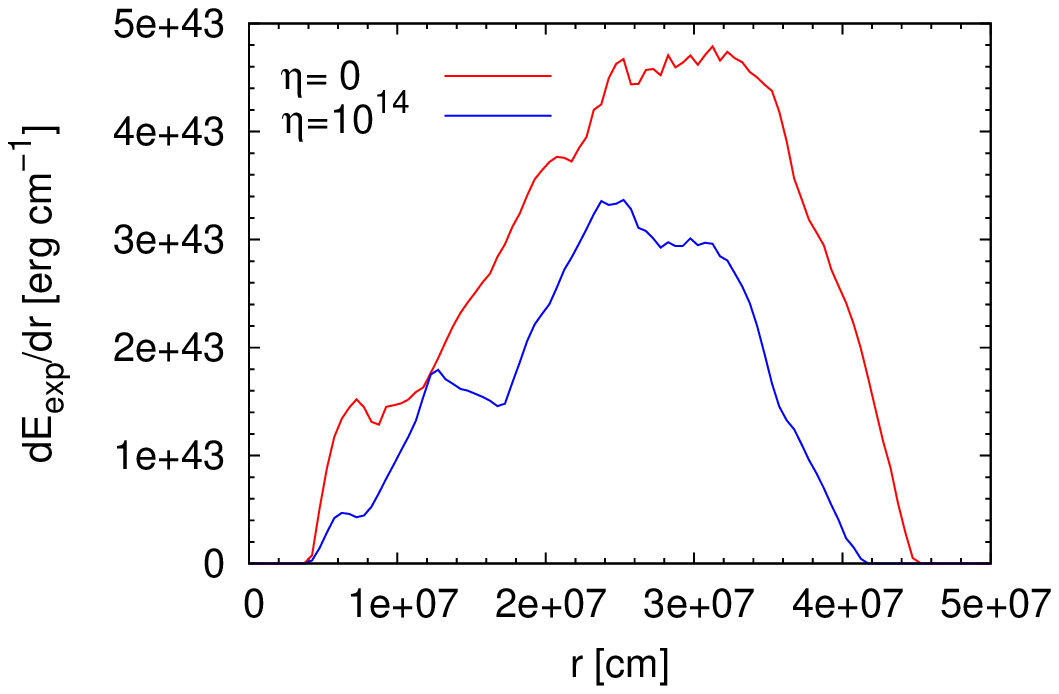}
\caption{Radial distributions of $dE_{\textrm{exp}}/dr$ at 155~ms
  in model B\textit{s}-$\Omega$-$\eta_{-\infty}$ and
  B\textit{s}-$\Omega$-$\eta_{14}$.} 
\label{fig.expr.55}
\end{center}
\end{figure}

\begin{figure}
\epsscale{1}
\plotone{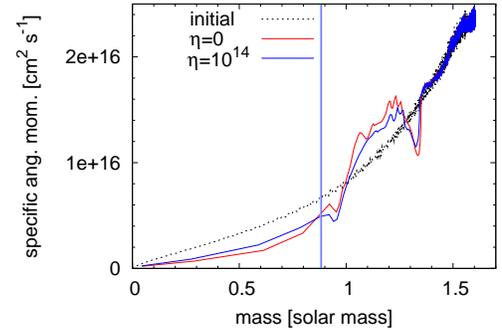}
\caption{Distributions of specific angular momentums in spherical
  mass coordinate at $t=155$~ms in model
  B\textit{s}-$\Omega$-$\eta_{-\infty}$ and 
  B\textit{s}-$\Omega$-$\eta_{14}$. The Black-dashed
  line shows the initial distribution shared by the two models. The
  magenta and cyan vertical lines show the mass that
  corresponds to $r=25$~km, respectively, in model $\eta_{-\infty}$ and
  $\eta_{14}$.}  
\label{fig.angmom.55}
\end{figure}

To know what causes the differences in the explosion energy, it may be
helpful to compare between model B\textit{s}-$\Omega$-$\eta_{-\infty}$
and B\textit{s}-$\Omega$-$\eta_{14}$ by the 
individual acceleration terms in the $r$-component of the equation of
motion written on the frame rotating around the pole with the angular
velocity $\Omega$; 
\begin{eqnarray}
\frac{D'v_r}{Dt}&=&
  -\frac{1}{\rho}\frac{\partial p}{\partial r}
  -\frac{\partial \Phi}{\partial r} 
  +\frac{1}{\rho c}\left(j_\theta B_\phi - j_\phi B_\theta\right)\nonumber\\
&&+\Omega^2 r\sin^2\theta,\label{eq.eom}
\end{eqnarray}
where $D'/Dt$ denotes the Lagrangian derivative on the rotating
frame. In the r.h.s. of Eq.~(\ref{eq.eom}), each term represents, from
left to right, the acceleration due to a pressure, gravity, magnetic
field, and rotation. In comparing the accelerations, we take
an angular average in the eruption-region and a time average during
$t=t_1$~-~$t_2$~ms, which are defined by 
\begin{eqnarray}\label{eq.acc}
\langle a \rangle (r)= \int^{t_2}_{t_1}\left[
   \frac{\int_{\textrm{erup}} a \rho \sin\theta d\theta}
        {\int_{\textrm{erup}}\rho \sin\theta d\theta}
                                   \right]
   dt\bigg/\left[t_2-t_1\right],
\end{eqnarray}
where each term in the r.h.s. of Eq.~(\ref{eq.eom}) is to be assigned
to $a$.
Fig.~\ref{fig.force.55} shows the radial distributions of
accelerations, angularly averaged in the eruption-region and
time averaged during $t=147$-$155$~ms, in model $\eta_{-\infty}$ and
$\eta_{14}$. The averages are taken inside
$\sim$140~km, the smallest shock radius at 147~ms in model
$\eta_{-\infty}$ and $\eta_{14}$. It is found that, in the both models, 
an acceleration is almost everywhere positive for $r\gtrsim 50$~km
(left-top panel). As shown in Fig.~\ref{fig.expr.55}, a matter
ejection also occurs roughly in this radial range, which implies that
the amplitude of an acceleration is a good measure for the resulting
magnitude of the explosion energy.
There, a pressure acceleration alone is always
smaller than a gravitational deceleration (left-bottom panel).
The right-bottom panel shows that it
is a magnetic and centrifugal acceleration that makes a matter
ejection possible.

In the right-top panel of Fig.~\ref{fig.force.55}, the differences in
accelerations between the two models are plotted, which shows that a
total acceleration is averagely larger in model~$\eta_{-\infty}$ for
$r\gtrsim 50$~km. It is likely that this causes a larger
$dE_{\textrm{exp}}/dr$ in model~$\eta_{-\infty}$ as observed in
Fig.~\ref{fig.expr.55}. The right-top panel also
indicates that a larger total acceleration in model
model~$\eta_{-\infty}$ is primary due to that in a magnetic and
centrifugal acceleration. Although, a pressure acceleration is smaller
in model model~$\eta_{-\infty}$, this is more than compensated by
them. Therefore, we conclude that a resistivity makes the explosion
less energetic due to a small magnetic and centrifugal acceleration.   

\begin{figure*}
\epsscale{1}
\plottwo{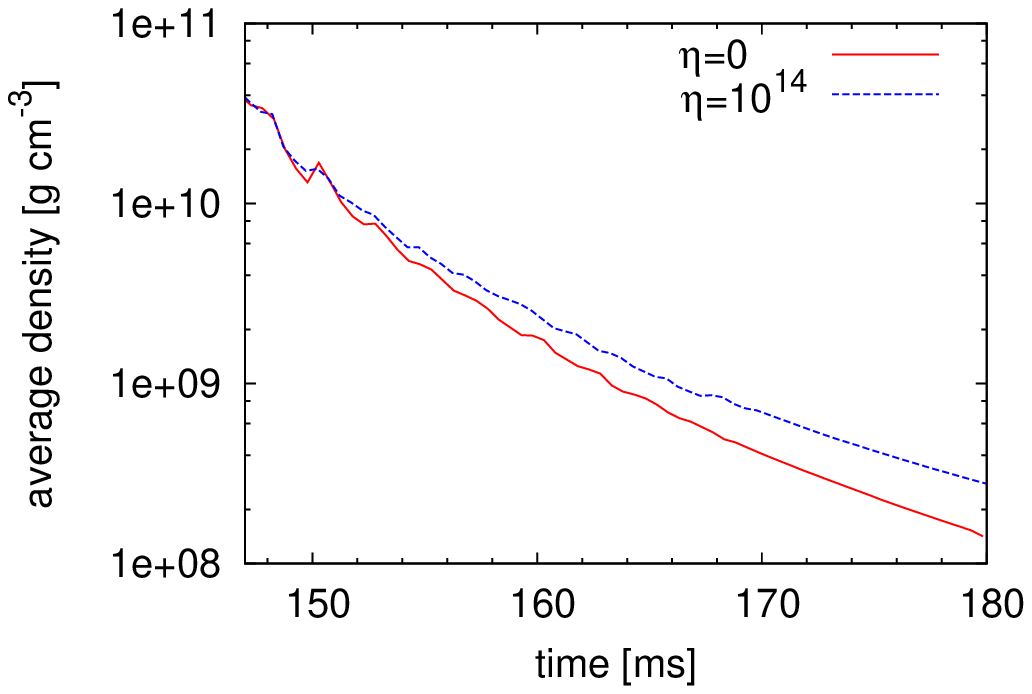}{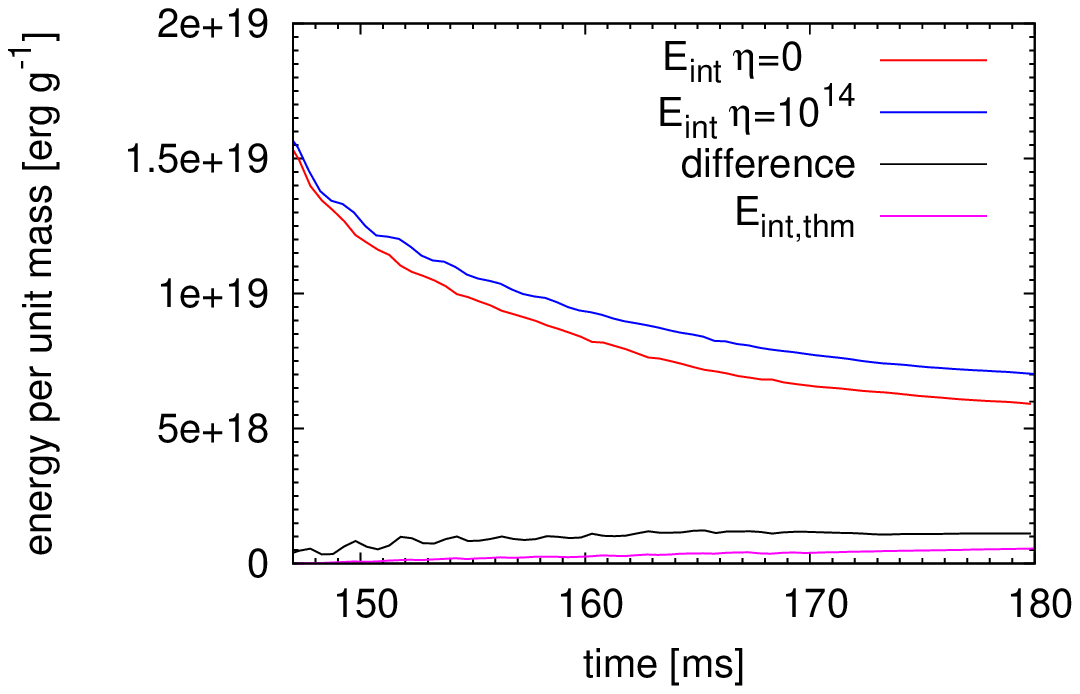}
\caption{\textit{Left}: Evolution of the average density for $r>50$~km
  in the
  eruption-region in model B\textit{s}-$\Omega$-$\eta_{-\infty}$ and 
  B\textit{s}-$\Omega$-$\eta_{14}$. \textit{Right}: Evolution of the
  average specific internal energy for $r>50$~km in the
  eruption-region in model $\eta_{-\infty}$ (red line), model
  $\eta_{14}$ (blue), and the difference between 
  them (black). The magenta line shows the time integrated Joule
  heating produced in the above region for model $\eta_{14}$. }
\label{fig.de1}
\end{figure*}

\begin{figure}
\epsscale{1}
\plotone{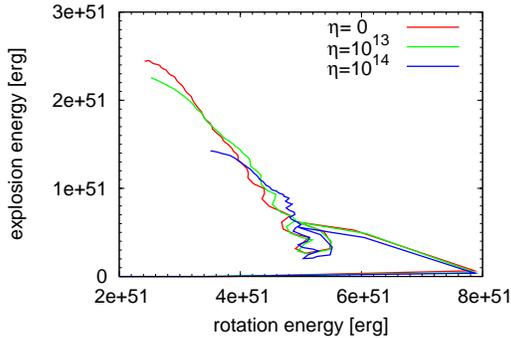}
\caption{Evolutions of the explosion energies in
  model-series B\textit{s}-$\Omega$ in terms of the remaining
  rotational energy. The evolution proceeds roughly in the
  counterclockwise direction.} 
\label{fig.erot-eexp.55}
\end{figure}

While a smaller magnetic acceleration in model $\eta_{14}$ will be
simply due to a weaker magnetic field as a result of a magnetic
diffusion, that of centrifugal acceleration is 
related to a less efficient angular momentum transfer owing to a weaker
magnetic stress. In Fig.~\ref{fig.angmom.55}, the distributions of
specific angular momentums at 155~ms is plotted in mass coordinate. It
is observed that a specific angular momentum in model $\eta_{14}$ is
larger than that of model $\eta_{-\infty}$ inside the radius of $\sim
25$~km, whereas it is smaller 
outside there. This is the consequence of a less efficient outward
angular momentum transfer in model $\eta_{14}$. Then, at a large
radius, a centrifugal acceleration in 
model $\eta_{14}$ is smaller than that in model $\eta_{-\infty}$.

We also examined reasons for a larger pressure acceleration in model
$\eta_{14}$ observed in the right-top panel of
Fig.~\ref{fig.force.55}. 
In the present situation, a pressure is a increasing function
of density and specific internal energy. Although a pressure also
depends on electron fraction, this dependence is very weaker than that
on the above two quantities. Hence, it is expected that the
density or specific internal energy volume-averaged over the
eruption-region are larger in model
D\textit{s}-$\Omega$-$\eta_{14}$. By calculating these two average 
values, we found that both the average density and
specific internal energy are larger for  model $\eta_{14}$ (see
Fig.~\ref{fig.de1}). A higher density in model $\eta_{14}$ implies that the
matter expansion rate is smaller, which is consistent with what is
observed in the left panels of Fig.~\ref{fig.vradbvec.55}. A larger
specific internal energy in model $\eta_{14}$ also may come from the
smaller expansion rate, but may be caused by Joule heating too. To
estimate an amount of thermal energy produced by Joule heating in model
D\textit{s}-$\Omega$-$\eta_{14}$, the total heating rate of the 
eruption-region, $\int_{\textrm{erup}}4\pi \eta j^2 /(\rho c^2)dm$, is
time-integrated from 147~ms when the infall-region begins to appear
clearly. Then it is divided by a mass of the expulsion region at each 
instant of time, and compared with the difference in average
specific internal energy between the two models. The result is shown in
the right-panel of Fig.~\ref{fig.de1}, which indicates that the
contribution form the Joule heating to the difference in the specific
internal energy is quite small around 150~ms.
Thus, it is not likely that a larger pressure acceleration in model
$\eta_{14}$ observed in the right-top panel of Fig.~\ref{fig.force.55}
is caused by Joule heating. A larger pressure acceleration seems to be
just due to a smaller expansion rate in model $\eta_{14}$. The panel
implies that only in a later 
phase, a larger specific internal energy in model $\eta_{14}$ may
be somewhat contributed by the Joule heating.
Note, however, that the thermal energy estimation made here may be crude,
since a part of thermal energy produced in one region may migrate into
another.

As we have seen, an inefficiency both in a magnetic field 
amplification and an angular momentum transfer leads to a weaker
explosion in a resistive model. From the energetical point of view, this
corresponds to an inefficiency in consuming the rotational energy as a
fuel. Then one may think that the explosion energy of the 
ideal model and a resistive model would be similar compared in terms of
the consumed rotational energy. If this is the case, the final
explosion energy in the three models, after all the available
rotational energy has been drained, would be comparable to each
other. According to Fig.~\ref{fig.erot-eexp.55} this is not 
ture, however. In this figure the evolutions of the explosion energies
in the three models are plotted against the remaining rotational
energy. Since the maximum rotational energy, which is reached at the
time of bounce, is almost same among the three models,
$E_{\textrm{rot}}\approx 7.9\times 10^{51}$~erg, each model will consume
roughly the same amount of rotational energy with a same position in
the abscissa. It is found that the explosion energy 
in a resistive model is smaller than that of the ideal model, even
though a same amount of rotational energy is expended. This implies
that a part of the rotational energy is wasted in locations where the
criterion for the explosion is not fulfilled. 

\subsubsection{Magnetic Field Amplification}\label{sec.mfa.55}
In this section we analyze a magnetic field amplification. In
Fig.~\ref{fig.ethm.55}, the angular distributions of the magnetic
energies per unit mass, averaged over $50$~km$<r<0.9\times r_{\textrm{sh}}$
at $t=145$~ms (2~ms after bounce) and $t=160$~ms (17~ms after bounce),
are shown for model B\textit{s}-$\Omega$-$\eta_{-\infty}$ and
B\textit{s}-$\Omega$-$\eta_{14}$. The left panel indicates
that the total magnetic energy is relatively stronger around the
pole at 145~ms in each model, reflecting the initial magnetic field
configuration (see Fig.~\ref{fig.binit}). Turning to the right panel,
it is found that, in each model, the contrast between the total
magnetic energy around the pole and that around $\theta\sim
40^\circ$-$70^\circ$ becomes stronger at 160~ms than at 145~ms.
In model $\eta_{14}$, the strong contrast in the total magnetic
energy at 160~ms is
mainly due to that in the toroidal magnetic energy, while in model
$\eta_{-\infty}$, the contrast both in the toroidal and poloidal
magnetic energies are responsible for that.

\begin{figure*}
\epsscale{1}
\plottwo{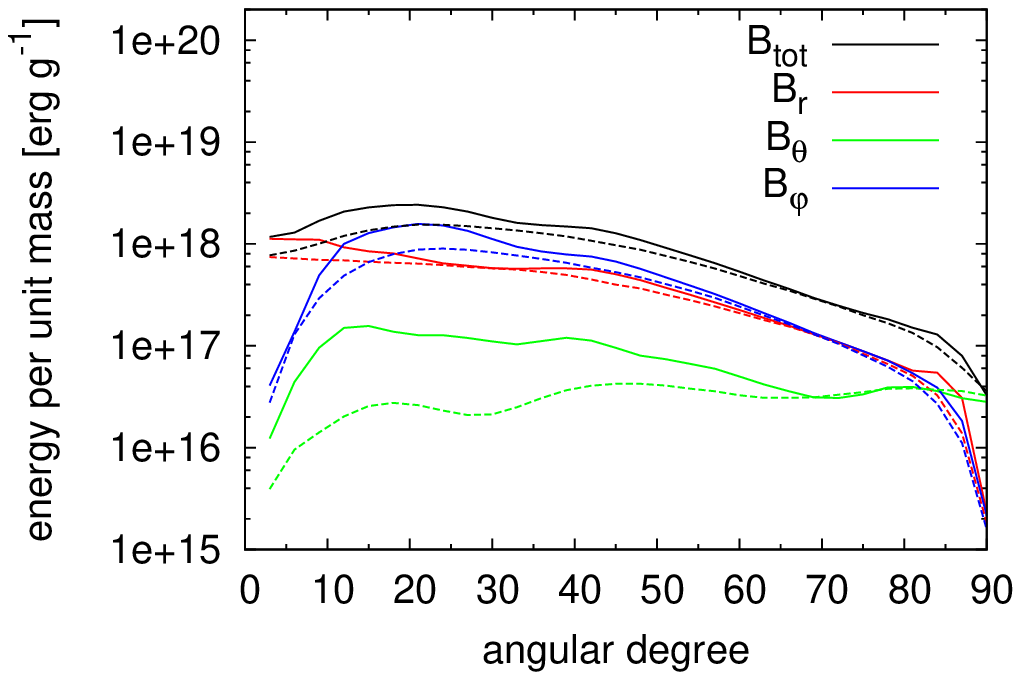}{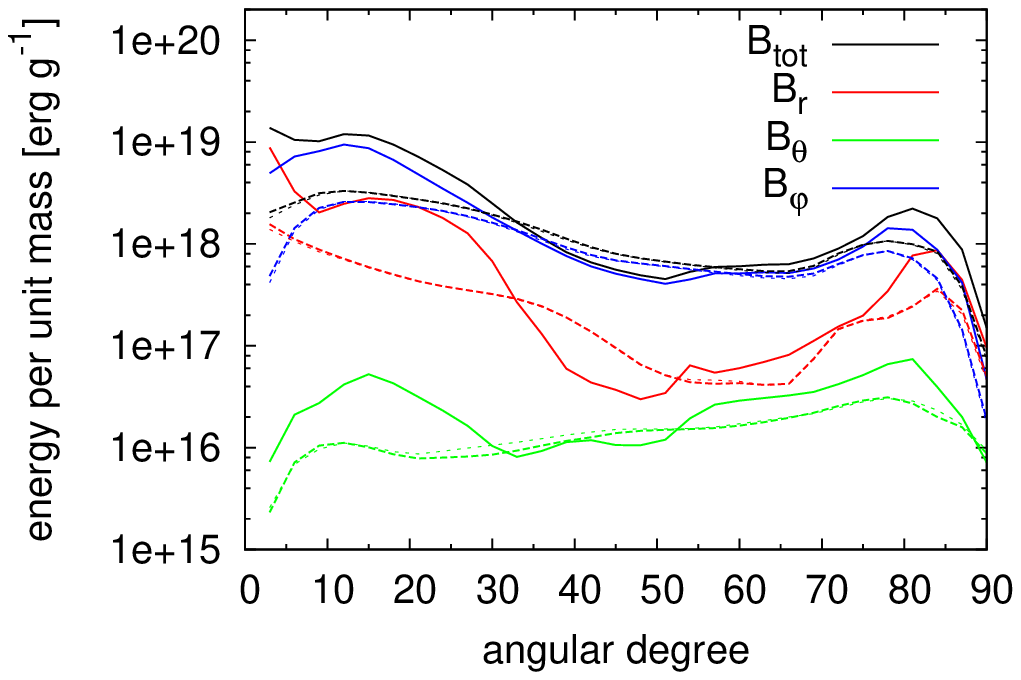}
\caption{Angular distributions of magnetic energies per unit mass
  averaged over $50$~km$<r<0.9\times r_{\textrm{sh}}$ at $t=145$~ms
  (2~ms after bounce; \textit{left}) and $t=160$~ms (17~ms after
  bounce, \textit{right}) in model
  B\textit{s}-$\Omega$-$\eta_{-\infty}$ (solid lines) and
  B\textit{s}-$\Omega$-$\eta_{14}$ (dashed lines). The dotted lines
  are for a different resolution run for model 
  B\textit{s}-$\Omega$-$\eta_{14}$ (see \S~\ref{sec.res}).}
\label{fig.ethm.55}
\end{figure*}

To understand how the contrast is strengthened, we follow the
evolution of magnetic energy per unit mass in two representative
volumes $V_{25.5}$ and $V_{58.5}$, where $V_{\theta_{\textrm{s}}}$ is
defined by 50~km$\le r\le 0.9\times r_{\textrm{sh}}$ and
$\theta_{\textrm{s}}-1^\circ.5\le\theta\le\theta_s+1^\circ.5$.
Fig.~\ref{fig.t-ethm.55} shows the evolution of the average
magnetic energy per unit mass in the two volumes. It is observed 
both in model $\eta_{-\infty}$ and $\eta_{14}$,
that the toroidal magnetic energy in the both volumes increases
around 150~ms, and keeps an almost constant value afterward. The
increase rate is higher in volume $V_{25.5}$. That is the contrast in
the toroidal magnetic energy is strengthened around 150~ms, and is
kept afterward. 
What is also found is that in model $\eta_{-\infty}$, the poloidal magnetic
energy in volume $V_{25.5}$ increases during $t\sim155$-$160$~ms,
while that in volume $V_{58.5}$ does not vary very much. It seems that
this makes the strong contrast in the poloidal magnetic energy in
model $\eta_{-\infty}$ at 160~ms shown in the right panel of
Fig.~\ref{fig.ethm.55}.

\begin{figure*}
\epsscale{1}
\plottwo{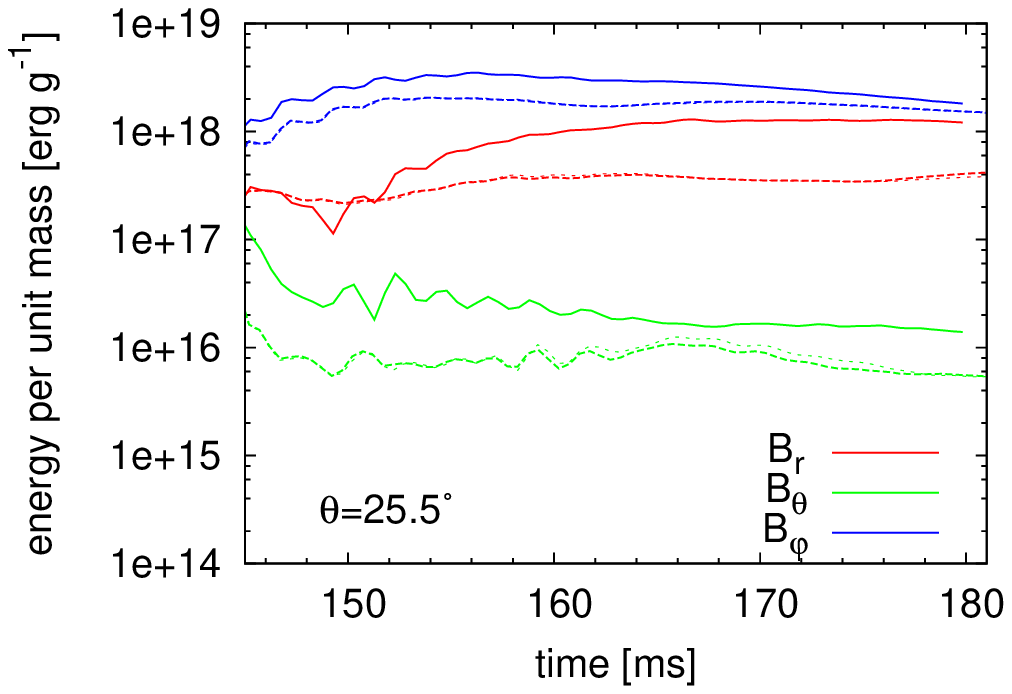}{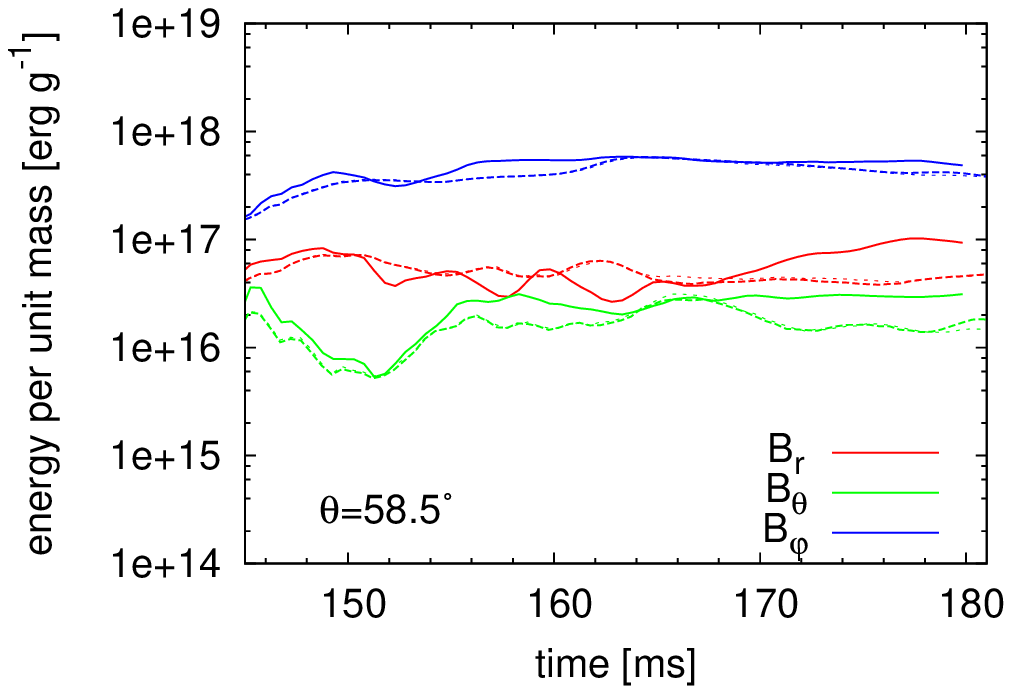}
  \caption{Evolutions of average magnetic energies per unit mass
    of volume $V_{25.5}$ (\textit{left}) and $V_{67.5}$
    (\textit{right}). The graphs are plotted from $t=145$~ms, 2~ms after
    bounce. The solid lines are for model
    B\textit{s}-$\Omega$-$\eta_{-\infty}$ while the dashed lines are
    for B\textit{s}-$\Omega$-$\eta_{14}$. The dotted lines are for a
    different resolution run for model B\textit{s}-$\Omega$-$\eta_{14}$
    (see \S~\ref{sec.res}). }   
 \label{fig.t-ethm.55}
\end{figure*}

In order to study the amplification mechanisms of magnetic field, we
write down the evolution equations of the average magnetic energies
per unit mass in volume $V_{\theta_{\textrm{s}}}$ with mass $M$,
$\mathcal{E}_{\{r,\theta,\phi\}}=[\int (B_{\{r,\theta,\phi\}}^2/8\pi)dV]/M$: 
\begin{eqnarray}
\frac{d\mathcal{E}_{\{r,\theta,\phi\}}}{dt}&=&
\phantom{+}
\dot{\mathcal{E}}_{\{r,\theta,\phi\},\textrm{adv}} +
\dot{\mathcal{E}}_{\{r,\theta,\phi\},\textrm{cmp}} +  
\dot{\mathcal{E}}_{\{r,\theta,\phi\},\textrm{shr}} 
\nonumber\\
&&
+
\dot{\mathcal{E}}_{\{r,\theta,\phi\},\textrm{rst}} +
\dot{\mathcal{E}}_{\{r,\theta,\phi\},\textrm{VM}},\label{eq.amp1}   
\end{eqnarray}
where the terms in r.h.s. mean, from the left, the change of
$\mathcal{E}_{\{r,\theta,\phi\}}$ due to 
an advection, compression, velocity shear along magnetic field lines,
resistivity, and the variation in the volume and mass. They are defined by
\begin{eqnarray}
\dot{\mathcal{E}}_{\{r,\theta,\phi\},\textrm{adv}}&=&
-\frac{1}{M}\int
\left[v_r \frac{\partial}{\partial r}
+\frac{v_\theta}{r} \frac{\partial }{\partial \theta}\right]
\frac{B_{\{r,\theta,\phi\}}^2}{8\pi}dV,
\nonumber\\[0.5pc]
\dot{\mathcal{E}}_{r,\textrm{cmp}}&=&
-\frac{1}{M}\int
\left[\frac{2 v_r}{r} + \frac{1}{r\sin\theta} 
\frac{\partial(\sin\theta v_\theta)}{\partial \theta}\right]
\frac{B_r^2}{4\pi}dV,
\nonumber\\[0.5pc]
\dot{\mathcal{E}}_{\theta,\textrm{cmp}}&=&
-\frac{1}{M}\int
\left[\frac{1}{r}\frac{\partial (r v_r)}{\partial r}
      +\frac{\cot\theta v_\theta}{r}\right]
\frac{B_\theta^2}{4\pi}dV,
\nonumber\\[0.5pc]
\dot{\mathcal{E}}_{\phi,\textrm{cmp}}&=&
-\frac{1}{M}\int
\left[\frac{1}{r}\frac{\partial (r v_r)}{\partial r}
      +\frac{1}{r}\frac{\partial v_\theta}{\partial \theta}\right]
\frac{B_\phi^2}{4\pi}dV,
\nonumber\\[0.5pc]
\dot{\mathcal{E}}_{r,\textrm{shr}}&=&
\hphantom{+[}
\frac{1}{M}\int
\frac{B_rB_\theta}{4\pi r}\frac{\partial v_r}{\partial \theta}dV,
\nonumber\\[0.5pc]
\dot{\mathcal{E}}_{\theta,\textrm{shr}}&=&
\hphantom{+[}
\frac{1}{M}\int
\frac{rB_rB_\theta}{4\pi}\frac{\partial (v_\theta/r)}{\partial r}dV,
\\
\dot{\mathcal{E}}_{\phi,\textrm{shr}}&=&
\hphantom{+[}
\frac{1}{M}\int
\left[
B_r\frac{\partial (v_\phi/r\sin\theta)}{\partial r}
+\frac{B_\theta}{r}\frac{\partial (v_\phi/r\sin\theta)}{\partial\theta}
\right]
\nonumber\\
&&\hspace{1pc}
\times\frac{r\sin\theta B_\phi}{4\pi}dV,
\label{eq.amp2}
\nonumber\\
\dot{\mathcal{E}}_{r,\textrm{rst}}&=&
\hphantom{+[}
\frac{\eta}{Mc}\int
\frac{B_r}{r\sin\theta}\frac{\partial(\sin\theta
  j_\phi)}{\partial\theta}dV,
\nonumber\\[0.5pc]
\dot{\mathcal{E}}_{\theta,\textrm{rst}}&=&
\hphantom{+[}
\frac{\eta}{Mc}\int
\frac{B_\theta}{r}\frac{\partial(r j_\phi)}{\partial r}dV,
\nonumber\\[0.5pc]
\dot{\mathcal{E}}_{\phi,\textrm{rst}}&=&
-\frac{\eta}{M}\int
\left[
 \frac{\partial(r j_\theta)}{\partial r}
-\frac{\partial j_r}{\partial \theta}
\right]
\frac{4\pi}{c}\frac{B_\phi}{r}dV,
\nonumber\\[0.5pc]
\dot{\mathcal{E}}_{\{r,\theta,\phi\},\textrm{VM}}&=&
\hphantom{+[}
 \frac{\ointop \left(B_{\{r,\theta,\phi\}}^2/8\pi\right)
 \mbox{\boldmath$v$}_{\textrm{srf}}\cdot d\mbox{\boldmath$S$}}{M}
\nonumber\\
&&   
-\frac{\int
  \left(B_{\{r,\theta,\phi\}}^2/8\pi\right)dV}{M^2}\frac{dM}{dt},
\nonumber
\end{eqnarray}
where $\mbox{\boldmath$v$}_{\textrm{srf}}$ is a surface velocity of 
volume $V_{\theta_{\textrm{s}}}$\footnote{These definitions are
  different from commonly used ones, where the terms on the r.h.s. of
  the induction equations, $\partial\mbox{\boldmath $B$}/\partial t =
  -(\mbox{\boldmath$v$}\cdot\nabla)\mbox{\boldmath$B$} +
  (\mbox{\boldmath$B$}\cdot\nabla)\mbox{\boldmath$v$} -
  \mbox{\boldmath$B$}(\nabla\cdot\mbox{\boldmath$v$})$, are
  interpreted as the changes of magnetic field due to advection,
  shear, and compression \citep[e.g.][]{bra05}. We carry out possible
  cancellations between these terms.}. 
Note that the resistive terms are 
written for a constant resistivity. By evaluating these terms we can
see which factors are essential for the magnetic field amplification
in a given volume. The results are shown in Fig.~\ref{fig.bamp.5500}
and~\ref{fig.bamp.5514}. 

We first focus on the amplifications of the toroidal magnetic energies
in model $\eta_{-\infty}$. In volume $V_{25.5}$, the toroidal
magnetic energy is primarily amplified by an 
advection (right-top panel of Fig.~\ref{fig.bamp.5500}). We found
that an radial advection dominates over an angular advection, i.e. the
amplification is due to an outward advection of a large
toroidal energy at small radii. As expected, a
velocity shear along poloidal magnetic field-lines, viz. a winding due
to a differential rotation, also substantially contributes to the
amplification. In volume $V_{58.5}$, the 
toroidal magnetic energy is also amplified due to an advection and
shear (see the right-bottom panel of Fig.~\ref{fig.bamp.5500}), but
the amplitude of each term is much smaller than in volume~$V_{25.5}$,
which seems due to a priori weaker magnetic field.

The amplification of the radial magnetic energy in volume $V_{25.5}$
is also dominated by a radial advection. The shear term, which is far
smaller than that of the advection term, seems to also play an important
role after $\sim 160$~ms, since it is comparable to a total
$\dot{\mathcal{E}}$ (see the left-top panel of Fig.~\ref{fig.bamp.5500}).
As seen in the left-bottom panel of Fig.~\ref{fig.bamp.5500}, the
amplification  mechanism of radial magnetic energy in the volume
$V_{58.5}$ is rather complex, contributed by several terms. As in the case
of a toroidal magnetic energy, an amplification of a radial magnetic
energy is also weaker in volume $V_{58.5}$ than in volume $V_{25.5}$.

Since the present model-series involves a rotation, the
magnetorotational instability (MRI) may occur
and may play an important role in a magnetic field 
amplification \citep{bal91,aki03,mas06}. Signs for MRI growth is found
in some of past MHD core-collapse simulations initially assuming a
magnetar-class magnetic field 
\citep{yam04,tak04,obe06,shi06}. Also, \citet{obe09} carried out simulations
of MRI with a rather weak initial magnetic field, which is still
stronger than that of ordinary pulsars by an order of magnitude, using
a local simulation box, and found that the MRI 
exponentially amplifies the seed magnetic field. Note however that the
effect of accretion is not considered in their local box simulations.
According to \citet{fog06}, the neutrino-driven convection in the gain
region can be stabilized or slowed down by accretion. This may also hold
for the MRI in the post-shock region. 

We investigated whether, in model $\eta_{-\infty}$
and $\eta_{14}$, there emerges a region where the criterion for the MRI
\citep{bal95} is satisfied and 
the growth timescale is short enough. In the left-top panel of
Fig.~\ref{fig.mri.5500}, we plot the distribution of a MRI linear
growth timescale, roughly estimated by $4\pi/|\varpi
d\Omega/d\varpi|$, in $\theta$-$r$ plane at $160$~ms (17~ms after
bounce) for model $\eta_{-\infty}$. It is shown 
that in a considerable part for $\theta\lesssim 40^\circ$, the growth
timescale of MRI is a few~ms to 10~ms, while in a larger $\theta$ the
growth timescale is averagely longer. We found that, in the above part, the
growth timescale of $\sim$10~ms is kept after bounce ($t=143$~ms) until the
end of the simulation. However, the right-top panel of
Fig.~\ref{fig.mri.5500} does not shows MRI-like field-line bending as
observed in some models in \citet{yam04} (model MF3 and MF8).  
This may be because the present model leads to a stronger matter
eruption in the radial direction than those models of \citet{yam04}
because of a rather mild initial rotation speed\footnote{It is known
  that a mildly 
  rapid rotation, $T/|W|\sim~0.5$~\%, is favorable for an energetic
  explosion (see \citet{yam04})}. In the present model, field-line
bending produced by the MRI may become invisible due to a dominant
radial flow, and the absence of that will not
necessarily mean the non-operation of the MRI. 
Note that the absence of field-line bending seems not due to a poor
spacial resolution: A field-line bending is observed even in the
computations of model-series B\textit{m}-$\Omega$, in which a magnetic
field is averagely weaker than in the present case and thus the
resolution for capturing MRI is poorer. In the present case, the
fastest growing wave length is resolved everywhere with several
10~numerical cells (see the bottom panels of
Fig.~\ref{fig.mri.5500}). Although the required number of numerical
cells to capture one wave length depends on numerical scheme, several 10
cells will be sufficient.

\begin{figure*}
\epsscale{1}
\plottwo{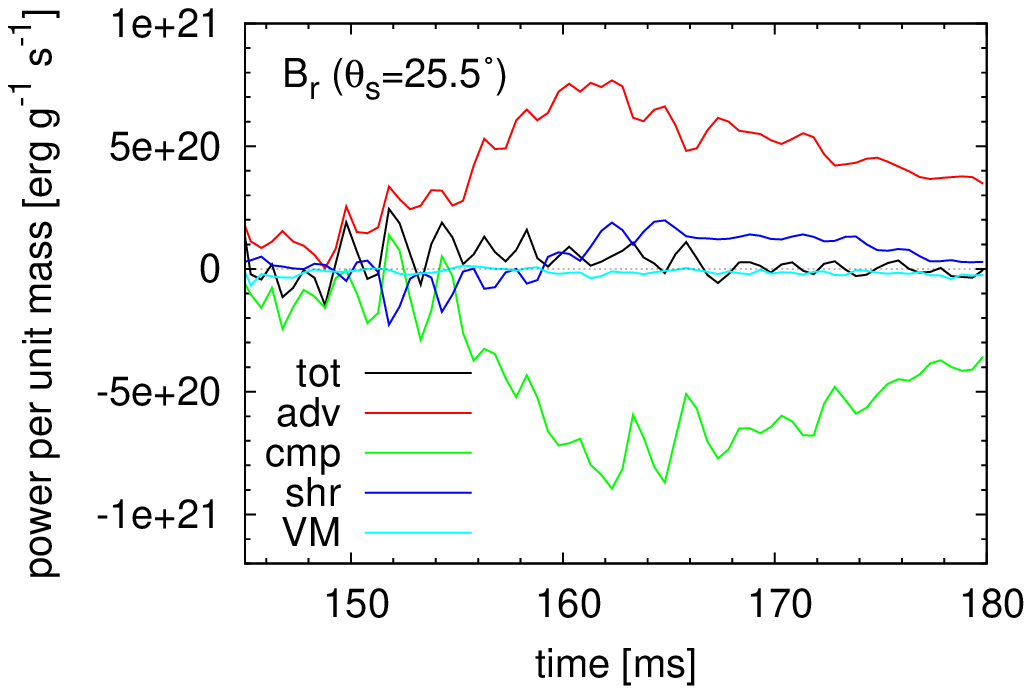}{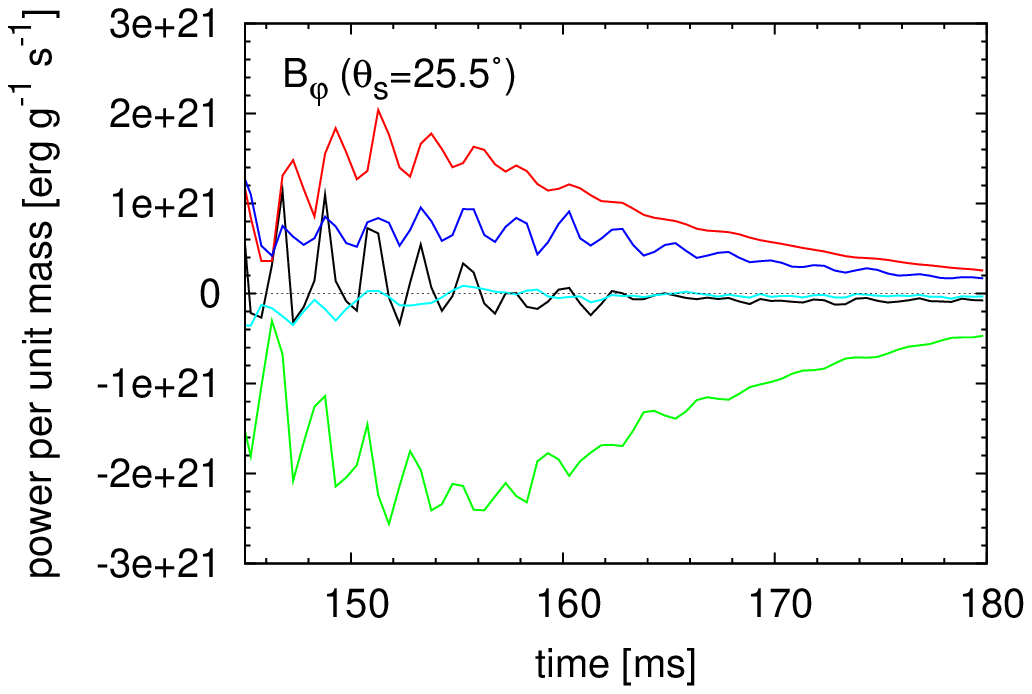}\\
\plottwo{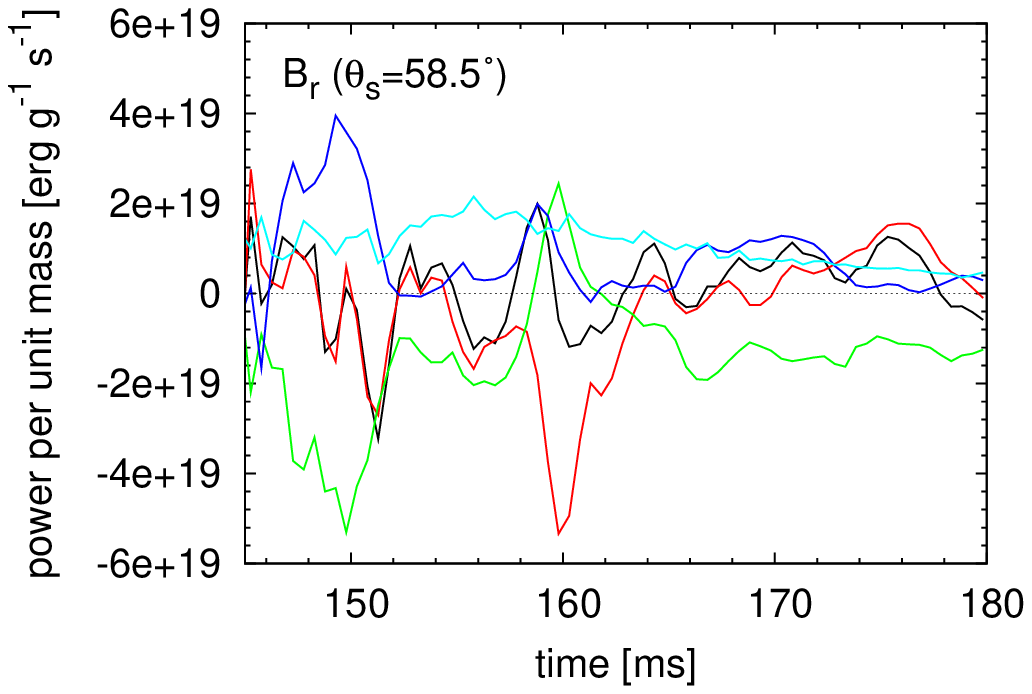}{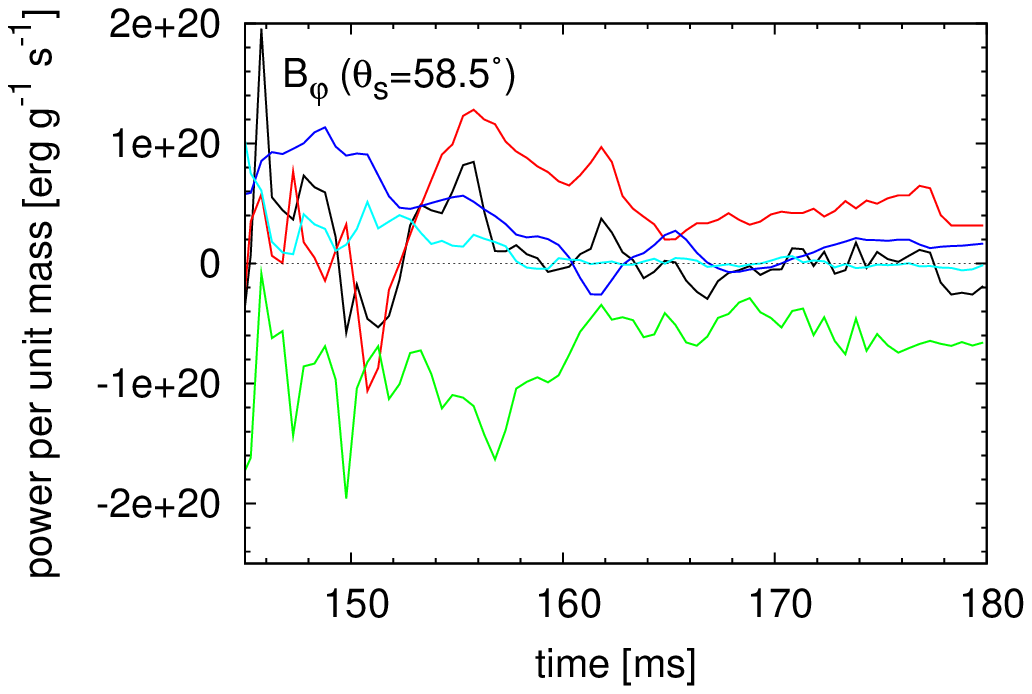}
\caption{Evolution of each $\dot{\mathcal{E}_r}$ and
  $\dot{\mathcal{E}_\phi}$ (see Eq.~(\ref{eq.amp2})) in
  volume $V_{25.5}$ and $V_{58.5}$ in model
  B\textit{s}-$\Omega$-$\eta_{-\infty}$.}
\label{fig.bamp.5500}
\end{figure*}
\begin{figure*}
\epsscale{1}
\plottwo{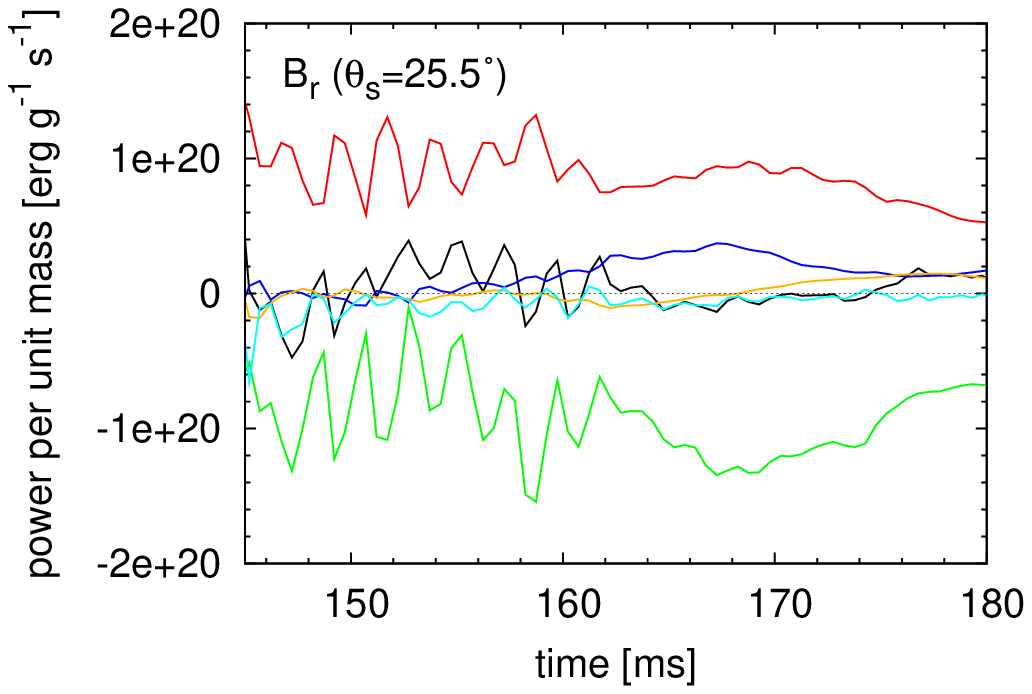}{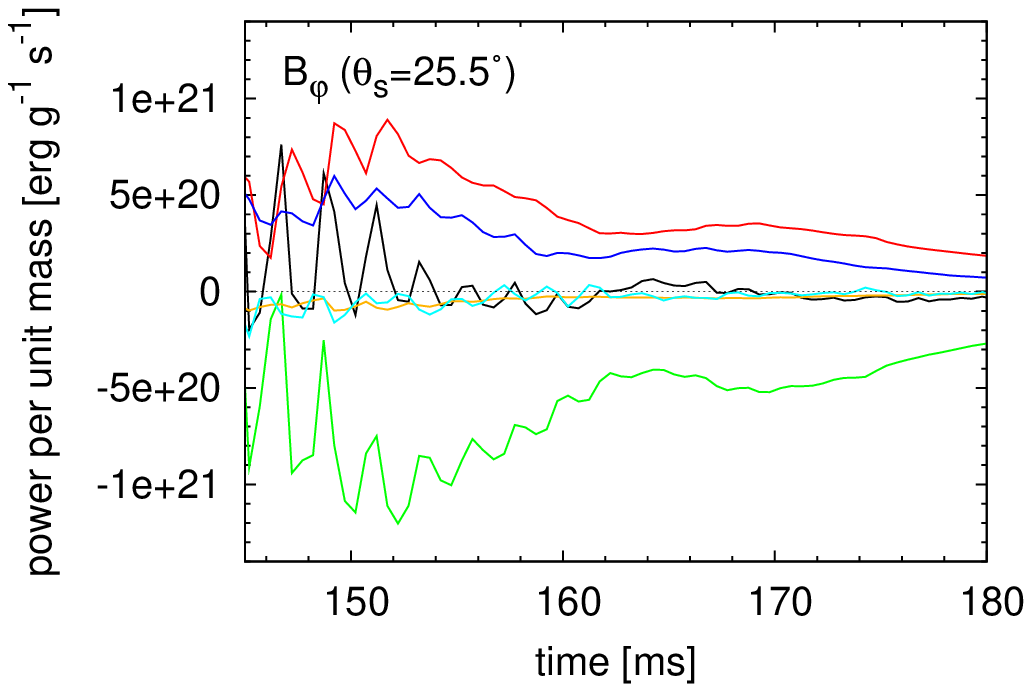}\\
\plottwo{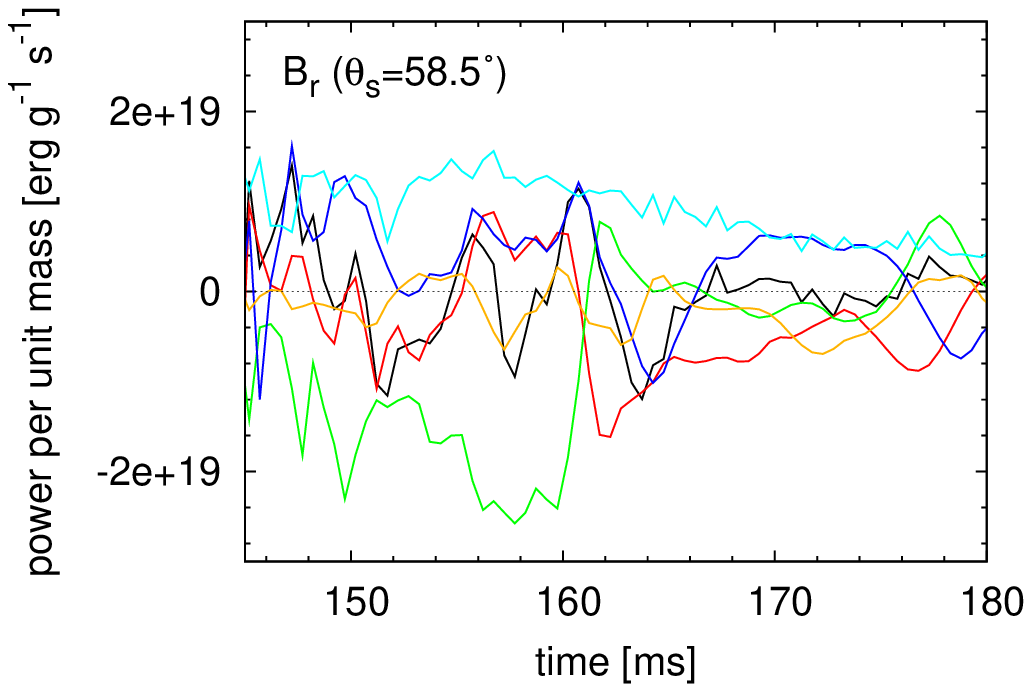}{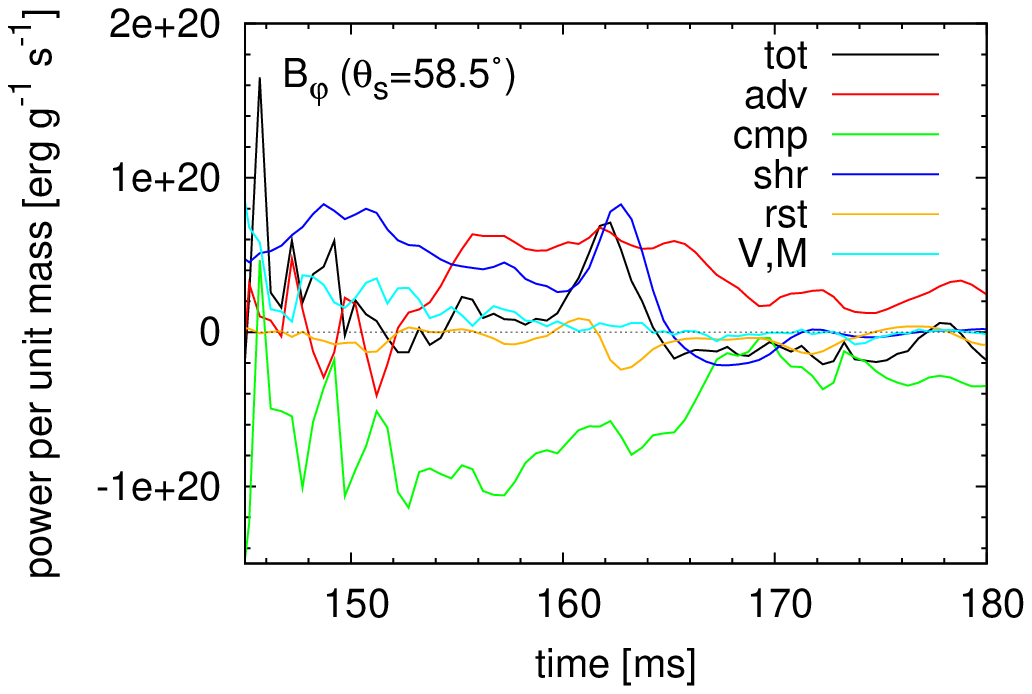}
\caption{Same as Fig.~\ref{fig.bamp.5500} but for
  B\textit{s}-$\Omega$-$\eta_{14}$. Note that the vertical scale of
  each panel in Fig.~\ref{fig.bamp.5500} and~\ref{fig.bamp.5514} are
  different except for the right-bottom panel.} 
\label{fig.bamp.5514}
\end{figure*}

\begin{figure*}
\epsscale{1}
\plottwo{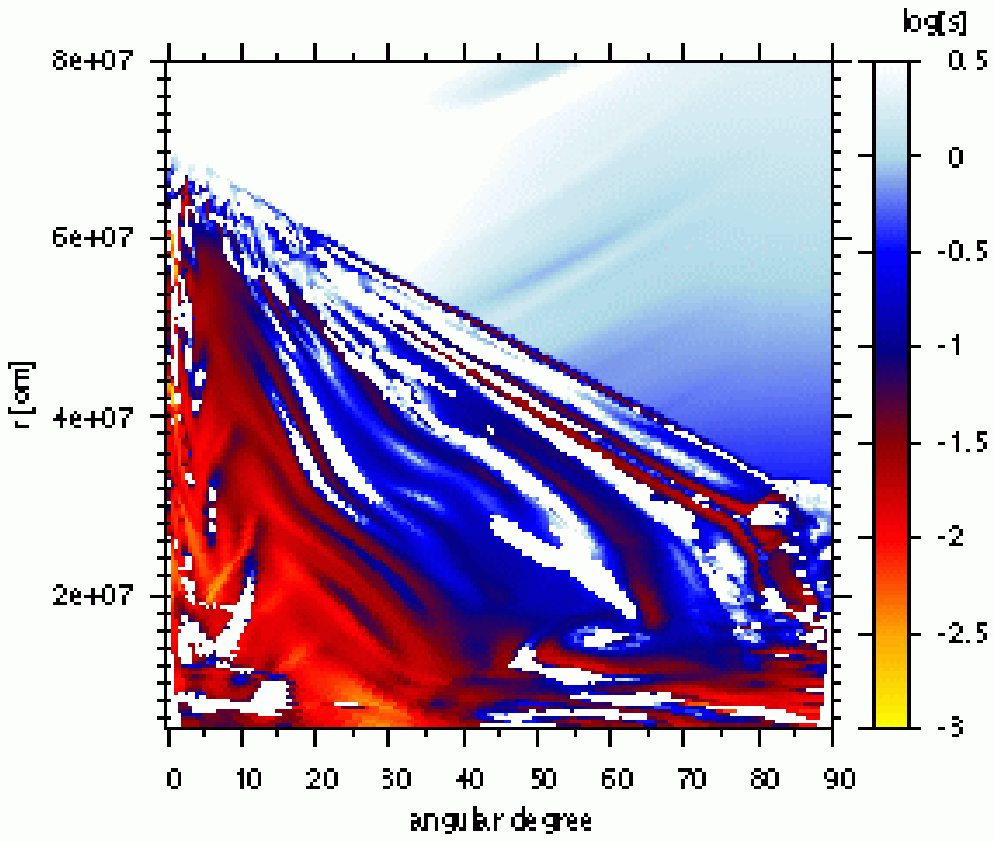}{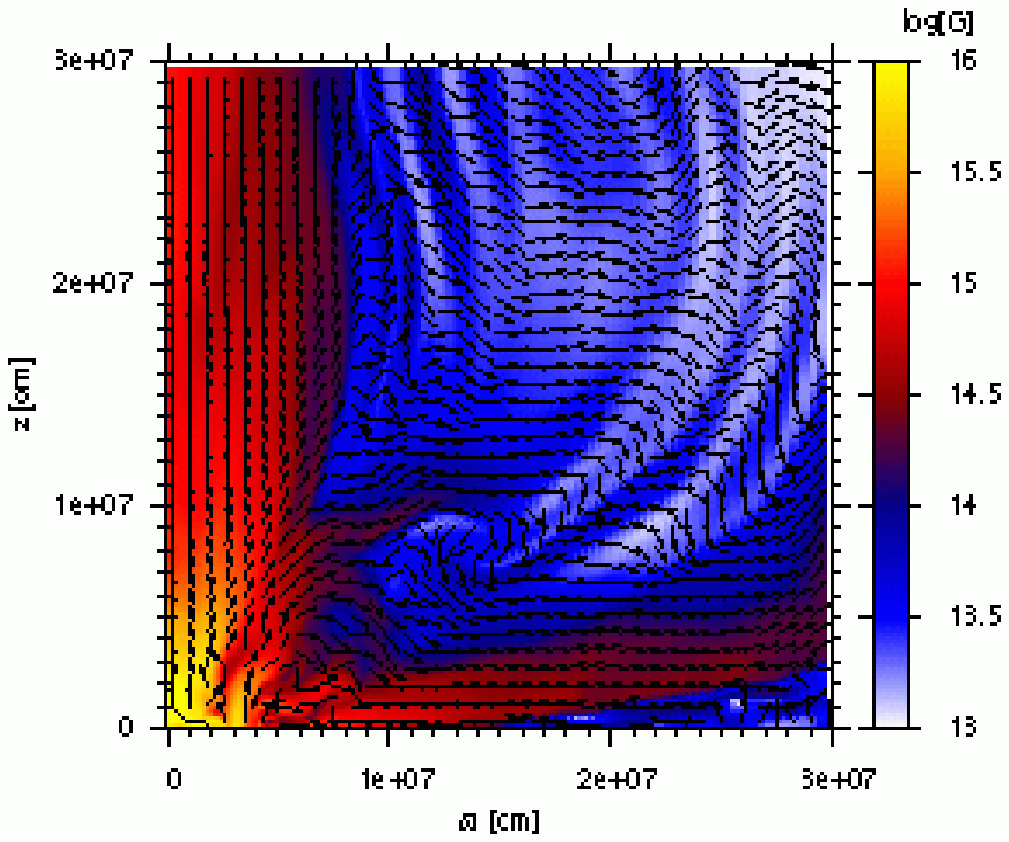}
\plottwo{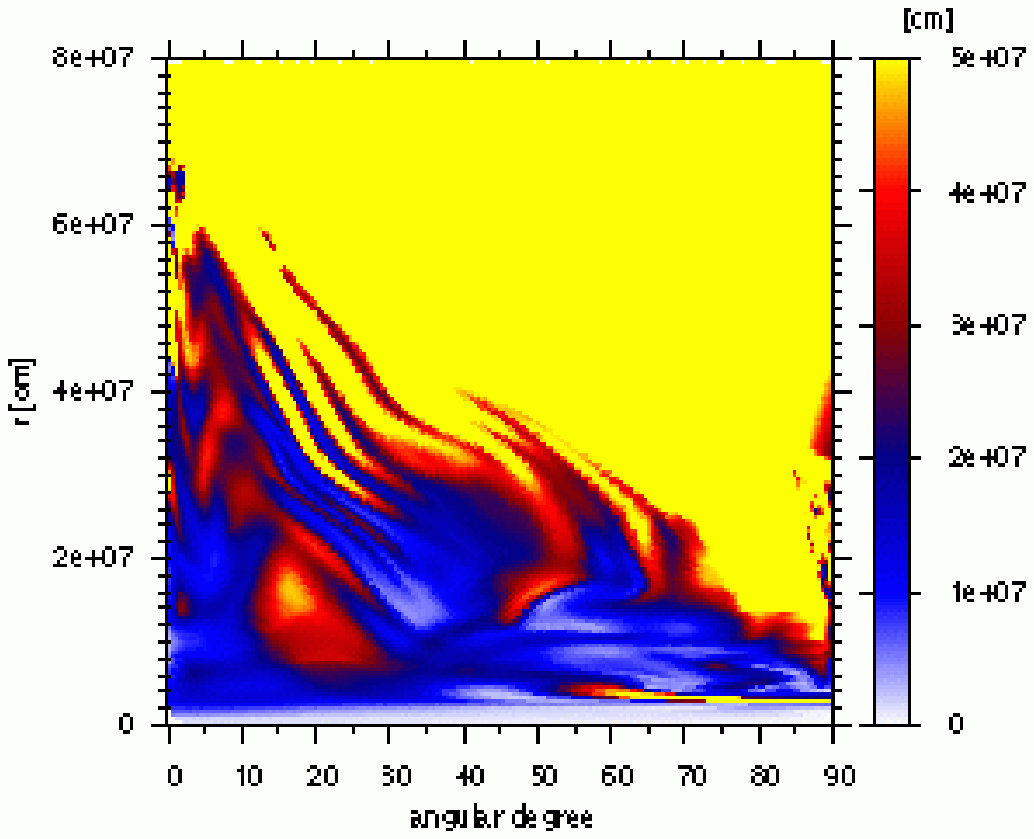}{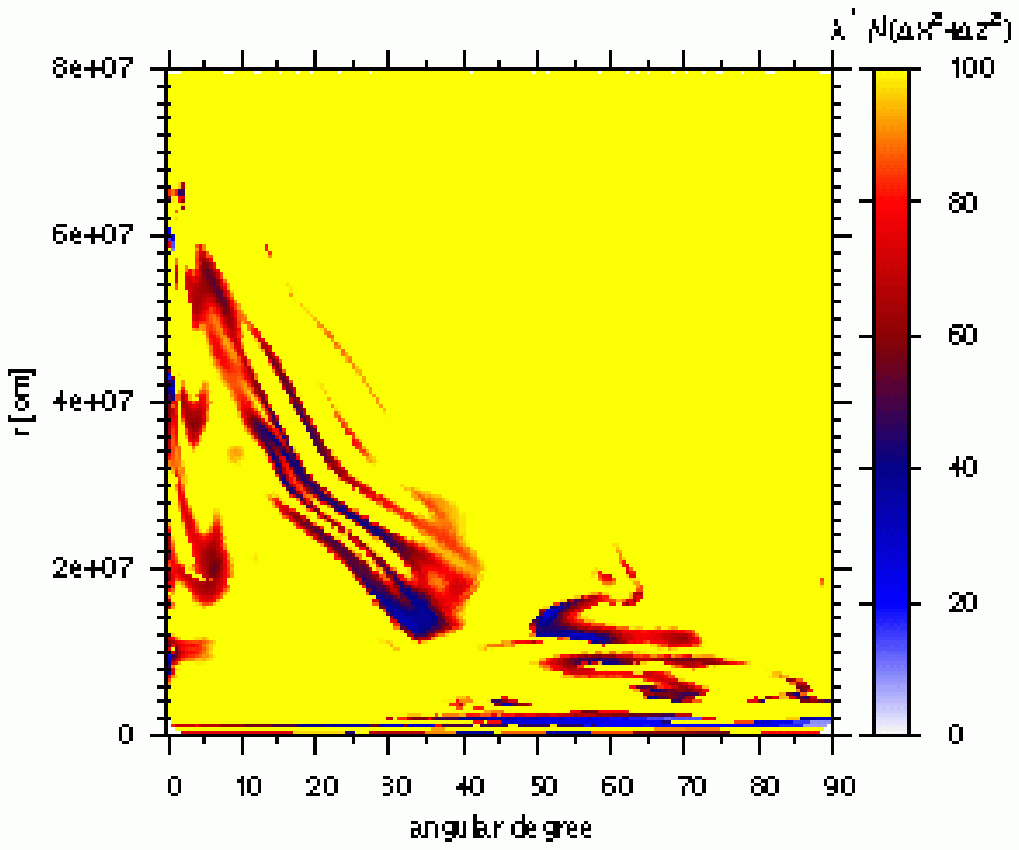}
  \caption{\textit{Left-top panel}: Distribution of MRI growth timescale
    in logarithmic scale in $\theta$-$r$ plane. White-colored
    regions include a location that is stable against the
    MRI. \textit{Right-top panel}: 
    Poloidal magnetic field vectors on top of a color map of 
    poloidal magnetic field strength. 
    \textit{Left-bottom panel}: Distribution of fastest growing wave
    length, $\lambda^*\sim 2\pi 
    c_{\textrm{A}}/\Omega$, where $c_{\textrm{A}}$ is the Alfv\'en
    velocity of a poloidal magnetic field. \textit{Right-bottom panel}: 
    Distribution of fastest growing wave length divided by
    numerical cell size, $\lambda^*/\sqrt{(\Delta x)^2+(\Delta
      z)^2}$. Yellow-colored region includes MRI-stable locations.
    These figures are depicted for model
    B\textit{s}-$\Omega$-$\eta_{-\infty}$ at $t=160$~ms.}
 \label{fig.mri.5500}
\end{figure*}

\begin{figure*}
\epsscale{1}
\plottwo{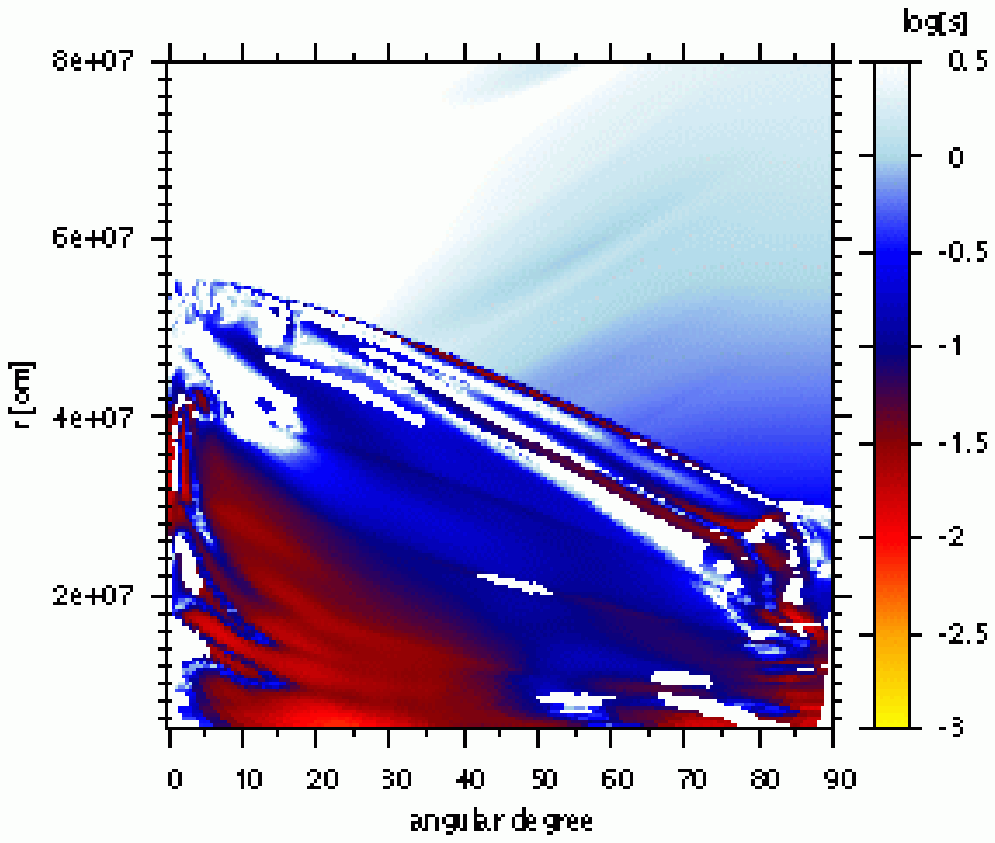}{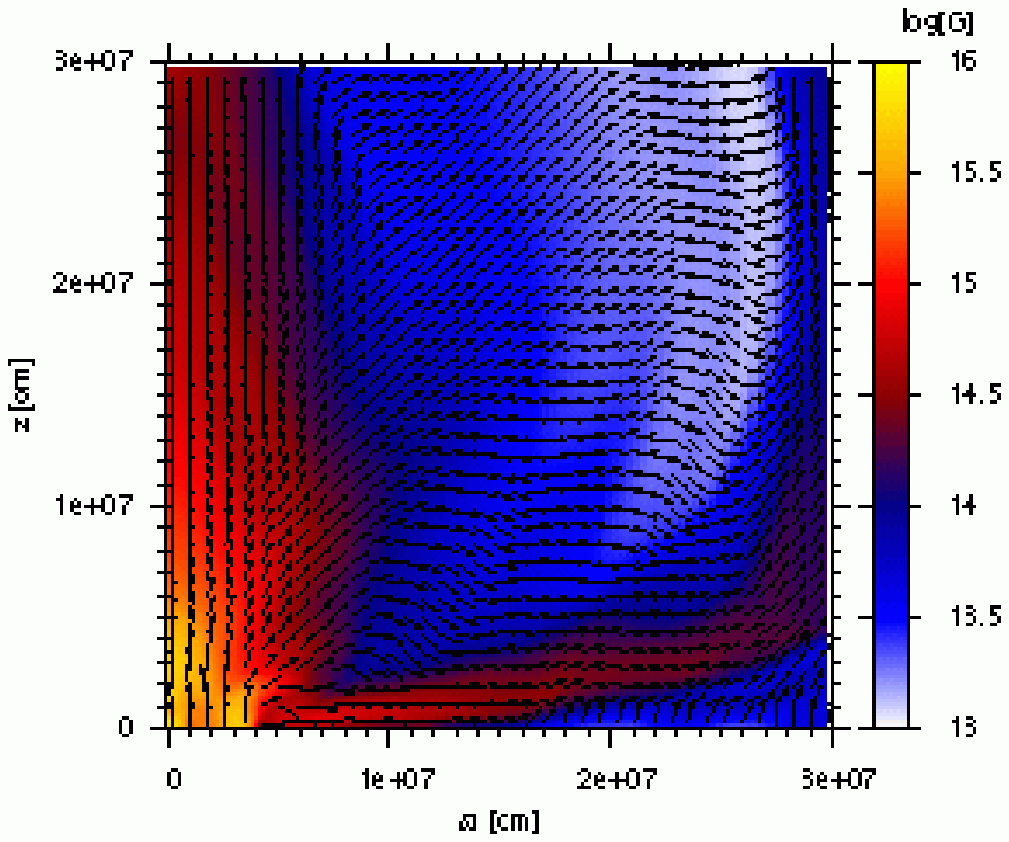}
\plottwo{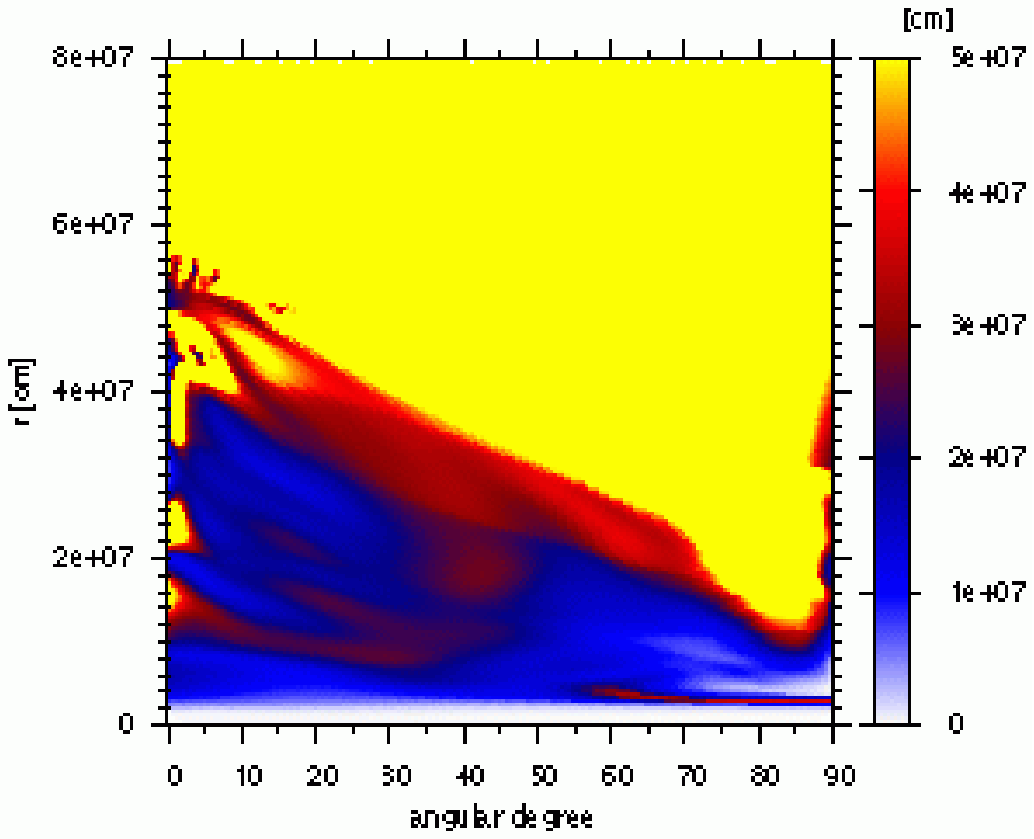}{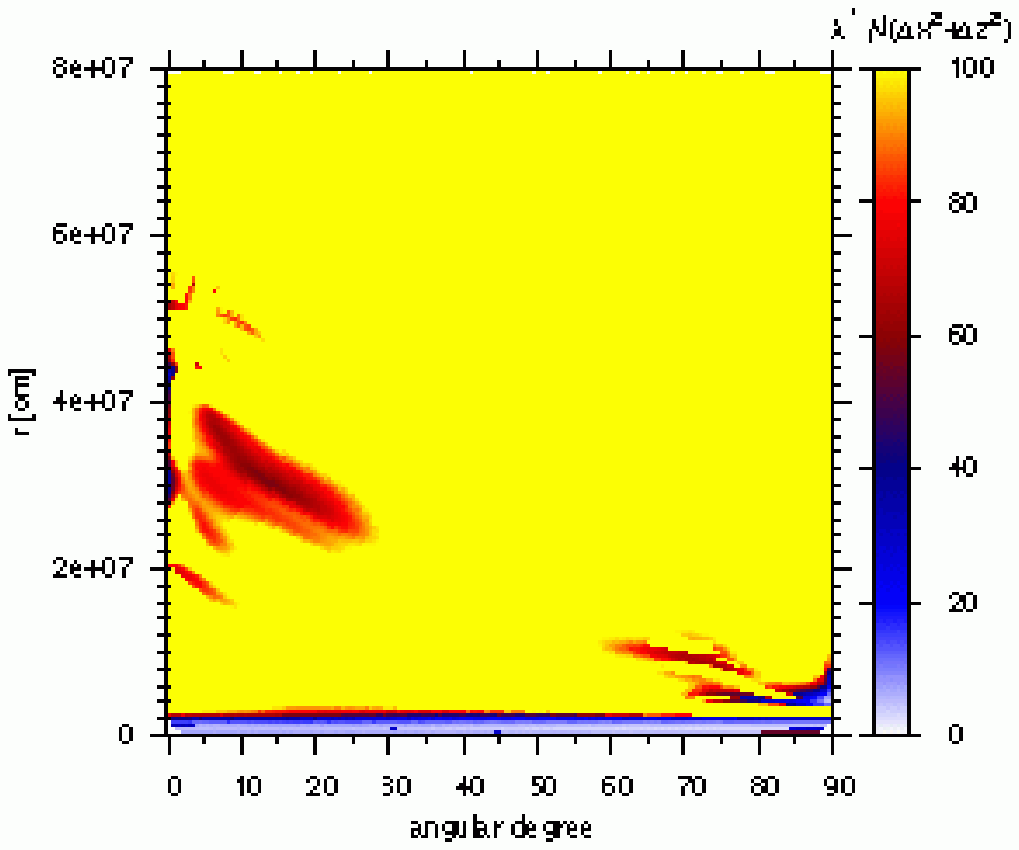}
  \caption{Same as Fig.~\ref{fig.mri.5500} but for model
    B\textit{s}-$\Omega$-$\eta_{14}$ at $t=160$~ms.} 
 \label{fig.mri.5514}
\end{figure*}

\begin{figure*}
\epsscale{0.5}
\plotone{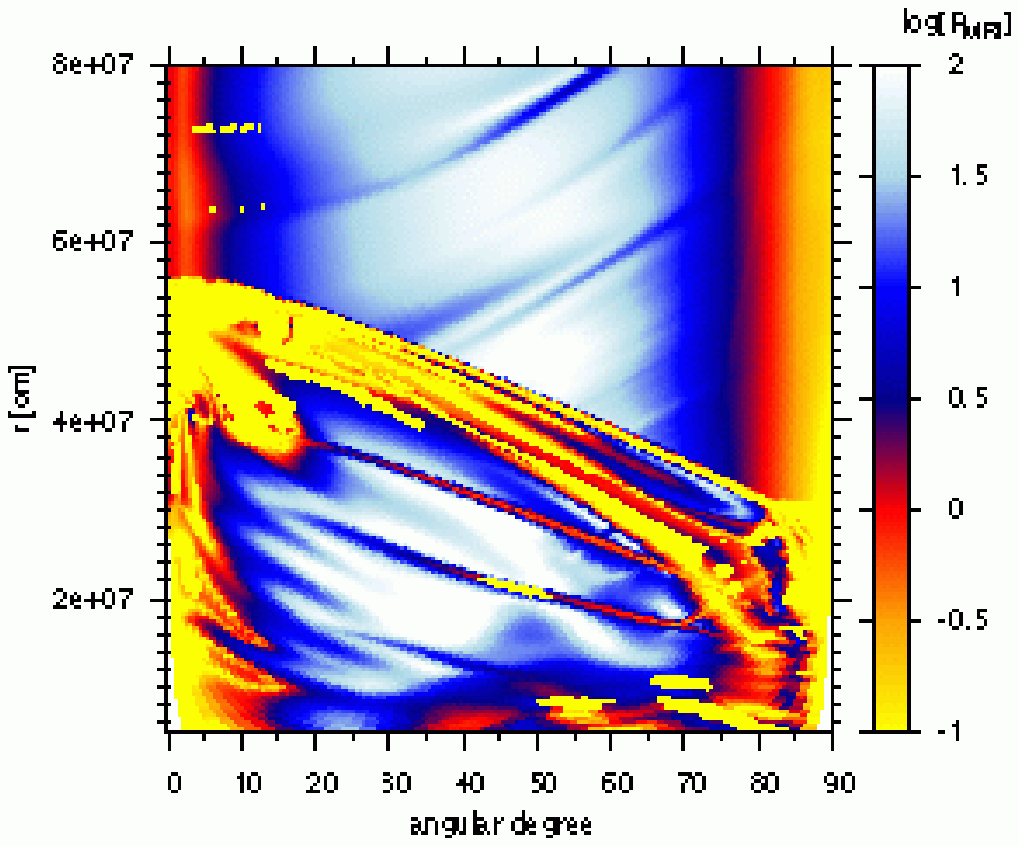}
  \caption{Distribution of
    $R_{\textit{MRI}}\equiv\tau_{\textrm{res}}/\tau_{\textrm{MRI}}$,
    in logarithmic 
    scale in $\theta$-$r$ plane in model
    B\textit{s}-$\Omega$-$\eta_{14}$ at $t=160$~ms.} 
 \label{fig.rmri.5514}
\end{figure*}

The growth of the MRI will lead to an increase of the shear term
in Eq.~(\ref{eq.amp1}), since the MRI produce a shear of velocity
along a magnetic field-line.
The left panel of Fig.~\ref{fig.bamp.5500} shows that the shear
term in volume $V_{25.5}$ becomes relatively large after $\sim
160$~ms ($\sim 20$~ms after bounce). Given that the MRI growth timescale
there is $\sim 10$~ms, it may be reasonable to consider that the
increase of the shear term is due to the operation of MRI. Even if
this is the case, however, it is unlikely that the MRI greatly
amplifies a magnetic field, since the radial magnetic energy is nearly
constant after $\sim 160$~ms (see left panel of Fig.~\ref{fig.t-ethm.55}).
The MRI seems at best to keep the strength of magnetic field in volume
$V_{25.5}$. Since the fastest MRI growth timescale is comparable at
each $\theta$ for $\theta\lesssim 40^\circ$, the situation will be
more or less similar in this angular range.

Meanwhile, in the angular range of $\theta\gtrsim 40^\circ$, even the
retention of magnetic field strength by the MRI will be modest, because a
growth timescale reaches $\sim 10$~ms only in some limited locations
(see the left-top panel of Fig.~\ref{fig.mri.5500}). Although, in volume
$V_{58.5}$, the shear term helps to increase or keep the radial
magnetic energy (the left-bottom panel of Fig.~\ref{fig.bamp.5500}),
this begins long before the MRI is expected to grow. 
Only in later time (e.g. after $\sim 160$~ms), the shear term might
contain some modest contribution from the MRI. 

The amplification mechanism in model~$\eta_{14}$ 
is qualitatively similar to that in model~$\eta_{-\infty}$, although
the amplitude of each $\dot{\mathcal{E}}$ is usually smaller due to
resistivity (see Fig.~\ref{fig.bamp.5514}). The left-top panel of
Fig.~\ref{fig.mri.5514} plotted for $t=160$~ms shows 
that in a considerable part for $\theta\lesssim 45^\circ$, the
MRI growth timescale is 10~ms to a few 10~ms, which is kept from the
time of bounce (143~ms) until the end
of the simulation. Fig.~\ref{fig.rmri.5514} displays the
distribution of resistive timescale divided by MRI growth timescale, 
$R_{\textit{MRI}}\equiv\tau_{\textrm{res}}/\tau_{\textrm{MRI}}$, at
160~ms. This indicates that the growth of the 
MRI is more or less hampered by a resistivity. Especially, in the
vicinity of the pole, there is almost no chance for the MRI growth. 
We found that, the growth of MRI is possible through the
simulation in the angular range of $20^\circ \lesssim \theta
\lesssim 30^\circ$, although a growth rate will be somewhat decreased
by a resistivity in some locations (blue-colored regions for $20^\circ
\lesssim \theta \lesssim 30^\circ$ in Fig.~\ref{fig.rmri.5514}). In
this model, the fastest growing wave length is resolved with several
10 numerical cells everywhere except for $r<20$~km (see bottom panels of
Fig~\ref{fig.mri.5514}). 

The left-top panel of Fig.~\ref{fig.bamp.5514} also shows that the
shear term in volume $V_{25.5}$ becomes large after $\sim 160$~ms. By
the same discussion done in the above, this is possibly due to the
MRI, but not important for a magnetic field amplification. Again, the
MRI at best may contribute to keep the radial magnetic energy in
model~$\eta_{14}$. 

In light of the above discussion, it can be said that the contrast in
magnetic energy per unit mass over $\theta$ observed both in
model~$\eta_{-\infty}$ and $\eta_{14}$ is strengthened or kept by an
outward advection of magnetic energy, winding of poloidal
magnetic field-lines, and possibly by the MRI, all of which
efficiently occur in a small-$\theta$ region.  

\subsubsection{Aspect Ratio}

\begin{figure*}[b]
\epsscale{1}
\plottwo{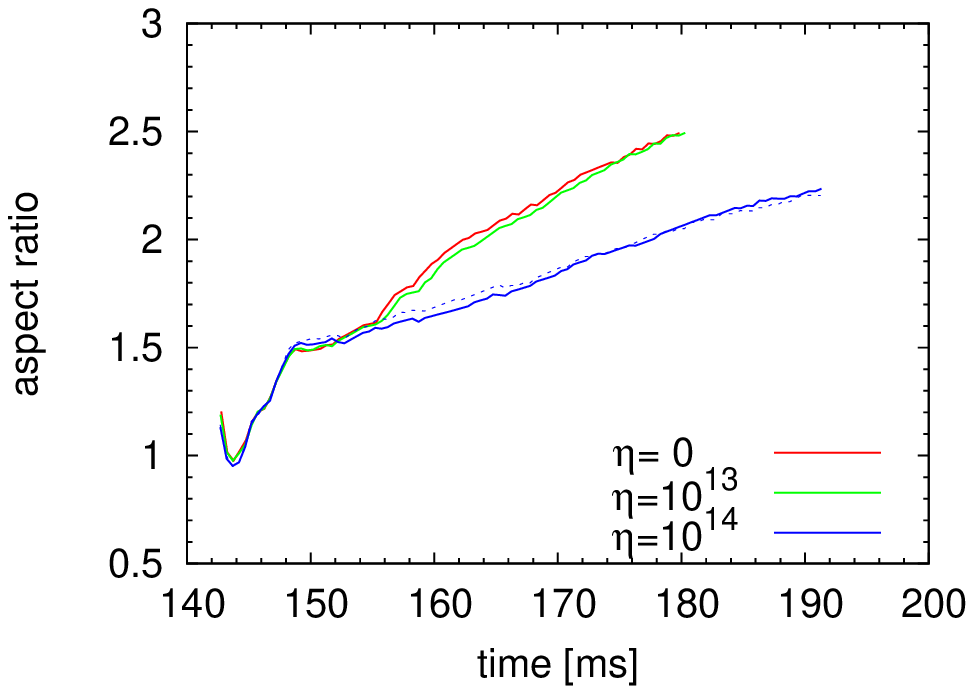}{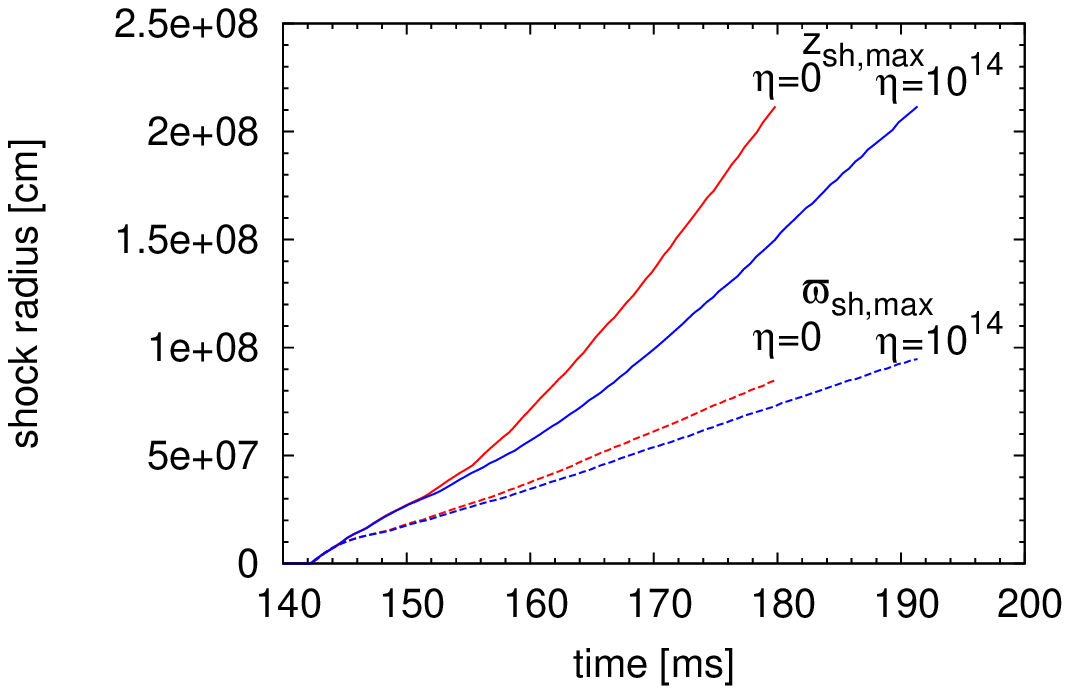}
\caption{Evolutions of aspect ratios (\textit{left panel}) and the
  maximum ejecta positions in $z$ and $\varpi$ (\textit{right panel})
  in model-series B\textit{s}-$\Omega$. The dotted line in the left
  panel is for a different resolution run for model 
  B\textit{s}-$\Omega$-$\eta_{14}$ (see \S~\ref{sec.res}).}
\label{fig.aspect.55}
\end{figure*}

\begin{figure*}
\epsscale{0.8}
\plotone{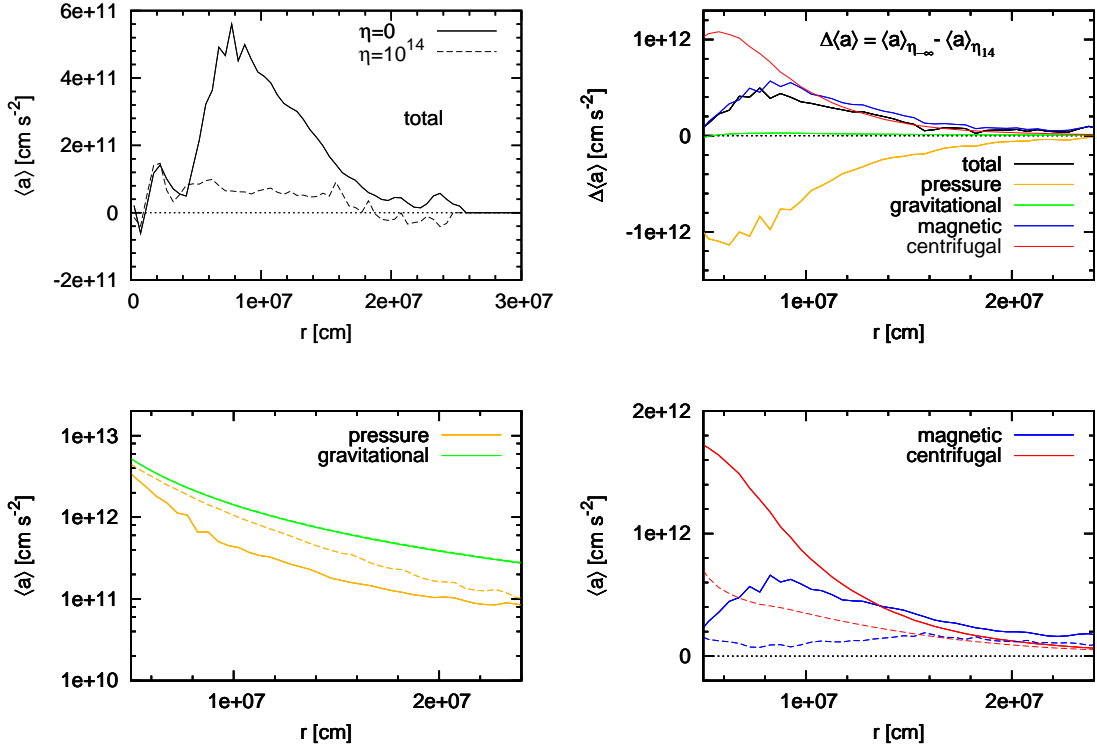}
\caption{Radial distributions of the accelerations in volume
  $V_{25.5}$, $\langle a \rangle$, time averaged 
  during $t=155-165$~ms, in models B\textit{s}-$\Omega$-$\eta_{-\infty}$ and
  B\textit{s}-$\Omega$-$\eta_{14}$. The right-top panel shows the  
  differences in the accelerations, $\Delta \langle a \rangle \equiv
  \langle a \rangle_{\eta_{-\infty}}-\langle a \rangle_{\eta_{14}}$,
  subtracting that of model $\eta_{14}$ from that of model
  $\eta_{-\infty}$. Colors and styles of lines are same as in
  Fig.~\ref{fig.force.55}}  
\label{fig.forcea.55}
\end{figure*}

\begin{figure*}
\epsscale{0.8}
\plotone{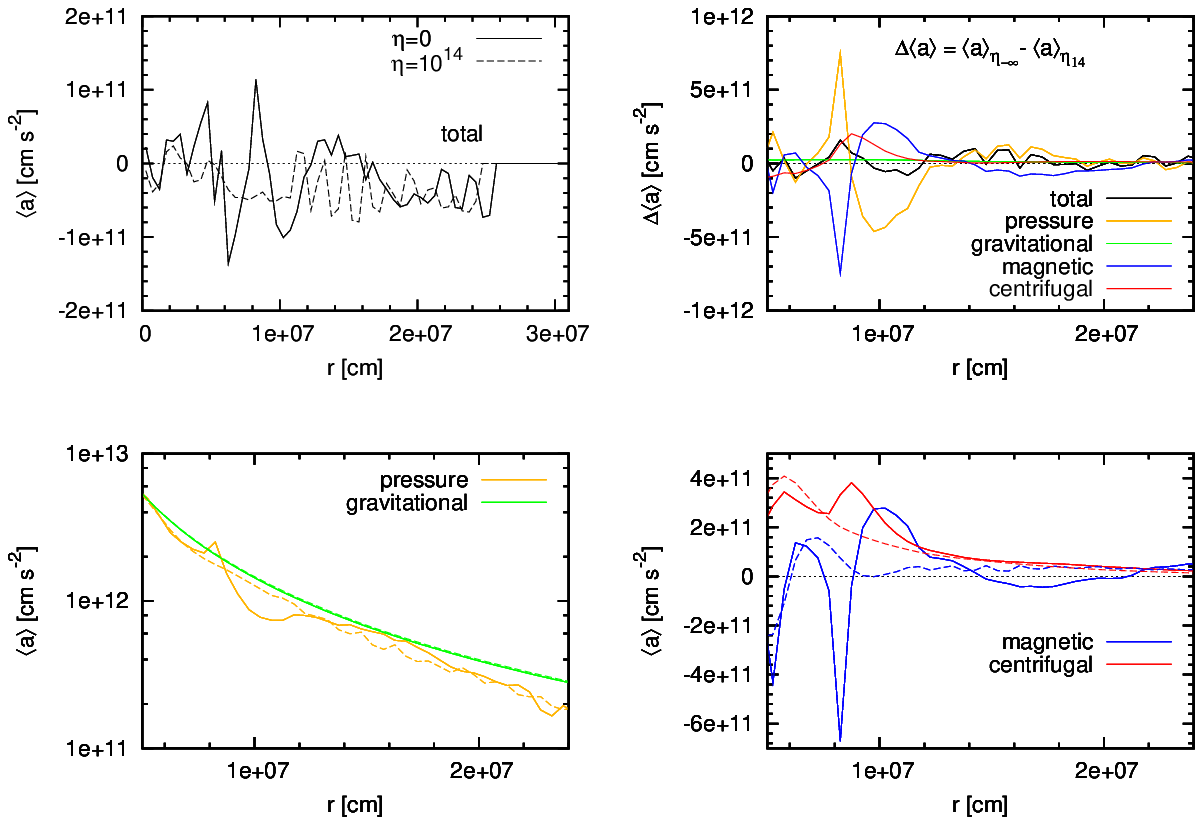}
\caption{Same as Fig.~\ref{fig.forcea.55}, but for volume $V_{58.5}$ }
\label{fig.forceb.55}
\end{figure*}

From Fig.~\ref{fig.vradbvec.55} it
is found that the shape of the shock surface is prolate both in model
B\textit{s}-$\Omega$-$\eta_{-\infty}$ and model
B\textit{s}-$\Omega$-$\eta_{14}$, in which the latter shows a
less prolate feature than the former. Defining the aspect ratio by
the maximum-$z$ position of ejected matters, $z_{\textrm{ej,max}}$,
divided by the maximum-$\varpi$ position, $\varpi_{\textrm{ej,max}}$, we
found that it exceeds two at the end of each simulation (see
Fig.~\ref{fig.aspect.55}). 

We first focus on the aspect ratio in model~$\eta_{-\infty}$.
Fig.~\ref{fig.aspect.55} indicates that the aspect ratio
becomes larger than unity soon after bounce ($t=143$~ms). Since a 
centrifugal force is relatively stronger around the 
equator, it hampers the collapse and weakens the bounce there. This is
one reason for the prolate matter ejection.
Figs.~\ref{fig.forcea.55} and~\ref{fig.forceb.55} shows the
radial distributions of the accelerations, time averaged during
$t=155$-$165$~ms, in volume $V_{25.5}$ and $V_{58.5}$,
respectively. It is found that in volume $V_{25.5}$, a matter is
greatly accelerated in the radial range of 50~km$\lesssim
r\lesssim$160~km (see solid line in the left-top panel of
Fig.~\ref{fig.forcea.55}). Meanwhile, in volume $V_{58.5}$, an
acceleration of matter is averagely smaller (see solid line in the
left-top panel of Fig.~\ref{fig.forceb.55}). This implies that an
acceleration of matter is larger for a smaller $\theta$, which causes
a further increase of the aspect ratio. Comparing the bottom
panels of the two figures, a larger magnetic and
centrifugal acceleration in volume $V_{25.5}$ seems responsible for
the larger acceleration. Reminding that a stronger magnetic field
leads to a larger amplitudes of these two accelerations (see
\S~\ref{sec.exp.55}), it is likely that a polarly concentrated
distribution of magnetic energy per unit mass (\S~\ref{sec.mfa.55}) is
essential to generate a large aspect ratio.

Comparing the aspect ratio among the three models in the left
panel of Fig.~\ref{fig.aspect.55}, it is the smallest in model
$\eta_{14}$, while that in models $\eta_{-\infty}$ and $\eta_{13}$
takes similar value. The right panel of
Fig.~\ref{fig.aspect.55} shows that the difference comes from the fact
that $z_{\textrm{ej,max}}$ is largely affected by resistivity, while
$\varpi_{\textrm{ej,max}}$ does not change so much. With this and the
above speculation that a prolate matter ejection is caused by a
polarly concentrated magnetic energy distribution, the smaller 
aspect ratio in model $\eta_{14}$ is readily understood. In
Fig.~\ref{fig.forcea.55}, it is shown that a total acceleration in
volume $V_{25.5}$ is smaller in model $\eta_{14}$ due to smaller
contribution from a magnetic and centrifugal acceleration. Meanwhile,
a total acceleration in volume $V_{58.5}$ is not very different
between the two models (see Fig.~\ref{fig.forceb.55}), because a
magnetic and centrifugal acceleration are not as important as in
volume $V_{25.5}$ due to a small magnetic energy per unit mass. This
leads to a large difference only in $z_{\textrm{ej,max}}$, and thus in
the aspect ratio.  

\subsubsection{Diffusion and Dissipation Sites}\label{sec.diff}

\begin{figure*}[t]
\epsscale{1}
\plottwo{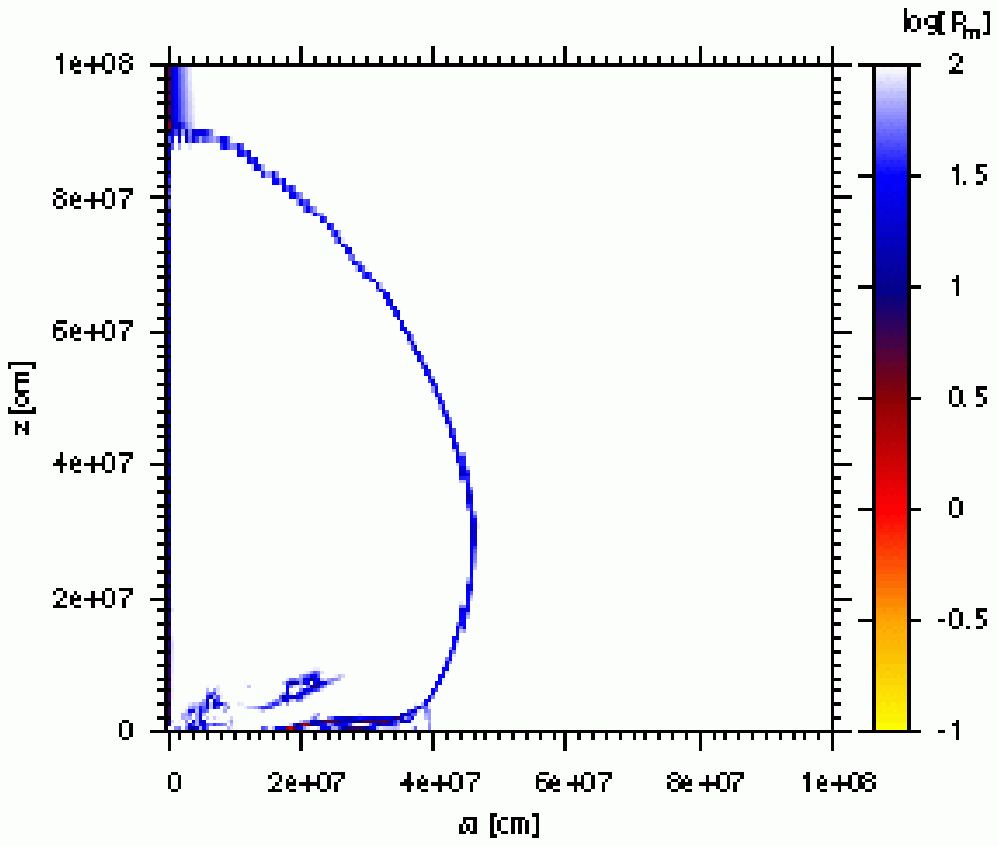}{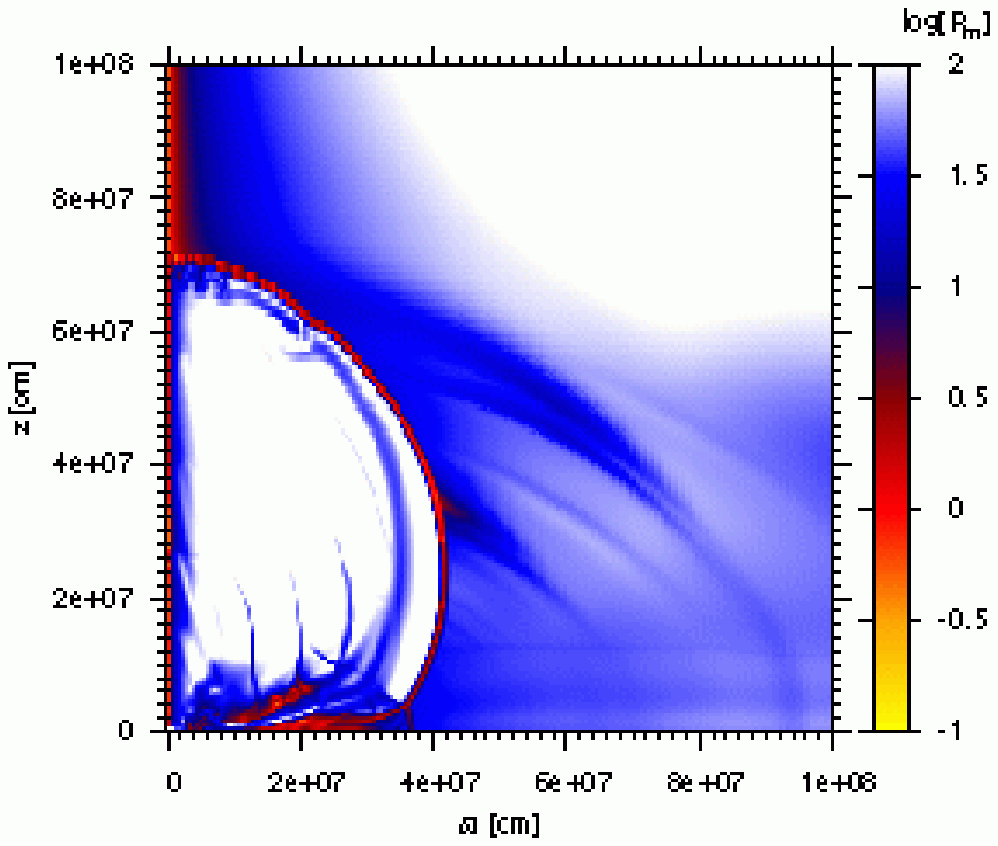}
  \caption{Distributions of magnetic Reynolds number $R_{\textrm{m}}$
    in logarithmic 
    scale at $t=164$~ms in model B\textit{s}-$\Omega$-$\eta_{13}$
    (left) and B\textit{s}-$\Omega$-$\eta_{14}$ (right).}  
 \label{fig.rme.55}
\end{figure*}

In this section, we will see in which sites a resistivity works
efficiently. In Fig.~\ref{fig.rme.55}, the distribution of magnetic
Reynolds number $R_{\textrm{m}}$, the ratio of the resistive timescale
to dynamical timescale, in model B\textit{s}-$\Omega$-$\eta_{13}$ and
B\textit{s}-$\Omega$-$\eta_{14}$ are displayed for $t=164$~ms. We
define the resistive and dynamical timescales, respectively, by
$L^2/\eta$ and
$|L\mbox{\boldmath$B$}|/|\mbox{\boldmath$v$}\times\mbox{\boldmath$B$}|$,
where the scale length of magnetic field is defined by
$L\equiv|c\mbox{\boldmath$B$}|/|4\pi\mbox{\boldmath$j$}|$. With these
definitions, the magnetic Reynolds number is also the ratio of the
size of the first term to second term in r.h.s. of Eq.~(\ref{eq.mhd.ohm}).
If we define a diffusion-dissipation site by the location where the
magnetic Reynolds number is $\lesssim$100, the right panel of
Fig.~\ref{fig.rme.55} shows that the diffusion-dissipation sites in
model $\eta_{14}$ are, around the pole, equator, the shock surface and
in the blue filaments inside the shock surface. In these sites, a
magnetic Reynolds number is 
small since the scale length of magnetic field is short: that of
toroidal field is short around the pole, while that of poloidal
field is short in the other sites. Comparing the right panels
of Fig.~\ref{fig.vradbvec.55}, an intense magnetic pressure dominance
seen around the pole in model $\eta_{-\infty}$ are not found in model
$\eta_{14}$. This implies that a diffusion and dissipation around the
pole are essential for producing the dynamical differences between model
$\eta_{-\infty}$ and $\eta_{14}$.

In the case of model $\eta_{13}$, a resistivity works efficiently also
around the pole, equator, and the shock surface, but the volumes of
these sites are much 
smaller than those in the former case due to a lower
resistivity. Besides, a magnetic Reynolds number there is generally at
least $\sim 30$, although in some very limited regions it reaches an
order of unity. Hence, model $\eta_{13}$ is rather close to the ideal
model. With 
this the explosion energy and aspect ratio in model $\eta_{13}$ are
not much different from those in model $\eta_{-\infty}$.

\subsubsection{Convergences of Results}\label{sec.res}
As a final remark for models-series B\textit{s}-$\Omega$, we mention
the convergences of results with respect to the size of the numerical
cells. As a representative model, we carry out another run for model 
B\textit{s}-$\Omega$-$\eta_{14}$ with a different spatial
resolution. Keeping the total number of cells, the size of the
inner-most cells is changed from $400$~m to $200$~m, where a cell size
increases outwards both in $\varpi$ and $z$-direction with the
constant ratio of 1.0069. With this 
distribution a resolution is higher than the original one for $z,
\varpi < 200$~km, and lower otherwise. In Figs.~\ref{fig.t-eng.55},
\ref{fig.t-exp.55}, \ref{fig.ethm.55}, \ref{fig.t-ethm.55}, and
\ref{fig.aspect.55}, 
the results obtained with this different resolution are shown for model
B\textit{s}-$\Omega$-$\eta_{14}$ by dotted lines. Although a little
deviations from 
the original results are observed in some figures, they seem not to
qualitatively and even quantitatively affect the discussions given
above. 

The error in the total energy conservation in the different
resolution run is 14~\% as mentioned at the beginning of \S~\ref{sec.result}.
That in the normal resolution run of model
B\textit{s}-$\Omega$-$\eta_{14}$, 26~\%, is roughly twice as large as
the different resolution one. Nonetheless, as we have seen, only
slight deviations are found between results of the two resolution
run. Hence, although the total energy error of 26~\% may be a little
too large, this will not spoil the results of the simulation. We
expect that this is also the case for the other models.

\subsection{Moderate Magnetic Field and Rapid Rotation --- Model Series
  B\textit{m}-$\Omega$}\label{result2}
The dynamical evolutions in model-series B\textit{m}-$\Omega$ is
qualitatively similar to those found in model-series B\textit{s}-$\Omega$, 
i.e. a winding of magnetic field-lines by a differential rotation
increases a magnetic and centrifugal acceleration, which play a key
role in a matter ejection. Owing to a weaker initial magnetic field,
the explosion occurs less energetically than in the
former model-series. The distributions of velocity and magnetic
field at $t=181$~ms in model B\textit{m}-$\Omega$-$\eta_{-\infty}$
and B\textit{m}-$\Omega$-$\eta_{14}$ are depicted in
Fig.~\ref{fig.vradbvec.45}. Compared to model-series
B\textit{s}-$\Omega$ with a same resistivity (see
Fig.~\ref{fig.vradbvec.55}), the present model-series shows a slower
outward velocity and an averagely weaker magnetic pressure. As in the
former model-series, a comparison between the upper and lower panels of
Fig.~\ref{fig.vradbvec.45} also indicates that an matter eruption
becomes weaker with the presence of resistivity, with which we
expect a lower explosion energy for a resistive model.

\begin{figure*}
\begin{center}
\epsscale{1}
\plottwo{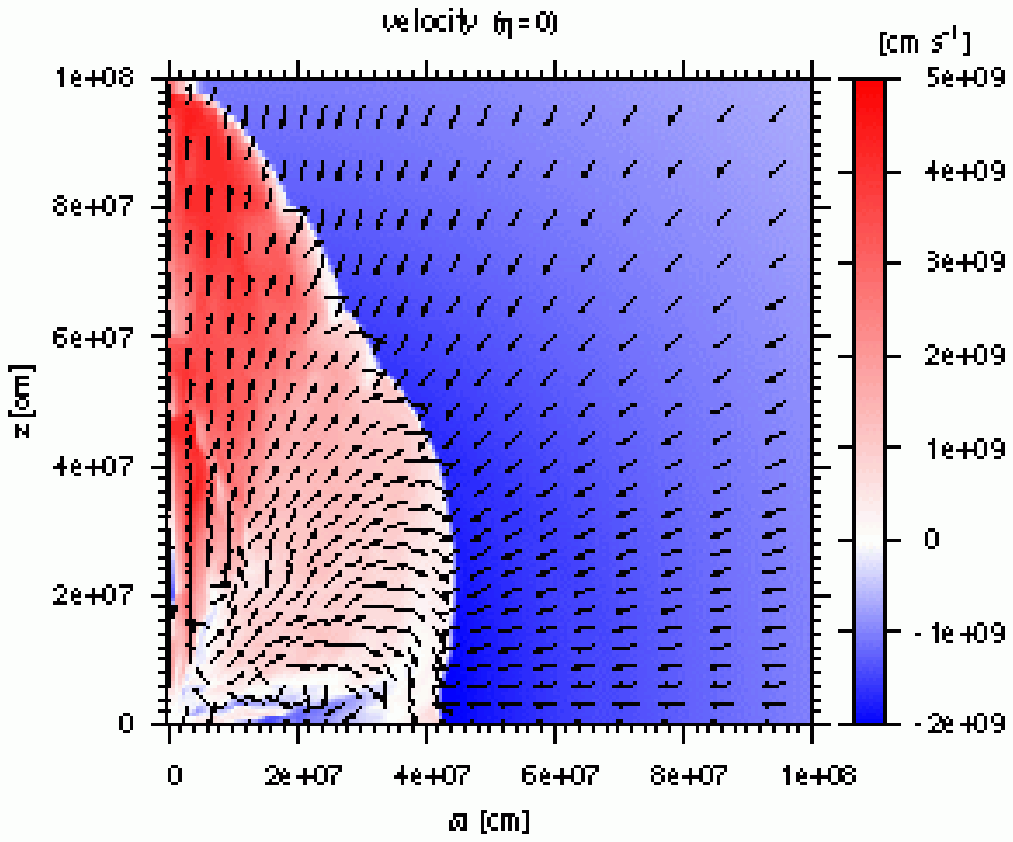}{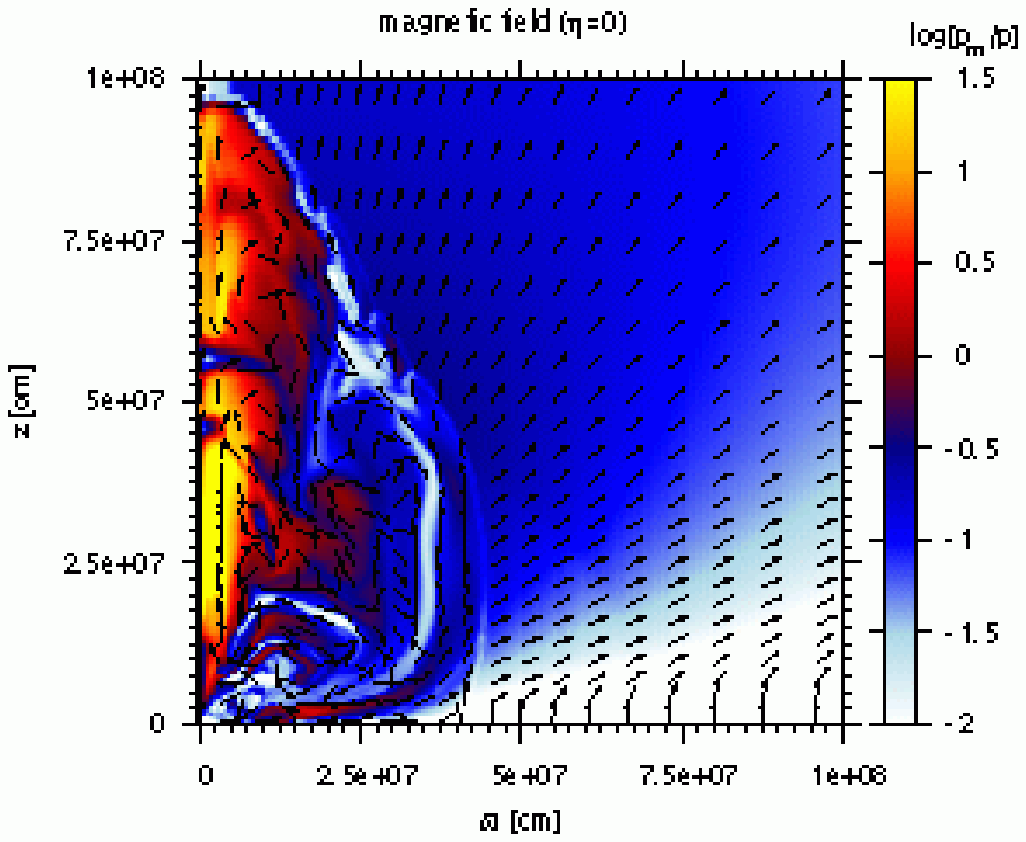}
\plottwo{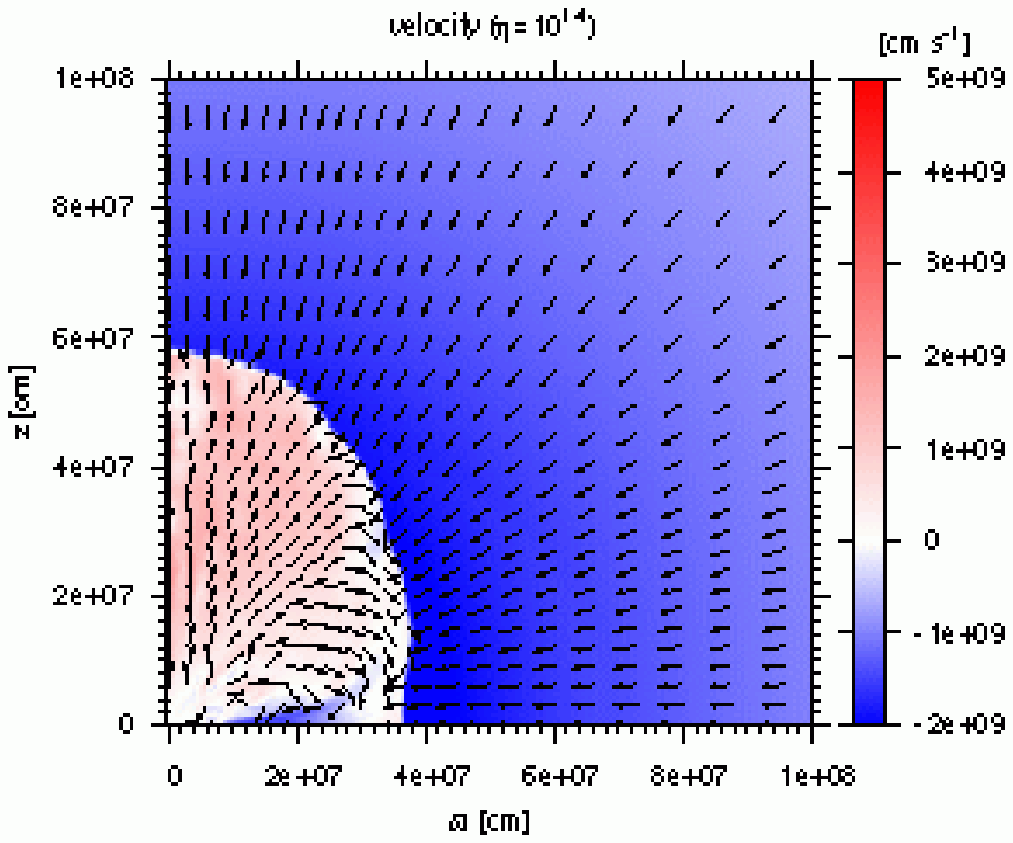}{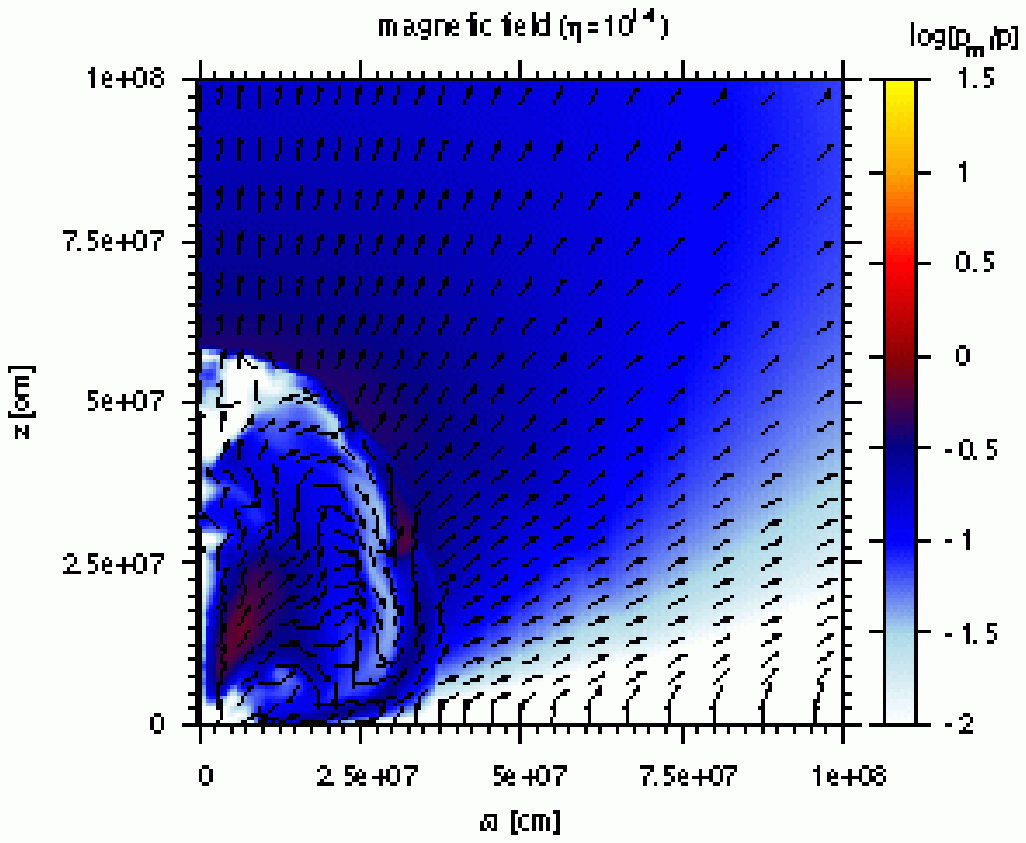}
  \caption{Same as Fig.~\ref{fig.vradbvec.55} but for model
    B\textit{m}-$\Omega$-$\eta_{-\infty}$ (upper panels) and
    B\textit{m}-$\Omega$-$\eta_{14}$ (lower panels) depicted at
  $t=181$~ms.} 
 \label{fig.vradbvec.45}
\end{center}
\end{figure*}

\subsubsection{Explosion Energy}
Fig.~\ref{fig.eexp1.45} plots the evolutions of the explosion
energies in the present model-series. The resulting explosion energy
in each model appears to be about factor three smaller than the
corresponding model in model-series B\textit{s}-$\Omega$. It is shown
that a resistivity also decreases the explosion energy as in the
former case. This can be understood same way as we have seen in
\S~\ref{sec.exp.55} for model-series B\textit{s}-$\Omega$.

\begin{figure*}
\epsscale{0.5}
\plotone{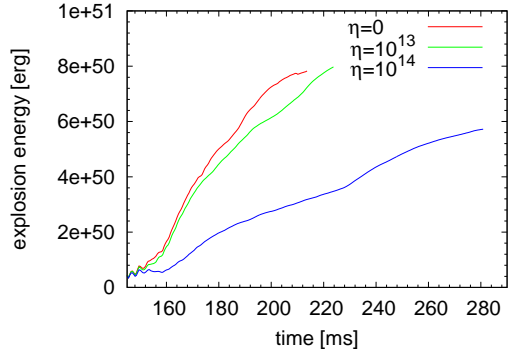}
  \caption{Evolutions of explosion energies in model-series
    B\textit{m}-$\Omega$.} 
 \label{fig.eexp1.45}
\end{figure*}

Fig.~\ref{fig.force.45} shows the radial distribution of
accelerations, angularly averaged in the eruption-region and time
averaged during 160-170~ms, according to Eq.~(\ref{eq.acc}). The
left-top panel shows that an 
acceleration of matter mainly takes place around $r\sim 100$~km. There,
although a pressure acceleration alone does not overcome an inward 
gravitational acceleration (see the left-bottom panel), the total
acceleration is positive with a help of a magnetic and centrifugal
acceleration (see the right-bottom panel).
From the right-top panel, it is found that a total acceleration there
is averagely larger in model $\eta_{-\infty}$, and that this is
primarily due a larger magnetic and centrifugal acceleration. 
In Fig.~\ref{fig.eexp2.45}, the distributions of
$dE_{\textrm{exp}}/dr$ at 170~ms are shown for model $\eta_{-\infty}$
and $\eta_{14}$. It is found that a matter eruption occurs for
$r\gtrsim 100$~km, and $dE_{\textrm{exp}}/dr$ is everywhere
larger in model $\eta_{-\infty}$. It is expected that a larger total
acceleration around $r\sim 100$~km is responsible for this.
Thus, it is likely that again in the present model-series, a
resistivity makes the explosion less energetic by decreasing a
magnetic and centrifugal acceleration as in model-series
B\textit{s}-$\Omega$. 

\begin{figure*}
\begin{center}
\epsscale{1}
\plotone{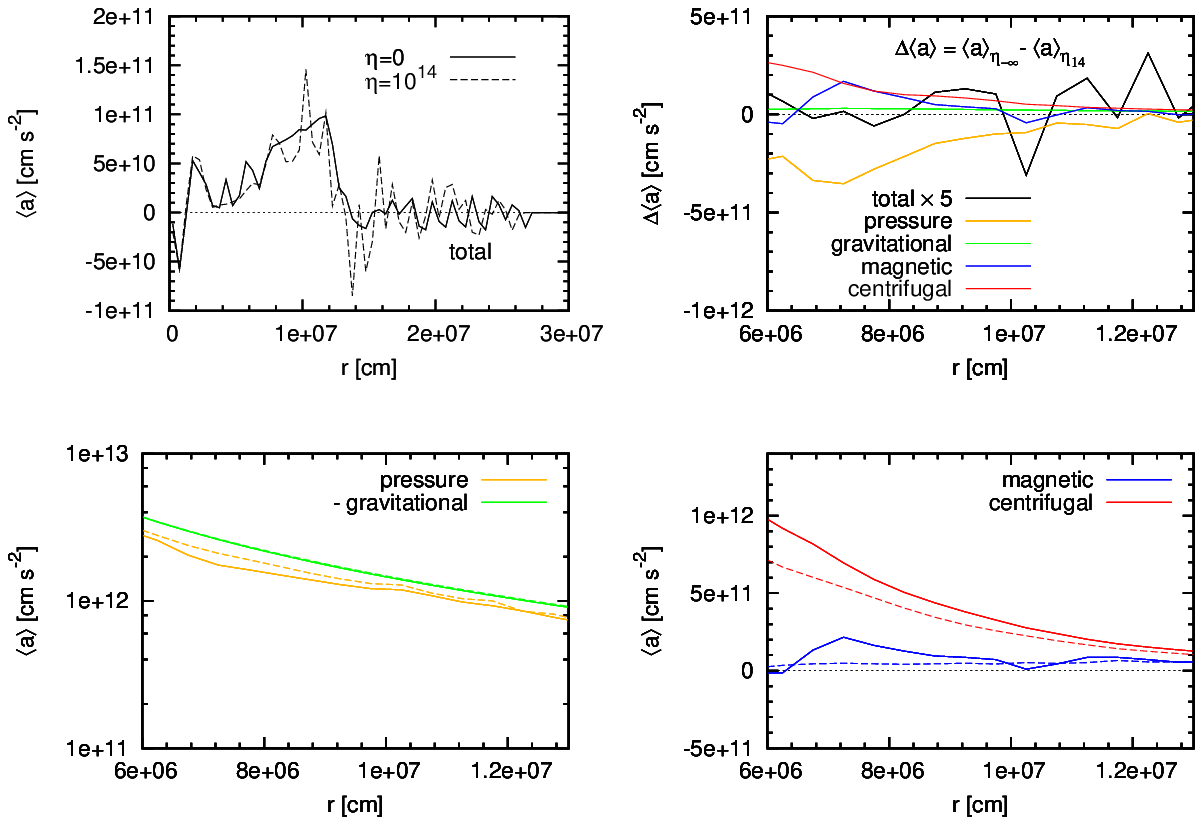}
  \caption{Same as Fig.~\ref{fig.force.55} but for model
    B\textit{m}-$\Omega$-$\eta_{-\infty}$ and
    B\textit{m}-$\Omega$-$\eta_{14}$. Radial accelerations are
    angularly averaged in the eruption-region and time averaged during
    $t=160-170$~ms. In the right-top panel the difference in the total
    acceleration is multiplied by 5.} 
 \label{fig.force.45}
\end{center}
\end{figure*}

\begin{figure*}
-\epsscale{0.5}
\plotone{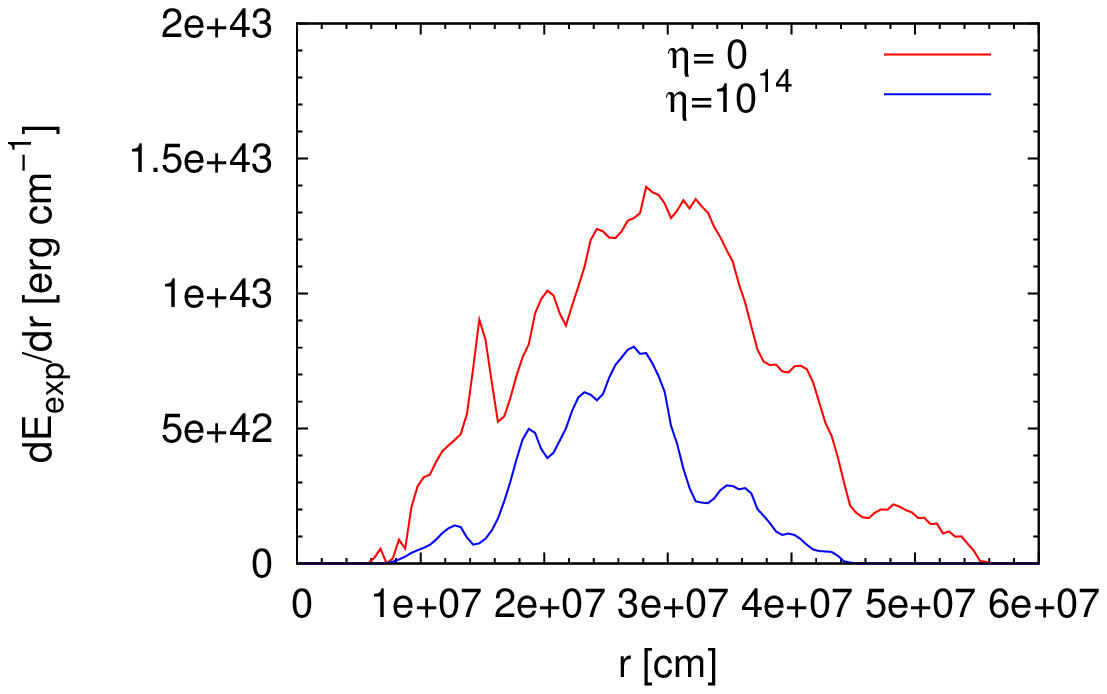}
\caption{Radial distributions of
    $dE_{\textrm{exp}}/dr$ at 170~ms in model
    B\textit{m}-$\Omega$-$\eta_{-\infty}$ and
    B\textit{m}-$\Omega$-$\eta_{14}$.}
\label{fig.eexp2.45}
\end{figure*}

\begin{figure*}
\epsscale{1}
\plottwo{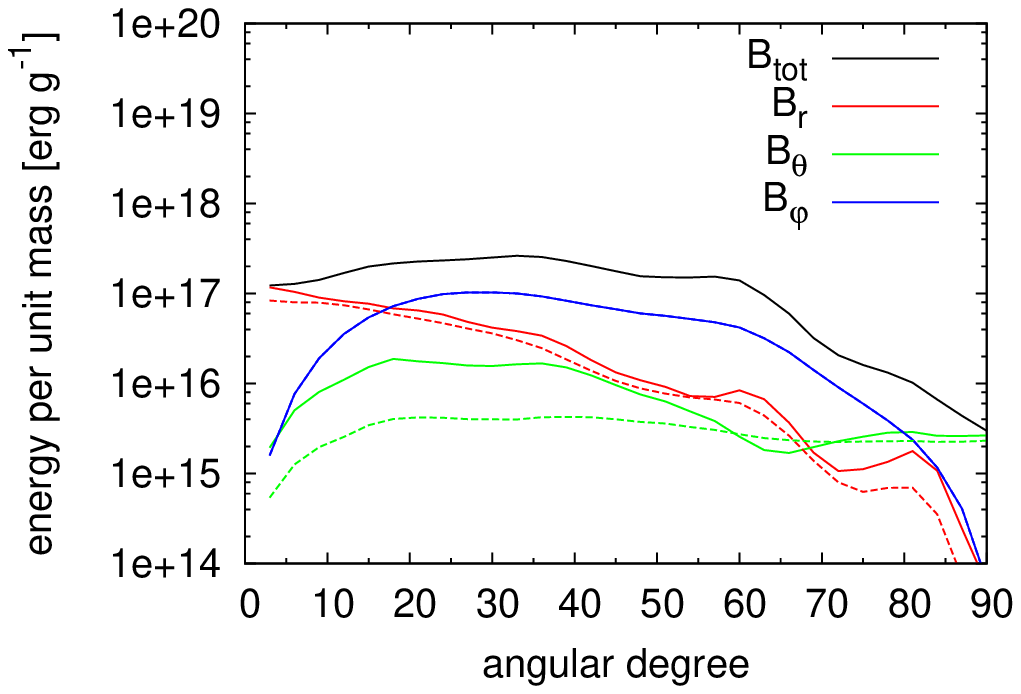}{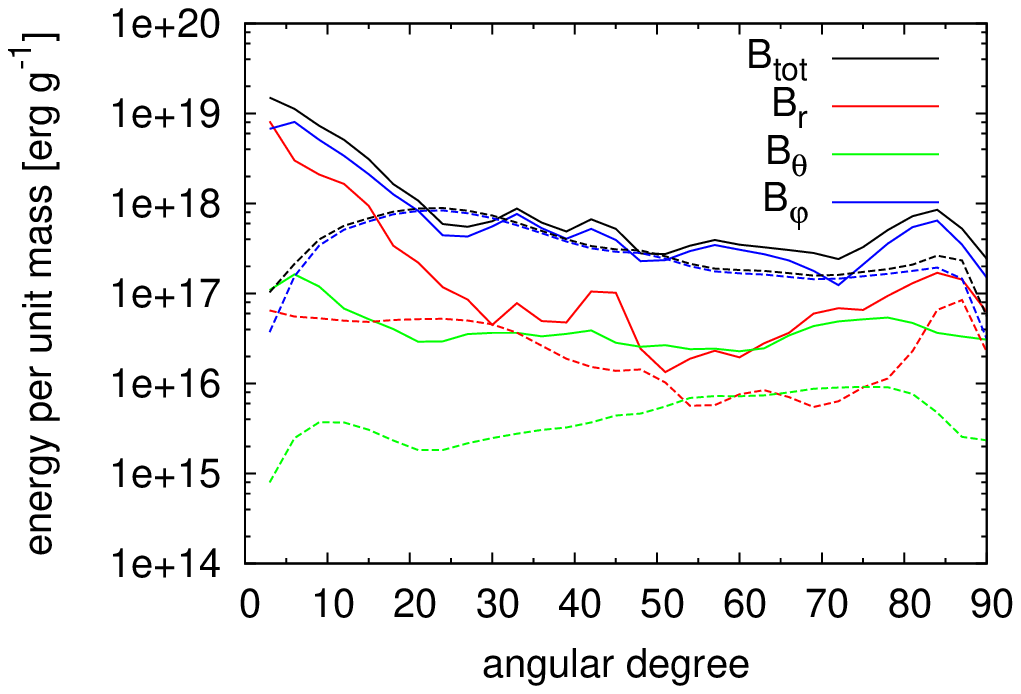}
\caption{Angular distributions of magnetic energies per unit mass
  averaged over $50$~km$<r<0.9\times r_{\textrm{sh}}$ at
  $t=143$~ms (2~ms after bounce; \textit{left}) and $t=180$~ms (39~ms
  after bounce; \textit{right}). The solid and dashed lines
  are for models 
  B\textit{m}-$\Omega$-$\eta_{-\infty}$ and
  B\textit{m}-$\Omega$-$\eta_{14}$, respectively.}  
\label{fig.ethm.45}
\end{figure*}

\begin{figure*}
\epsscale{1}
\plottwo{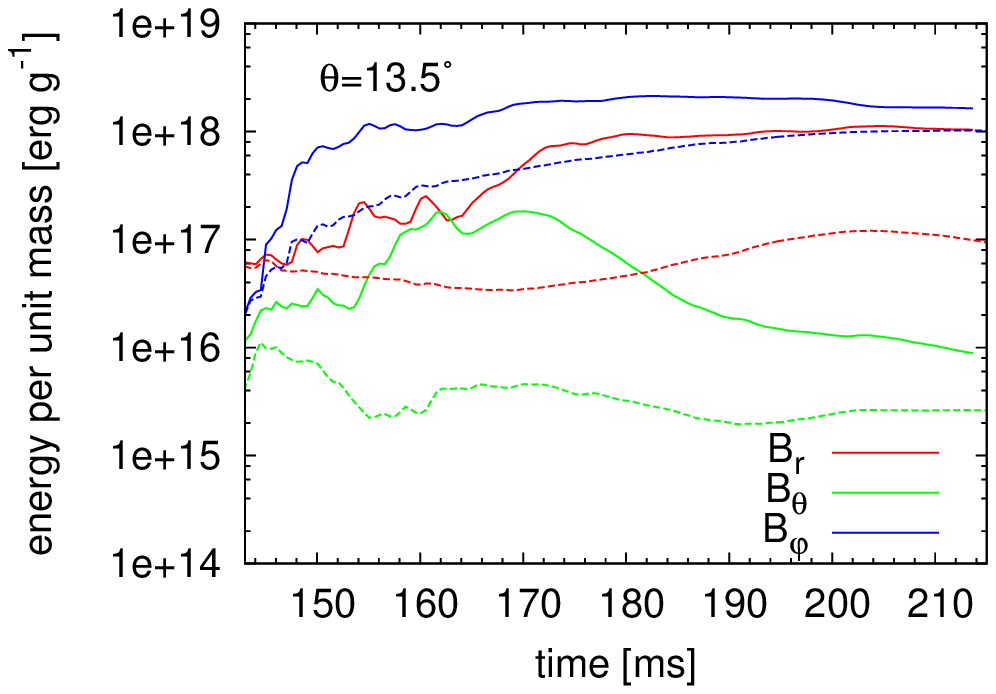}{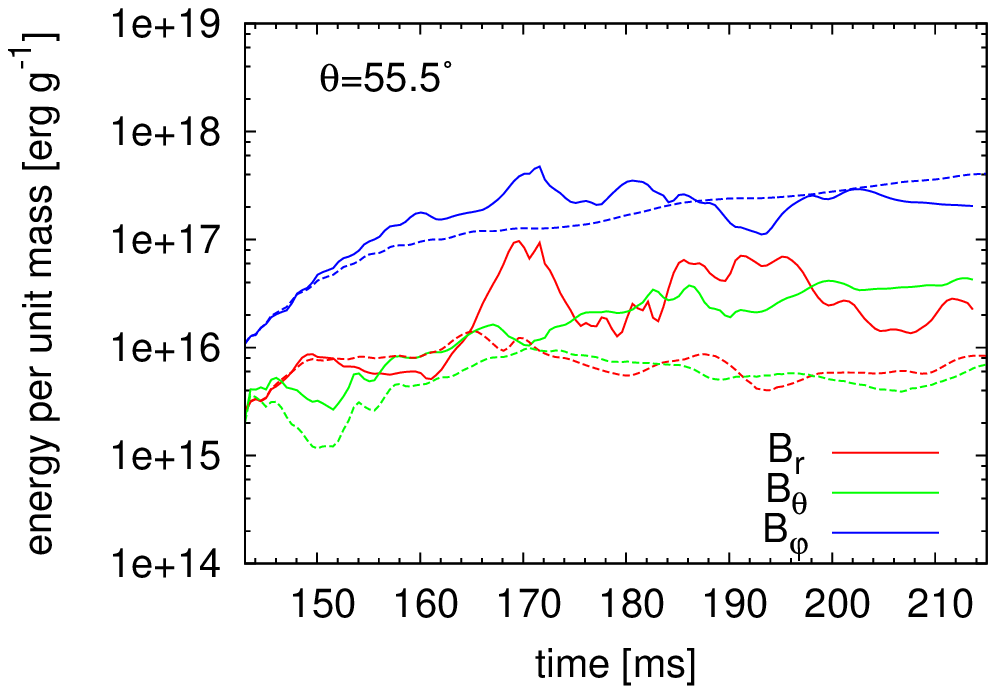}
  \caption{Evolutions of the average magnetic energies per unit
    mass of volume $V_{13.5}$ (\textit{left}) and $V_{55.5}$
    (\textit{right}). The solid lines are for model
    B\textit{m}-$\Omega$-$\eta_{-\infty}$ while the dashed lines are for 
    B\textit{m}-$\Omega$-$\eta_{14}$. The graphs are plotted from
    $t=143$~ms (2~ms after bounce).}  
 \label{fig.t-ethm.45}
\end{figure*}

\subsubsection{Magnetic Field Amplification}\label{sec.mfa.45}
A magnetic field amplification in the present model-series goes on in
a similar way as in the stronger magnetic field case
B\textit{s}-$\Omega$. Fig.~\ref{fig.ethm.45} shows the distributions
of magnetic energies per unit mass, averaged over $50$~km$<r<0.9\times
r_{\textrm{sh}}$ at $t=143$~ms (2~ms after bounce) and
$t=180$~ms (39~ms after bounce), for model
B\textit{m}-$\Omega$-$\eta_{-\infty}$ and
B\textit{m}-$\Omega$-$\eta_{14}$. Again, it is observed that the
contrast in a magnetic energy per unit mass
over $\theta$ becomes stronger at the latter time. The evolutions of
the magnetic energies per unit mass in two representative volumes
$V_{13.5}$ and $V_{55.5}$ are shown in Fig.~\ref{fig.t-ethm.45}, which
indicates that the difference between them becomes larger from 143~ms
to 214~ms.

Here, we also evaluated each term in the evolution
equations of magnetic energies per unit mass (see Eqs.~(\ref{eq.amp1})
and (\ref{eq.amp2})). The results are shown in
Fig.~\ref{fig.bamp.4500} and \ref{fig.bamp.4514}. It is found
both in model $\eta_{-\infty}$ and $\eta_{14}$ that the amplification
mechanisms in volume~$V_{13.5}$ are similar to those found in
volume~$V_{25.5}$ of model-series B\textit{s}-$\Omega$,
i.e. the toroidal and radial magnetic energy are mainly amplified or
kept due to
an advection and shear (see \S~\ref{sec.mfa.55}). As in the former
model-series, we found here that the amplification is weaker in a
volume with a lager $\theta_{\textrm{s}}$, viz. weaker in volume $V_{55.5}$.

\begin{figure*}
\epsscale{1}
\plottwo{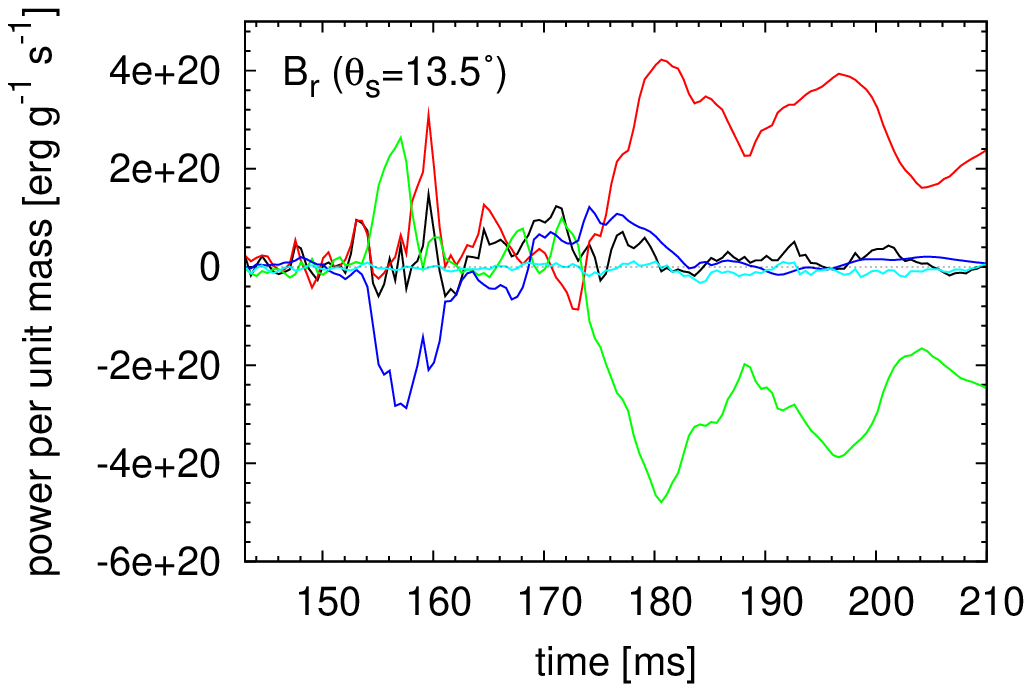}{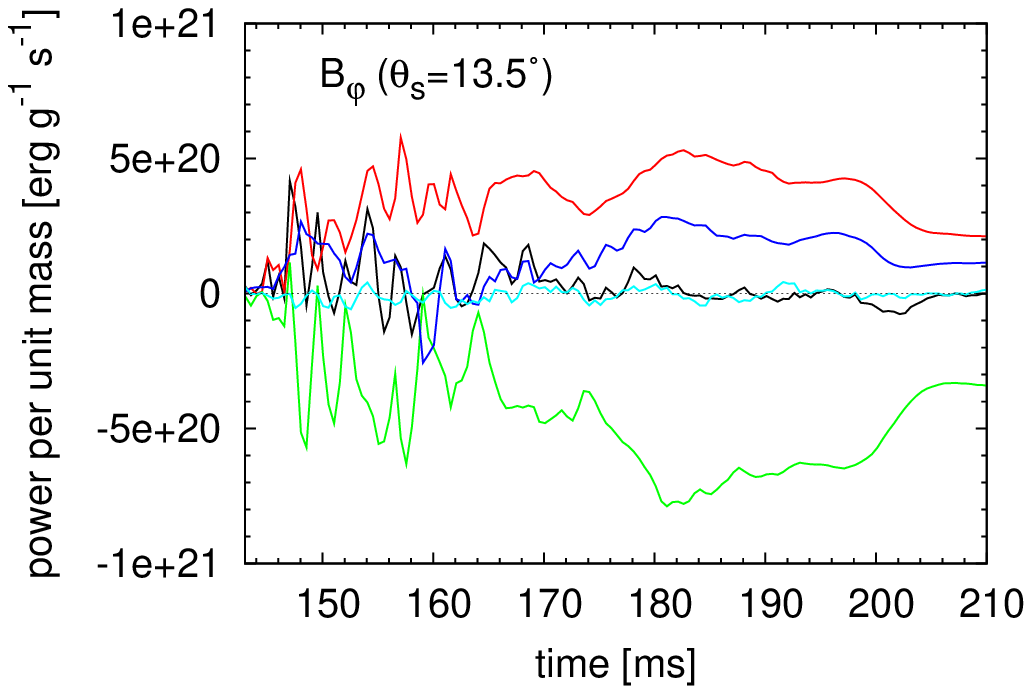}\\
\plottwo{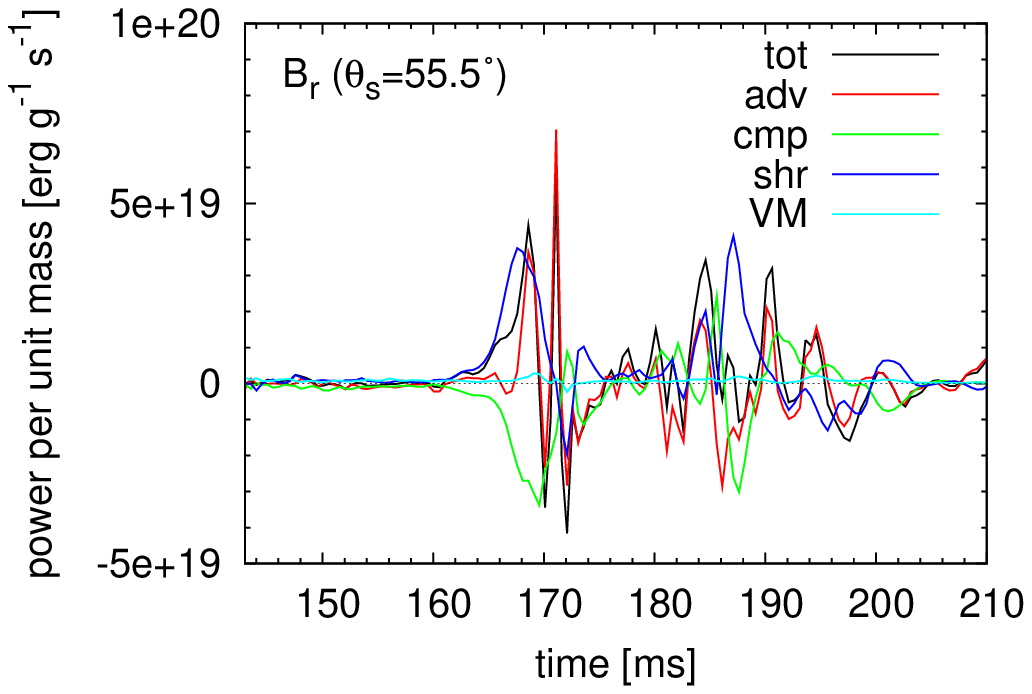}{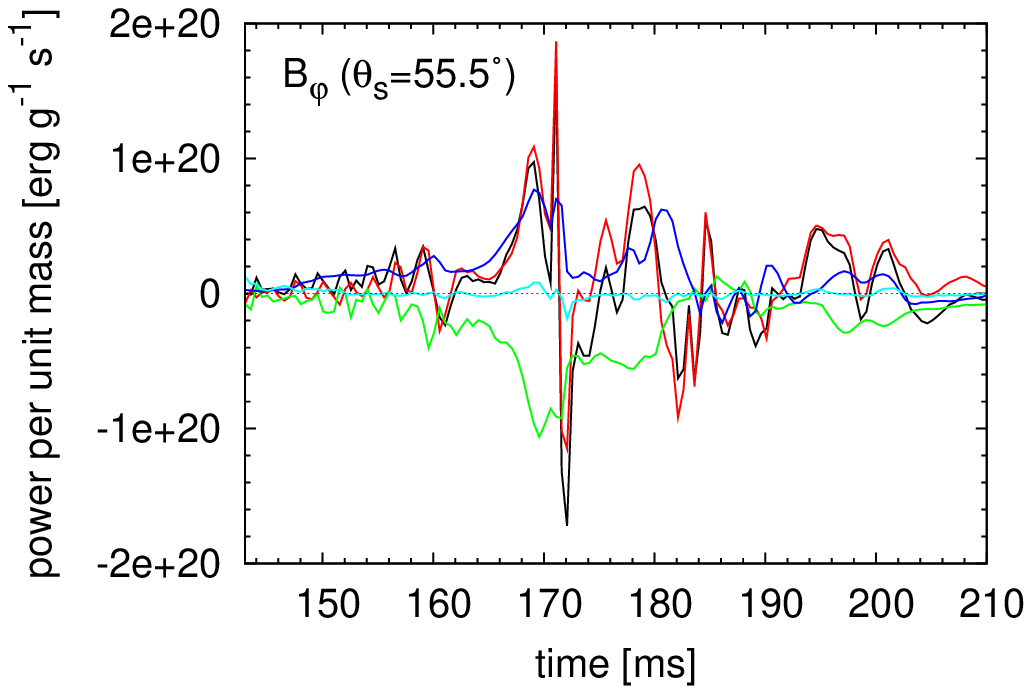}
\caption{Evolution of each $\dot{\mathcal{E}_r}$ and
  $\dot{\mathcal{E}_\phi}$ (see Eq.~(\ref{eq.amp2})) in a volume
  $V_{13.5}$ and $V_{55.5}$ in model
  B\textit{m}-$\Omega$-$\eta_{-\infty}$.}
\label{fig.bamp.4500}
\end{figure*}
\begin{figure*}
\epsscale{1}
\plottwo{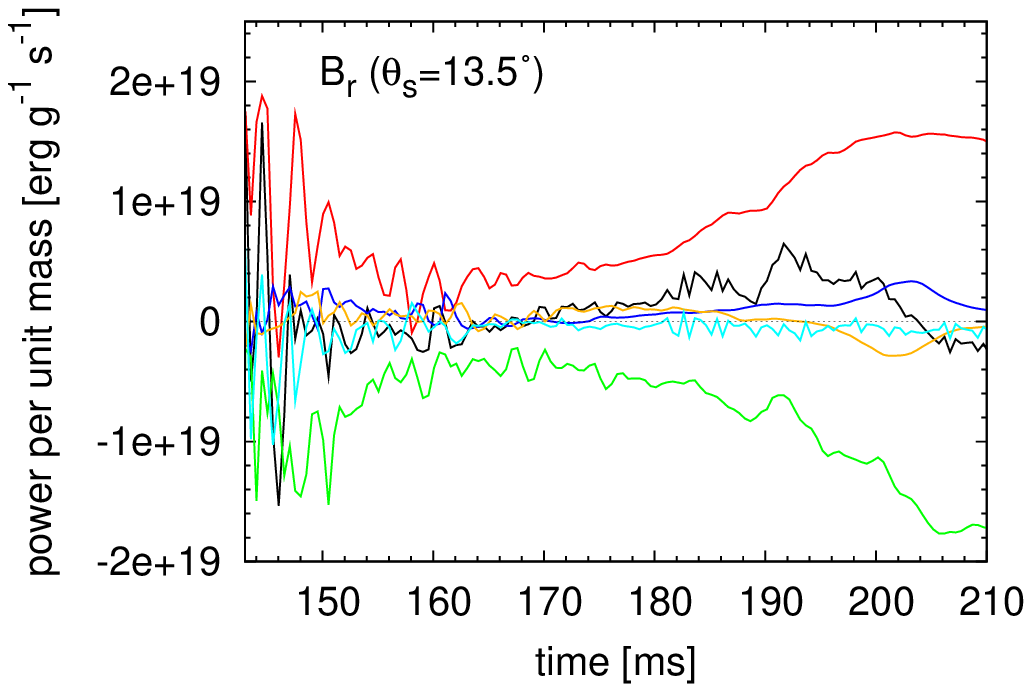}{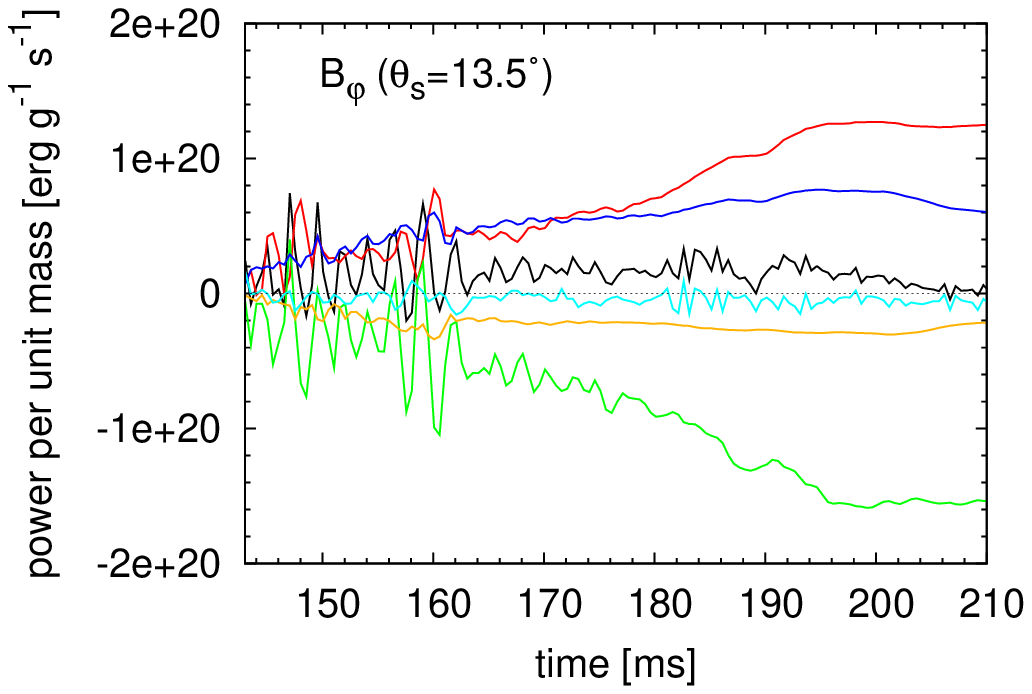}\\
\plottwo{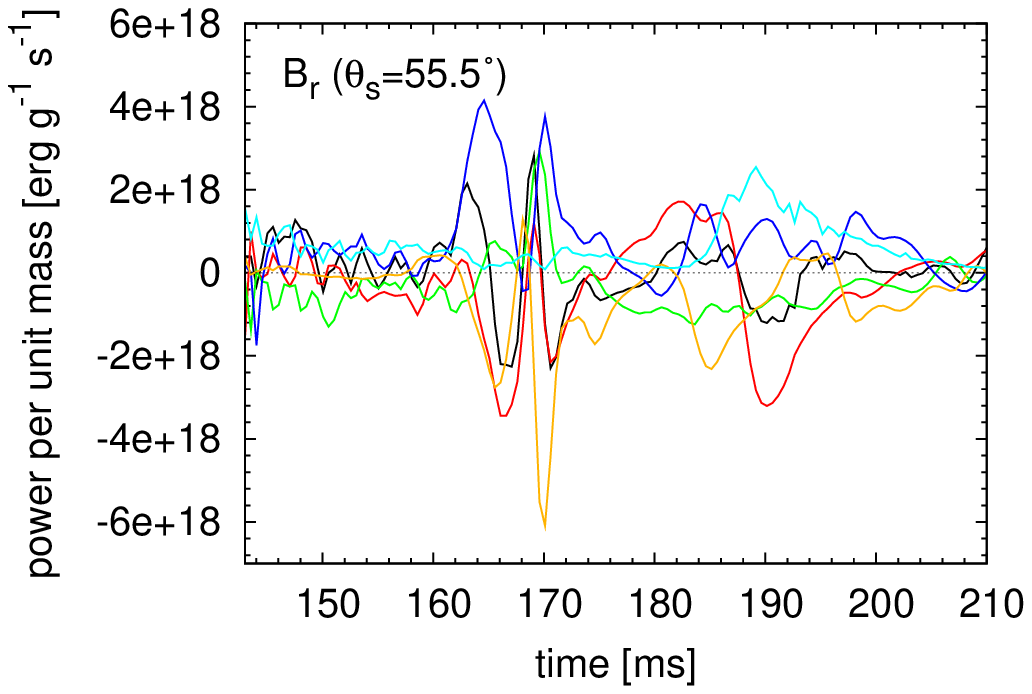}{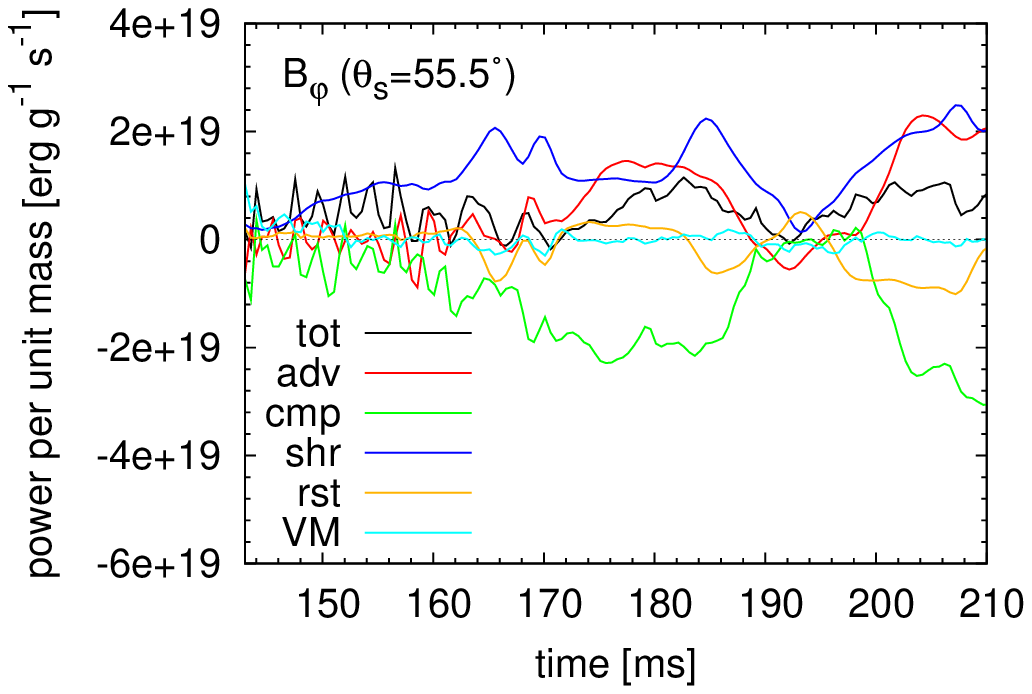}
\caption{Same as Fig.~\ref{fig.bamp.4500} but for model
  B\textit{m}-$\Omega$-$\eta_{14}$. Note that the vertical scale of
  each panel in Fig.~\ref{fig.bamp.4500} and~\ref{fig.bamp.4514} are
  different.} 
\label{fig.bamp.4514}
\end{figure*}

\begin{figure*}
\epsscale{1}
\plottwo{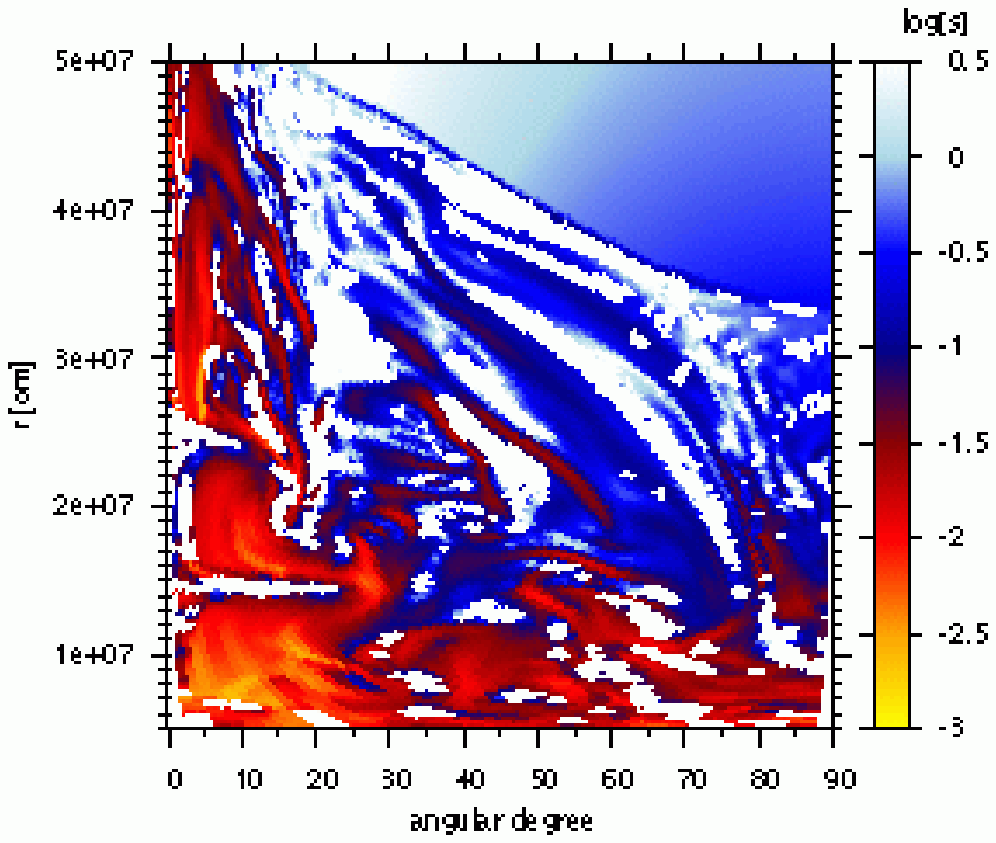}{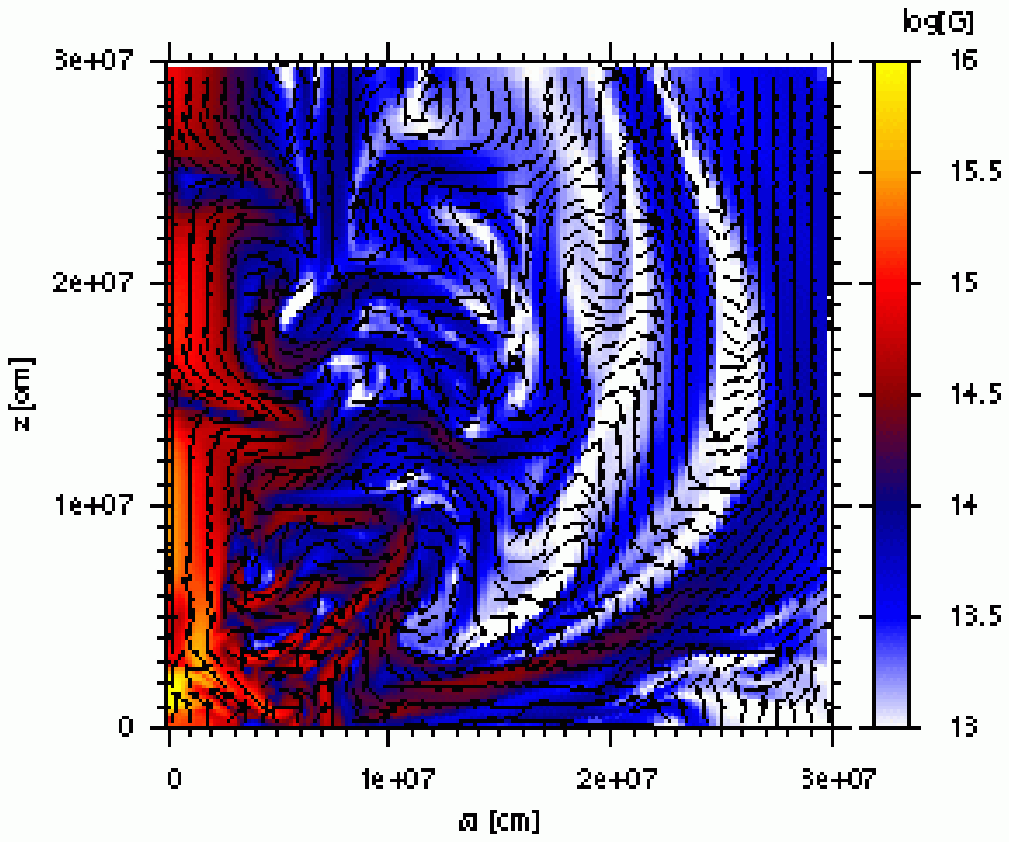}
\plottwo{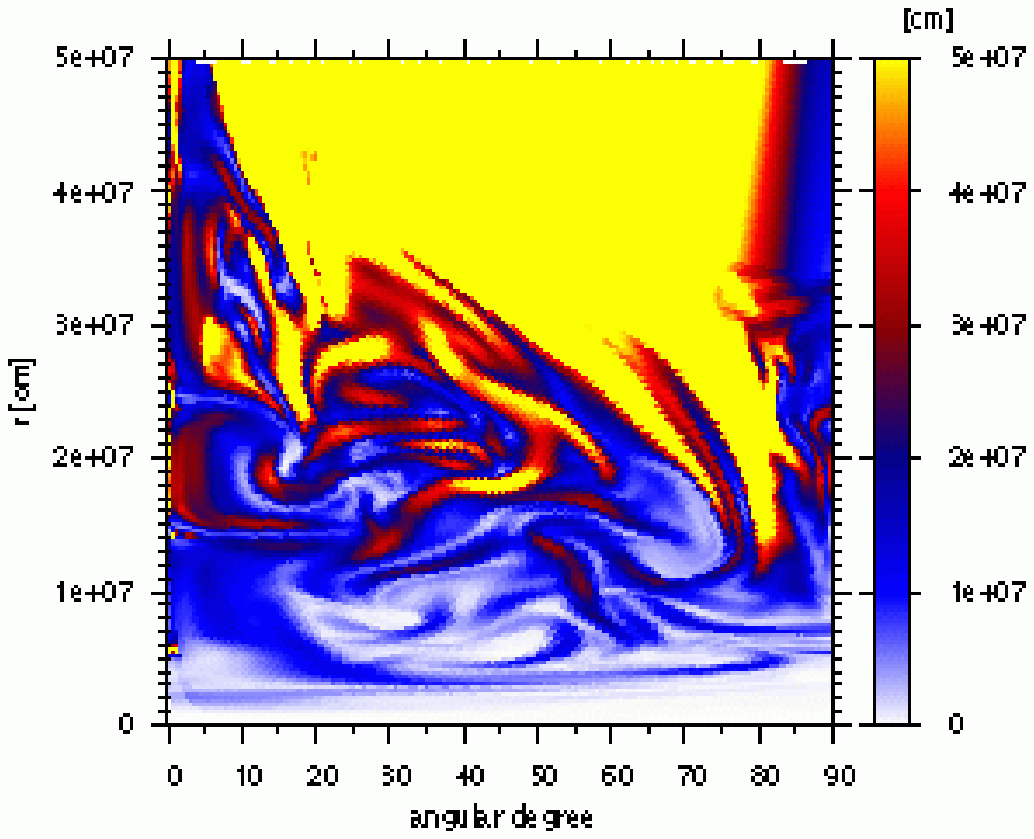}{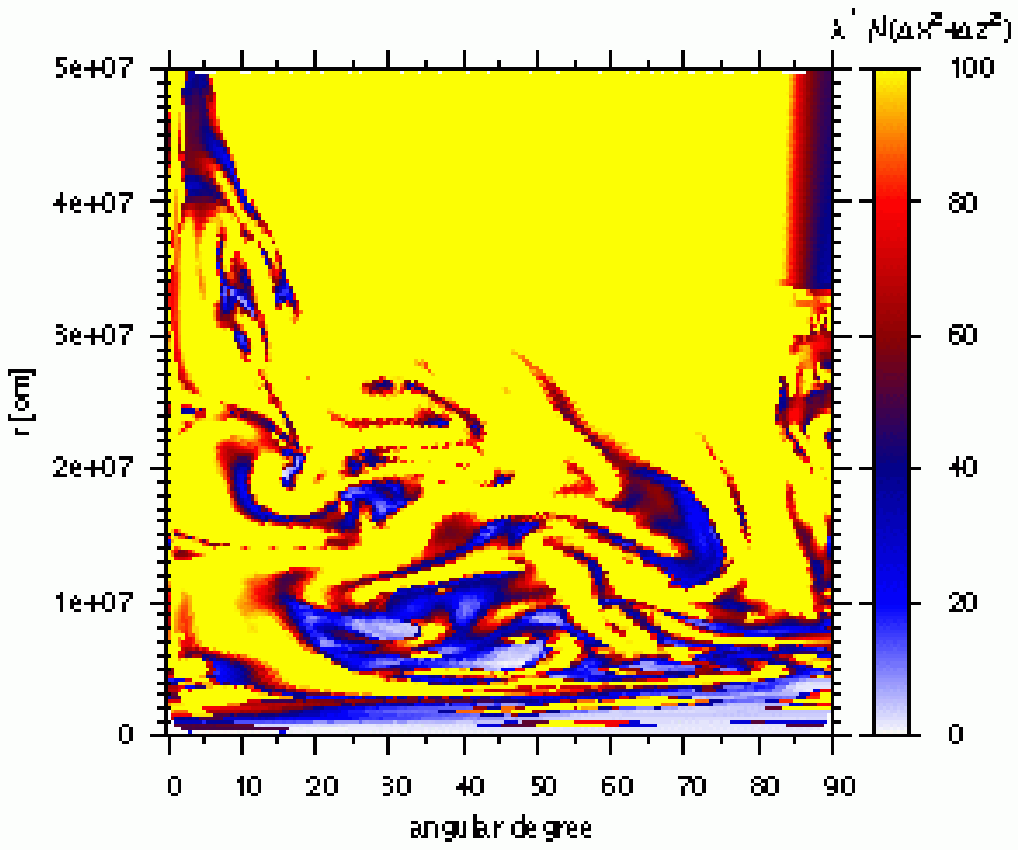}
  \caption{\textit{Left-top panel}: Distribution of MRI growth time
    in logarithmic scale in $\theta$-$r$ plane. White-colored
    regions include a location that is stable against the
    MRI. \textit{Right-top panel}:
    Poloidal magnetic field vectors on top of a color map of 
    poloidal magnetic field strength in logarithmic scale. 
    \textit{Left-bottom panel}: Distribution of fastest growing wave
    length, $\lambda^*\sim 2\pi 
    c_{\textrm{A}}/\Omega$, where $c_{\textrm{A}}$ is the Alfv\'en
    velocity of a poloidal magnetic field. \textit{Right-bottom panel}: 
    Distribution of fastest growing wave length divided by
    numerical cell size, $\lambda^*/\sqrt{(\Delta x)^2+(\Delta z)^2}$.
    Yellow-colored region includes MRI-stable locations.
    These figures are depicted for model
    B\textit{m}-$\Omega$-$\eta_{-\infty}$ at $t=170$~ms. }
 \label{fig.mri.4500}
\end{figure*}

\begin{figure*}
\epsscale{1}
\plottwo{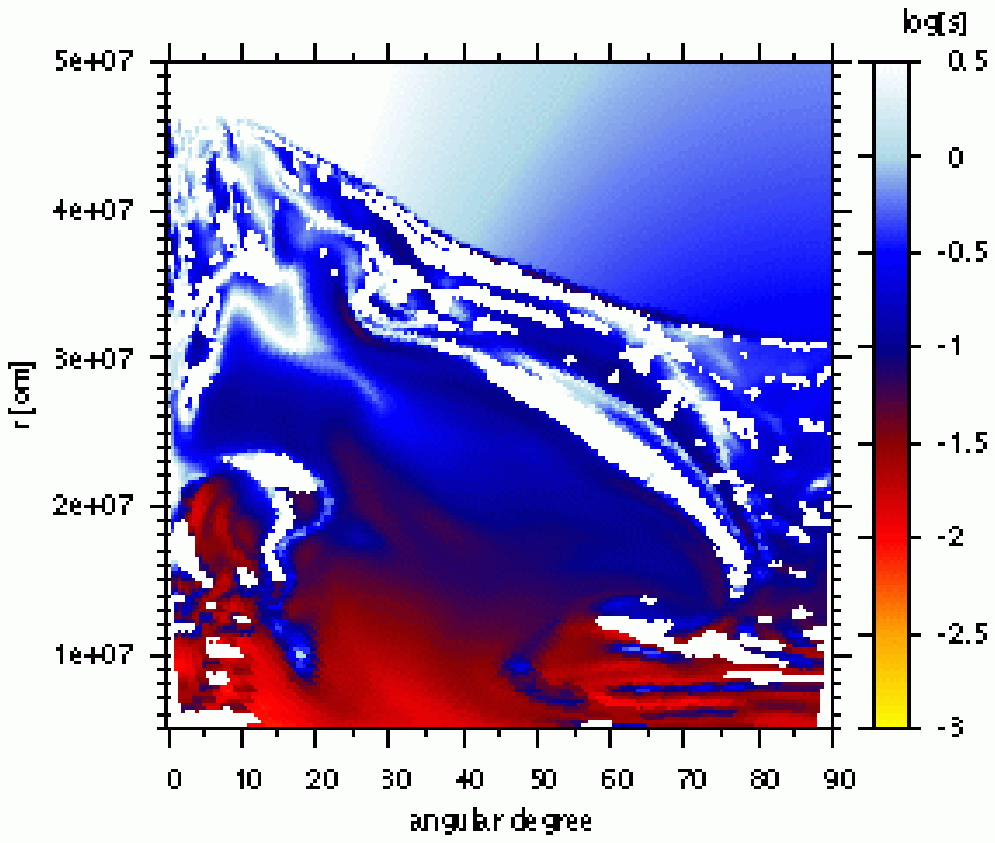}{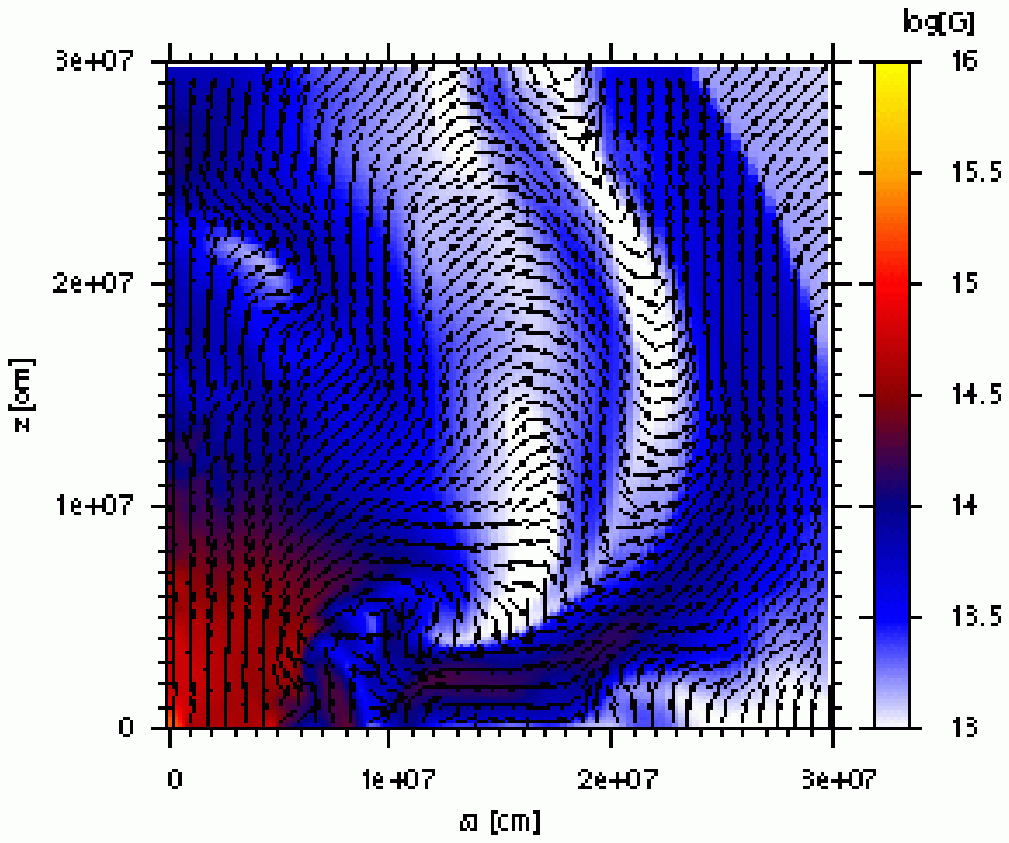}
\plottwo{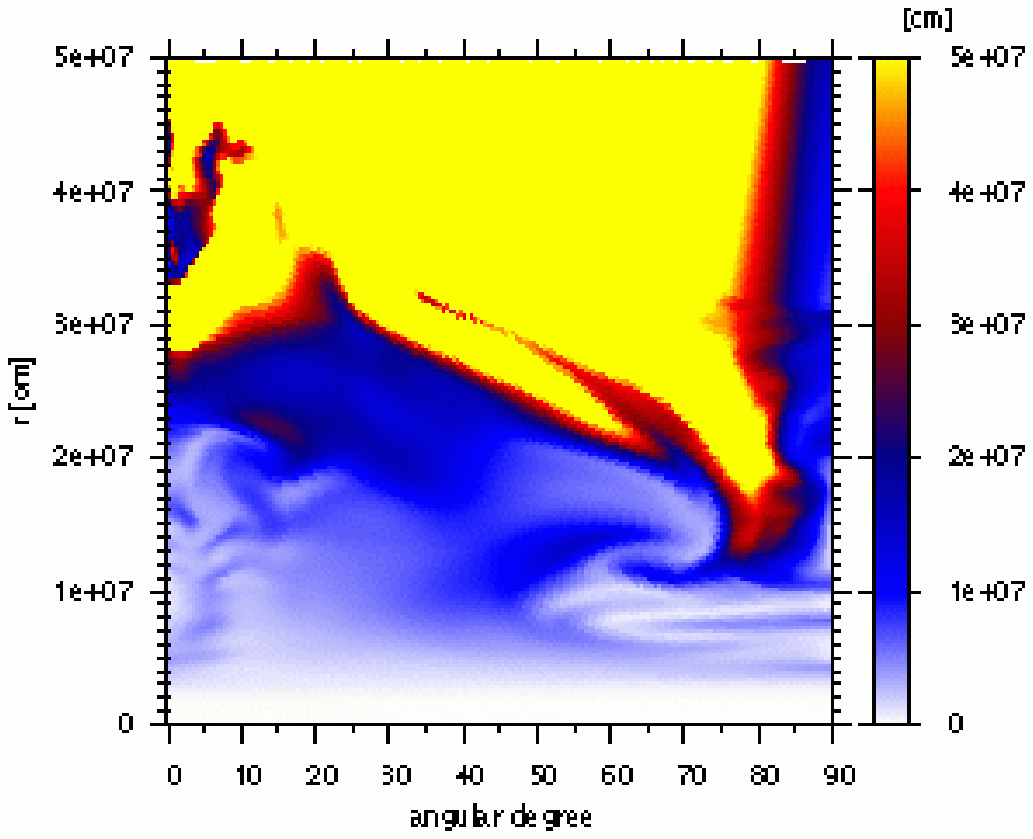}{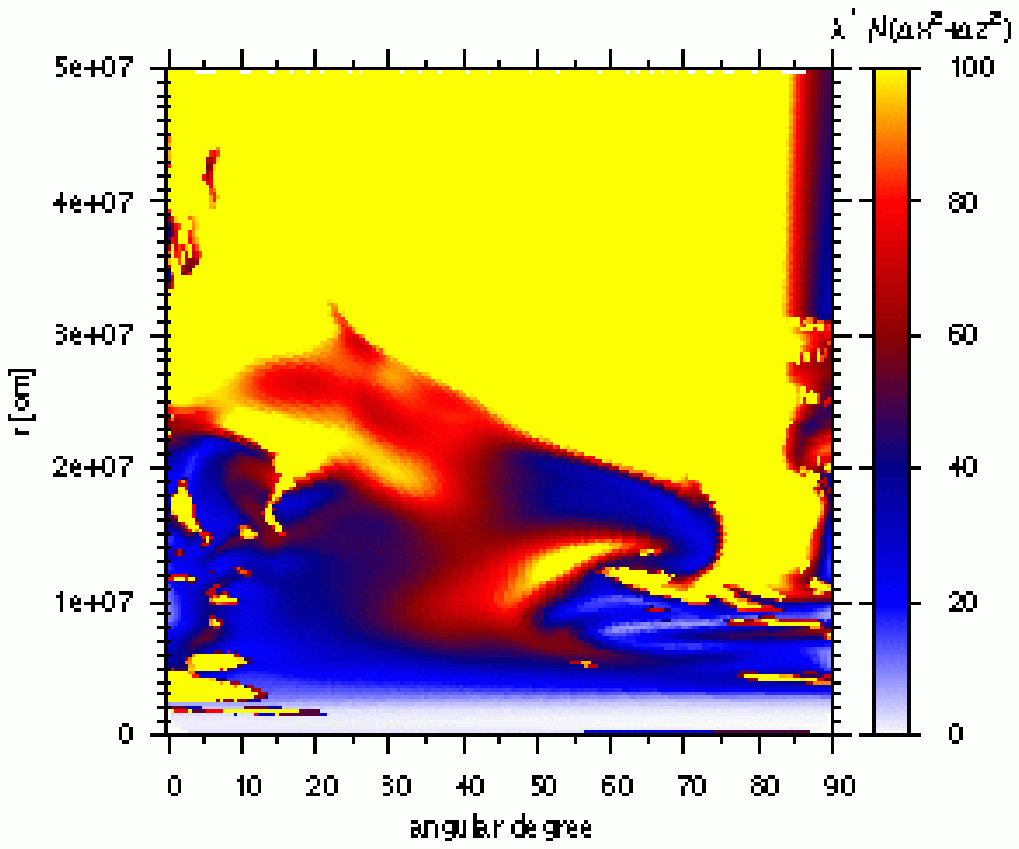}
  \caption{Same as Fig.~\ref{fig.mri.4500} but for model
    B\textit{m}-$\Omega$-$\eta_{14}$ at $t=170$~ms.} 
 \label{fig.mri.4514}
\end{figure*}

\begin{figure*}
\epsscale{0.5}
\plotone{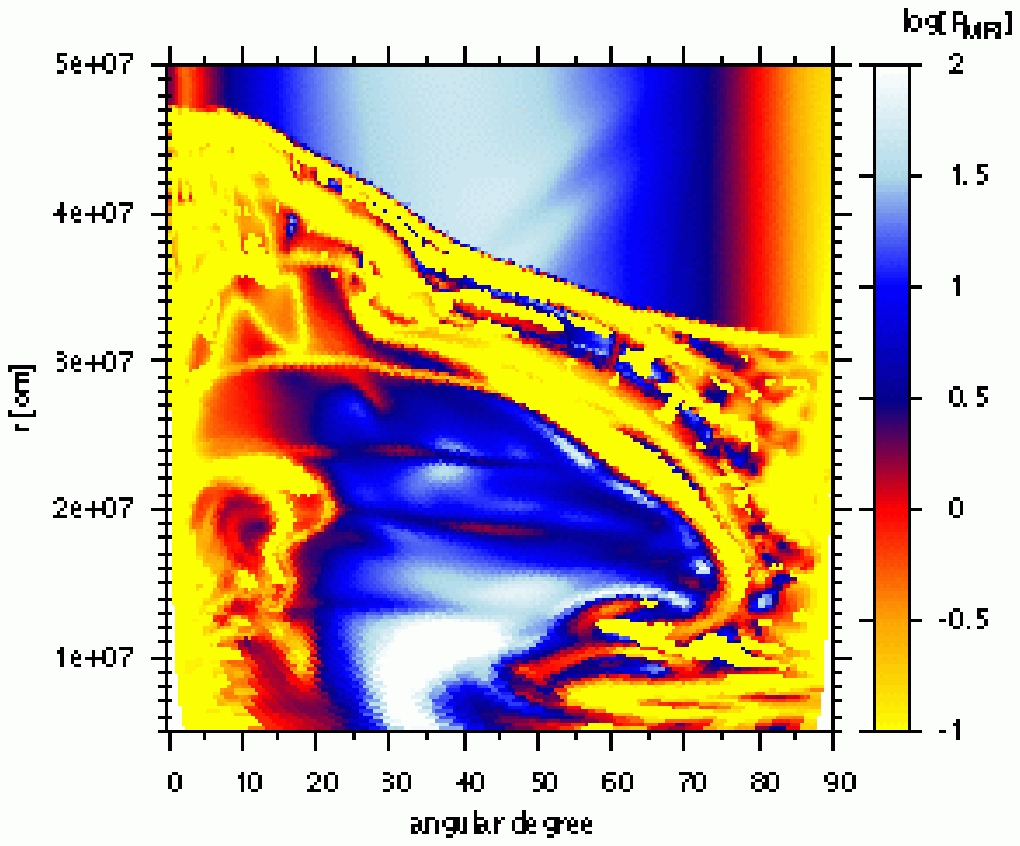}
  \caption{Same as Fig.~\ref{fig.rmri.5514}, but for model
    B\textit{m}-$\Omega$-$\eta_{14}$ at $t=170$~ms.} 
 \label{fig.rmri.4514}
\end{figure*}

The left-top panel of  
Fig.~\ref{fig.mri.4500} shows the distributions of MRI growth timescale
in $\theta$-$r$ plane for model B\textit{m}-$\Omega$-$\eta_{-\infty}$ at
$t=170$~ms (29~ms after bounce). It is found that a MRI growth timescale is
short, $\sim$ a few~ms to $10$~ms, in a considerable part for
$r\lesssim 20^\circ$, while for a larger $\theta$, a growth timescale
is averagely much longer. We found that a short growth timescale
mentioned above is kept from 148~ms ($7$~ms after bounce)
until the end of the simulation. In this model, the growth of MRI is
also assured by magnetic field-line bending, which start appearing
shortly after $150$~ms. The right-top panel of Fig.~\ref{fig.mri.4500}, which
depicts the structure of poloidal magnetic field for $t=170$~ms,
shows apparent field-line bending especially around the pole, where a
growth timescale is averagely short. 

In \S~\ref{sec.mfa.55}, we have seen that in both model
B\textit{s}-$\Omega$-$\eta_{-\infty}$ and
B\textit{s}-$\Omega$-$\eta_{14}$, the shear term for volume
$V_{25.5}$ starts increasing around the time when the growth of the MRI is
expected, and thus we speculated that the increase is caused by the MRI. 
In the present model, B\textit{m}-$\Omega$-$\eta_{-\infty}$, the shear
term for volume $V_{13.5}$ does not behaves like that. As seen in the
left-top panel of Fig.~\ref{fig.bamp.4500}, it has a large negative
value around $\sim 155$-$160$~ms and a large positive value around
$\sim 170$-$180$~ms. With this rather complicated behavior, it is
difficult to guess when the MRI gives a substantial contribution to the
shear term. Nonetheless, we could at least state that the MRI does
not play a crucial role in amplifying a magnetic field. This is
because the radial magnetic energy has already grown strong enough by the
period of $\sim 170$-$180$~ms, while only during this period of time,
the shear term works to amplify the magnetic field (see
Fig.~\ref{fig.t-ethm.45} and \ref{fig.bamp.4500}). 

In model~$\eta_{14}$, a region of short growth timescale is
not as limited to a small $\theta$ as in model~$\eta_{-\infty}$,
albeit a growth timescale is averagely longer, typically a few
10~ms. As seen in the left-top panel of Fig.~\ref{fig.mri.4514}, 
although a small-$\theta$ region is more favorable site for the MRI, a
large-$\theta$ region is still subjected to 
a fast MRI. However, as noted in \S~\ref{sec.mfa.55}, we should take
into account that the growth of MRI may be suppressed by resistivity.
In Fig.~\ref{fig.rmri.4514}, the distribution of
$R_{\textit{MRI}}$, resistive timescale divided by MRI growth timescale, is
depicted at $t=170$~ms for model~$\eta_{14}$. It is shown
that $R_{\textit{MRI}}$ is  
smaller than unity in the vicinity of the pole and equator, which
means no growth of MRI there. We found that, for
$20^\circ\lesssim\theta\lesssim 40^\circ$, $R_{\textit{MRI}}$ is
always larger than unity in a considerable parts of the fast MRI
regions. Hence, the MRI still 
seems to work also in model $\eta_{14}$. In the right-top panel of
Fig.~\ref{fig.mri.4514}, MRI-like field-line bending is also found
around the pole, albeit less prominent than in the ideal model.

In model $\eta_{14}$, the shear term for volume $V_{13.5}$ becomes large at
late time, $\sim 190$~ms (the left-top panel of
Fig.~\ref{fig.bamp.4514}). Since, as well as in the former models, the 
radial magnetic energy does not largely increase after this time (the
left panel of Fig.~\ref{fig.t-ethm.45}), the shear term and thus the
MRI does not seem to very important for a amplification of a magnetic
field, although they may play some role in keeping the strength of a
magnetic field. 

Compared with model-series B\textit{s}-$\Omega$, the computations of
present model-series have poorer spatial resolution for capturing MRI
(see the right-bottom panels of Fig.~\ref{fig.mri.4500} and
\ref{fig.mri.4514}). Especially, in model
B\textit{m}-$\Omega$-$\eta_{-\infty}$, the fastest growing wave length
is resolved with less than 10 numerical cells in some locations even for
$r>50$~km, where a magnetic field plays a dynamically important role.
However, since these locations are limited to small areas, we do not
expect that the dynamics would drastically change, if the growth of
the MRI were fully resolved there.

Comparing the right panels of Fig.~\ref{fig.ethm.55} and
\ref{fig.ethm.45}, the distribution of magnetic energy in model
B\textit{m}-$\Omega$-$\eta_{-\infty}$ is more concentrated towards
small $\theta$ than in model B\textit{s}-$\Omega$-$\eta_{-\infty}$.
Although the two figures are depicted at different times, the above
mentioned feature is still observed for the distributions compared at
the same physical time after bounce (e.g. 37~ms after bounce for the
both models). This may be explained as
follows. Because a magnetic field of model 
B\textit{m}-$\Omega$-$\eta_{-\infty}$ is 
weaker, a strong matter eruption supported by magnetic force is
restricted to a small region in the vicinity of the pole, where a
magnetic field is relatively strong. Then due to an amplification of
magnetic field by a radial advection, the contrast between a magnetic
energy in the vicinity of the pole and the other region becomes
stronger and stronger. 

\subsubsection{Aspect Ratio}

\begin{figure*}
\epsscale{1}
\plottwo{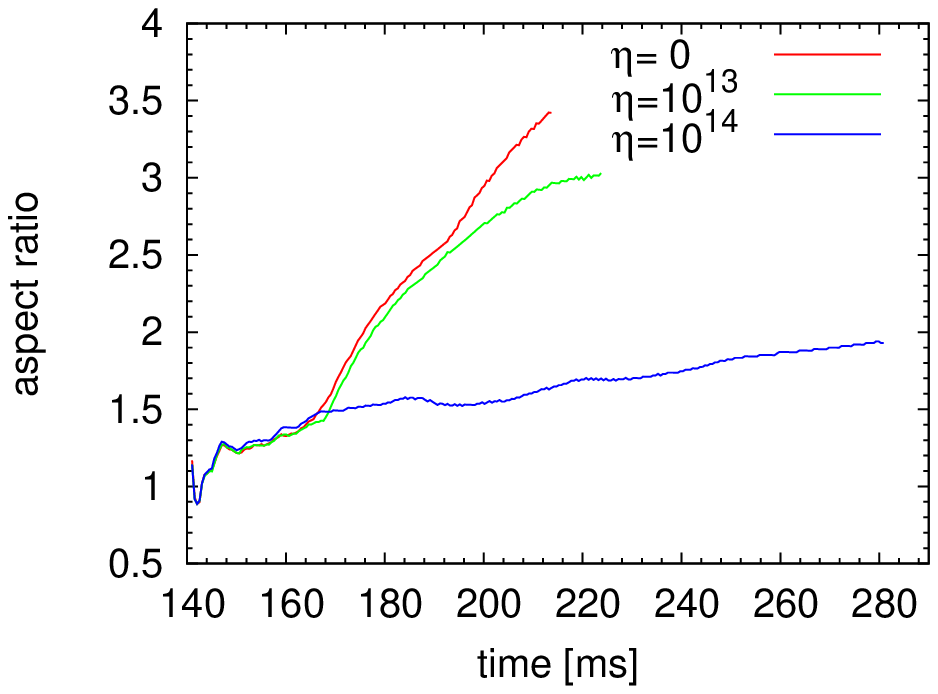}{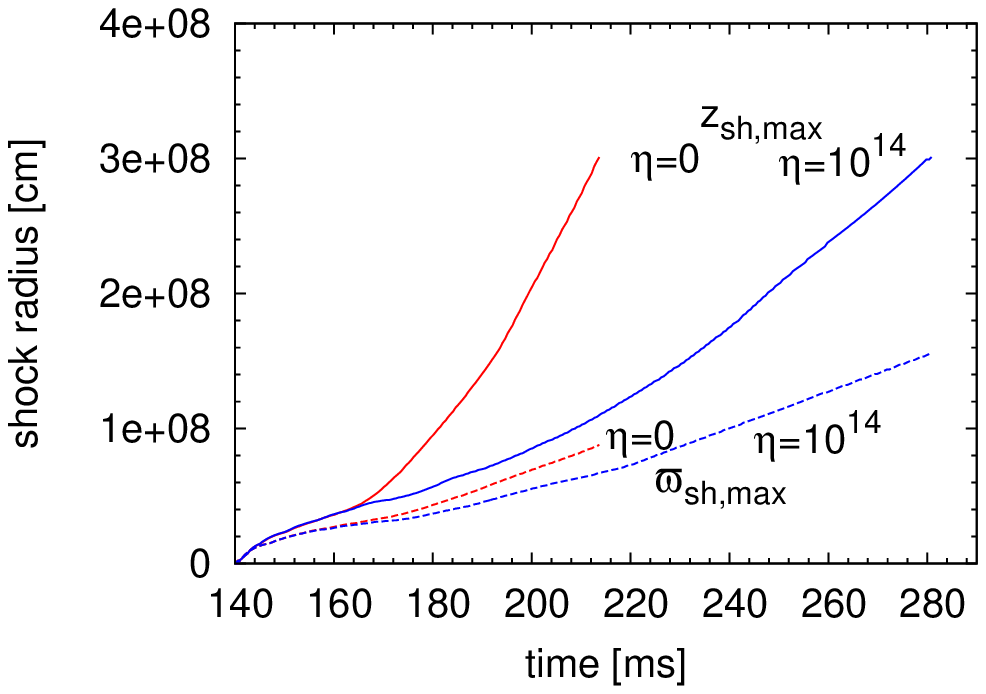}
\caption{Evolutions of aspect ratios (\textit{left panel}) and the
  maximum ejecta positions in $z$ and $\varpi$ (\textit{right panel})
  in model series B\textit{m}-$\Omega$.}
\label{fig.aspect.45}
\end{figure*}

The aspect ratio of a shock surface is also become smaller for a larger
resistivity in models-series B\textit{m}-$\Omega$ (see left panel of
Fig.~\ref{fig.aspect.45}). As seen in right panel of
Fig.~\ref{fig.aspect.45}, the difference in the aspect ratio is attained due
to that in $z_{\textrm{ej,max}}$ as in the case of models-series
B\textit{s}-$\Omega$. The same discussion as before will 
explain this feature, i.e. a stronger magnetic field and thus a larger
magnetic and centrifugal acceleration in a small $\theta$ compared
with in a large $\theta$ make a shock 
surface prolate, which is, however, less standout with the presence of a
resistivity. A trait of the present model-series is a larger impact of
resistivity, i.e. the aspect ratio is larger in model
B\textit{m}-$\eta_{-\infty}$ than in model B\textit{s}-$\eta_{-\infty}$,
while it is smaller in model B\textit{m}-$\eta_{14}$ than in model
B\textit{s}-$\eta_{14}$. 

The larger aspect ratio in model B\textit{m}-$\Omega$-$\eta_{-\infty}$
than in B\textit{s}-$\Omega$-$\eta_{-\infty}$ would be related 
to the distribution of the magnetic energy, which is more notably
concentrated towards small $\theta$ in the former model, as discussed in
\S~\ref{sec.mfa.45}. The smaller aspect ratio in model
B\textit{m}-$\Omega$-$\eta_{14}$ than in B\textit{s}-$\Omega$-$\eta_{14}$
may be understood as follows. Although the distributions of magnetic
energy is similar in their shape between the two models (compare the
right panels of Fig.~\ref{fig.ethm.55} and \ref{fig.ethm.45}), its
magnitude 
is averagely weaker in model B\textit{m}-$\Omega$-$\eta_{14}$,
reflecting the initial field strength, and a role of magnetic field in
erupting matter is less important than in
B\textit{s}-$\Omega$-$\eta_{14}$. As a result, the shape of 
a shock surface is not very much affected by magnetic field, 
and then the aspect ratio becomes smaller.

\subsection{Very Strong Magnetic Field and No Rotation --- Model Series
  B\textit{ss}-$\mho$}\label{sec.result.50}
The dynamical evolutions found in model-series
B\textit{ss}-$\mho$ are qualitatively different from those in the former
two model-series. Without rotation, neither the generation of a toroidal
magnetic field nor an angular momentum transfer occurs. However, a
magnetic field still seems to play some role. The initially strong
magnetic field affects the dynamics even during the collapse. A
typical evolution proceeds as follows. Although a total magnetic energy
per unit mass is a priori smaller around the equator, that of the
$B_\theta$ is conversely larger there. A strong $B_\theta$ attenuates a
matter infall around the equator, and leads to a weak bounce in the 
lateral direction. Due to the weak bounce, outgoing matters are soon
prevailed by falling matters, and then the infall-region
forms around the equator. Meanwhile, in the
other part, a bounce occurs strong enough to form the eruption-region,
and with a help of a magnetic force, especially a magnetic pressure
of $B_{\theta}$, the shock surface propagates outwards to reach 3000~km at
the end of the simulation. Fig.~\ref{fig.vradbvec.50} shows
the distributions of velocity and magnetic field in model
B\textit{ss}-$\mho$-$\eta_{-\infty}$ and 
B\textit{ss}-$\mho$-$\eta_{14}$ at $t=195$~ms (40~ms after bounce). In
each of the left panels, the formation of the 
eruption-region and infall-region is well observed. Each right panel
shows that a magnetic pressure is mildly important in the
eruption-region. 
As shown in Fig.~\ref{fig.t-eng.50} the magnetic energies in the two
models are $\sim 10^{51}$~erg. This implies that a magnetic field
has a potential to somewhat boost the explosion. Although no
significant difference between model $\eta_{-\infty}$ and
$\eta_{14}$ can be found in Fig.~\ref{fig.vradbvec.50}, as we will
see soon later, a resistivity still plays a certain role in the
dynamics of the present model-series.

\begin{figure*}
\begin{center}
\epsscale{1}
\plottwo{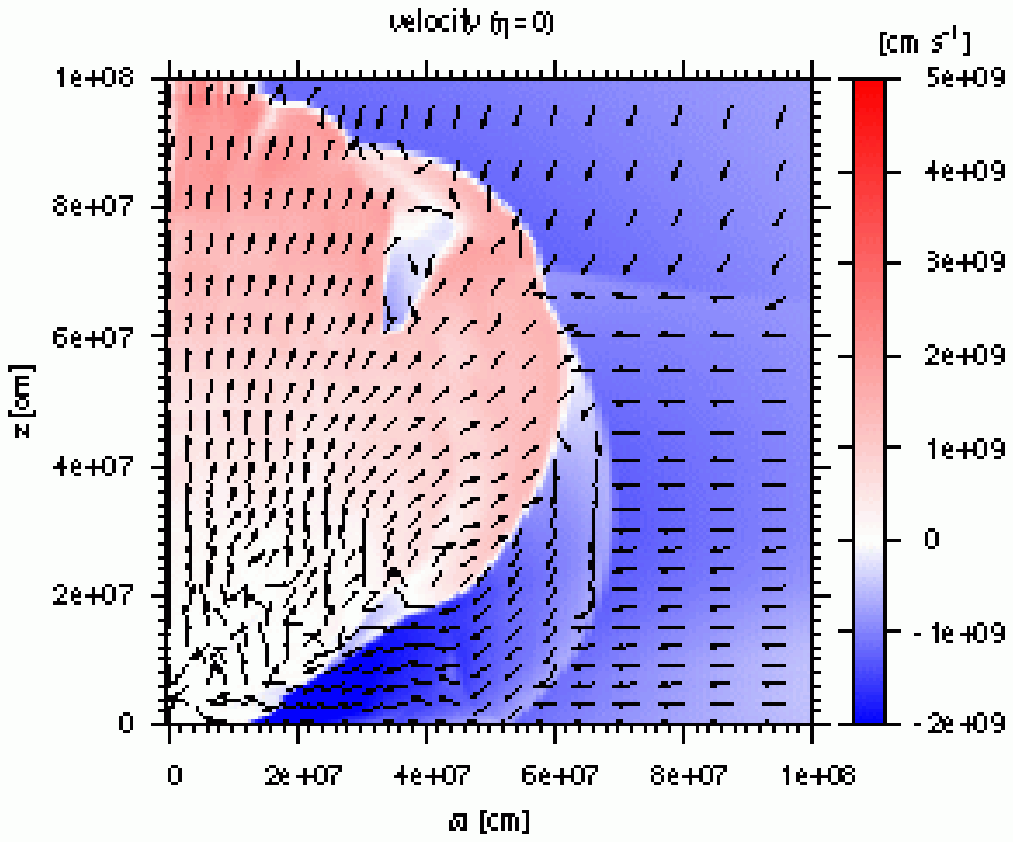}{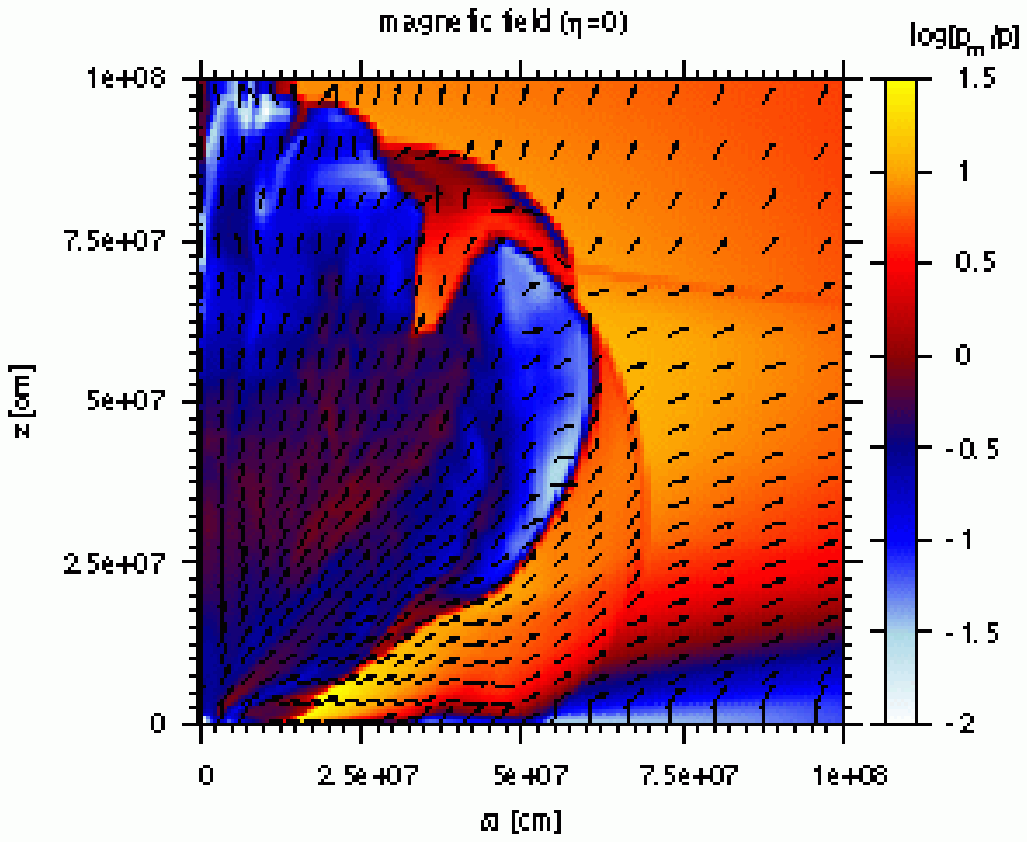}
\plottwo{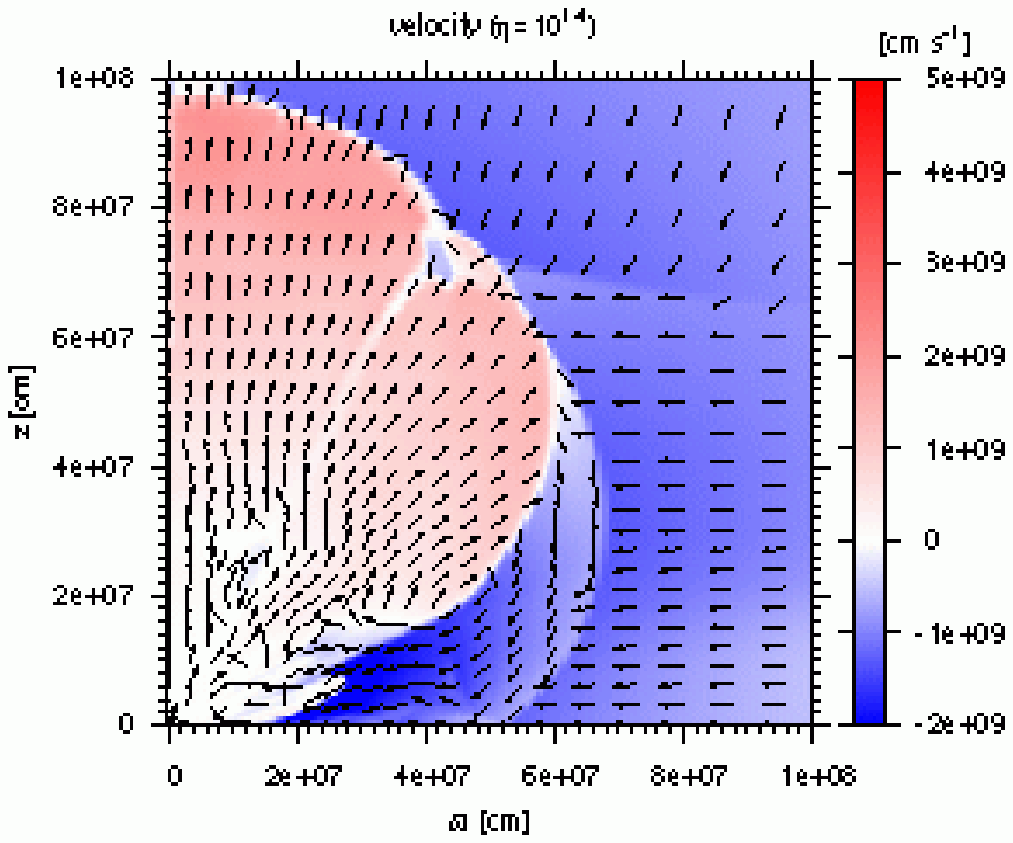}{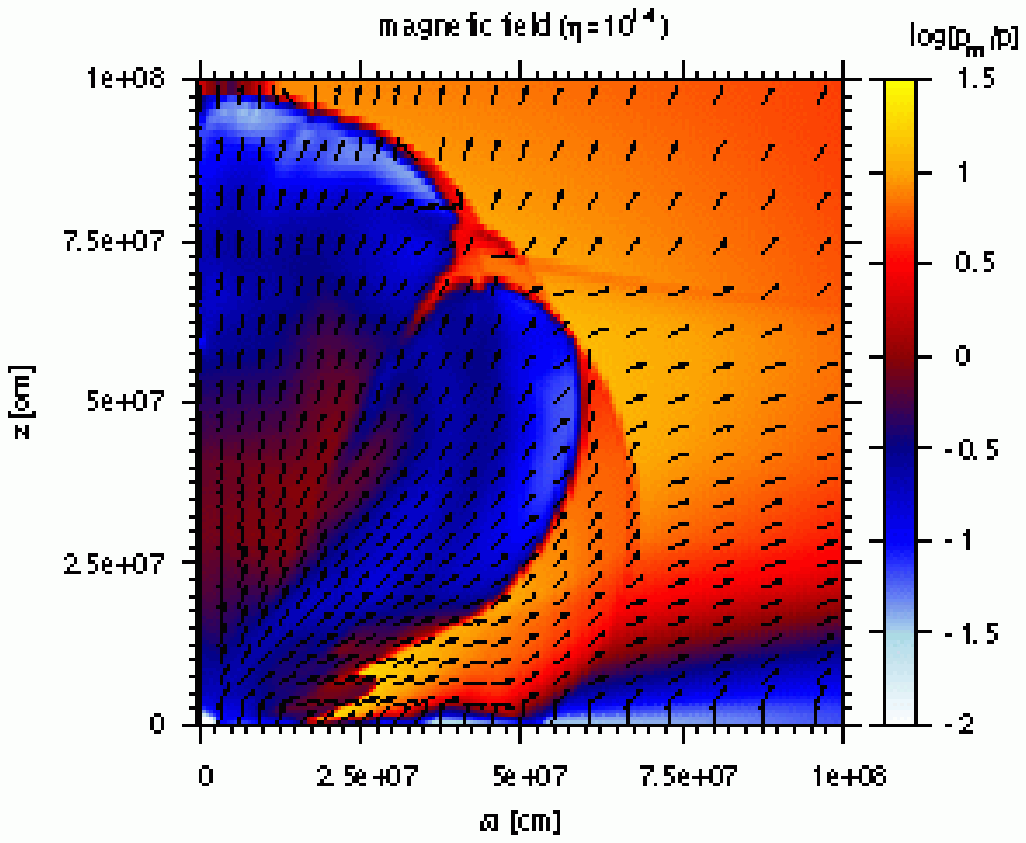}
  \caption{Same as Fig.~\ref{fig.vradbvec.55} but for model
    B\textit{ss}-$\mho$-$\eta_{-\infty}$ (upper panels) and
    B\textit{ss}-$\mho$-$\eta_{14}$ (lower panels) depicted at
  $t=195$~ms.} 
 \label{fig.vradbvec.50}
\end{center}
\end{figure*}

\begin{figure}
\epsscale{1}
\plotone{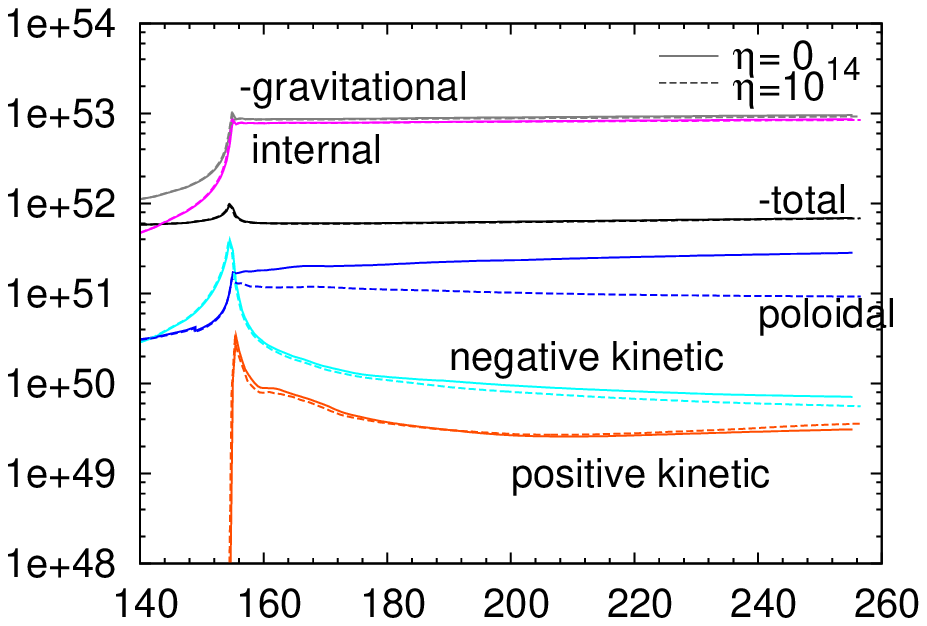}
  \caption{Evolutions of the total (black lines), internal (magenta),
    gravitational (gray), positive kinetic (orange),
    negative kinetic (cyan), and poloidal magnetic (blue) energy
    integrated over the whole numerical domain. The total
    and gravitational energy are multiplied by $-1$. The solid and dashed 
    lines are drawn for model B\textit{ss}-$\mho$-$\eta_{-\infty}$ 
    and B\textit{ss}-$\mho$-$\eta_{14}$, respectively.}
 \label{fig.t-eng.50}
\end{figure}

\subsubsection{Explosion Energy and Diffusion of Magnetic Field}
In Fig.~\ref{fig.exp1.50}, the evolution of explosion energies in
model-series B\textit{ss}-$\mho$ are plotted. As shown there, the
explosion energies are found to be $\sim 10^{50}$~erg, one order of
magnitude smaller than that of a canonical supernova. The figure also
indicates that the models involving resistivity produce a larger
explosion energy compared with the ideal model, which is opposite
to what is found in the model-serieses involving rotation.

\begin{figure}
\epsscale{1}
\plotone{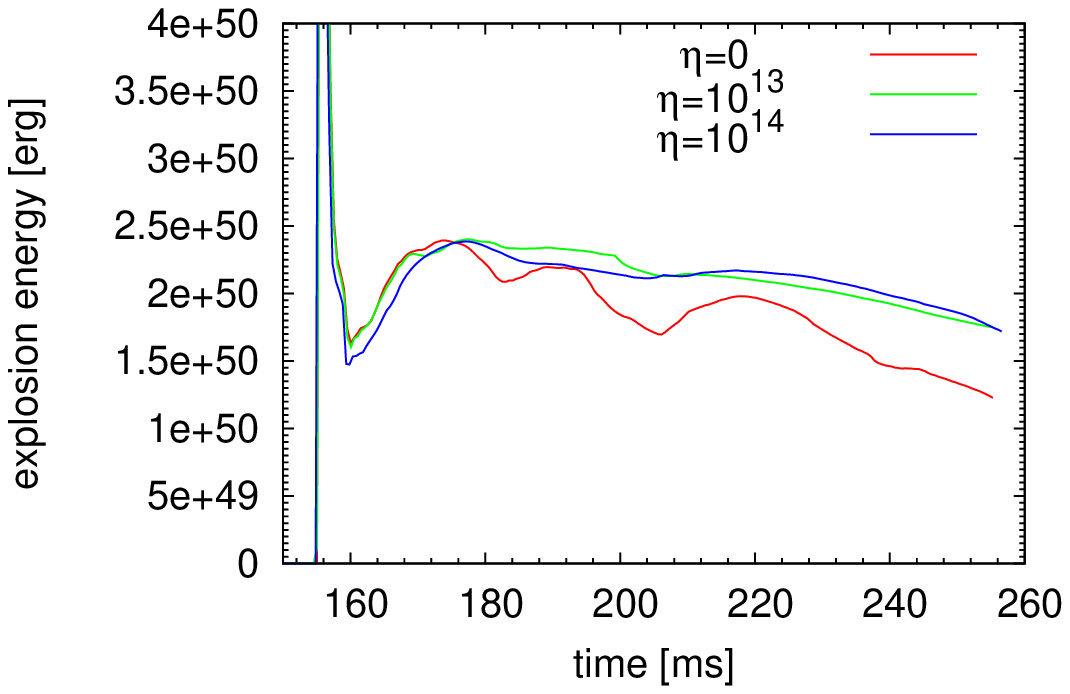}
  \caption{Evolutions of explosion energies in model-series
    B\textit{ss}-$\mho$.} 
 \label{fig.exp1.50}
\end{figure}

In Fig.~\ref{fig.exp2.50}, we plot the radial
distributions of $dE_{\textrm{exp}}/dr$ at $t=201$~ms in model
B\textit{ss}-$\mho$-$\eta_{-\infty}$ and
B\textit{ss}-$\mho$-$\eta_{14}$, which shows that the explosion
energy in model $\eta_{14}$ is grater than that in
model $\eta_{-\infty}$ at most radial range. We
examined whether a radial acceleration in model $\eta_{14}$ is
lager than in model $\eta_{-\infty}$, but found no substantial
difference between them. We also compared the two models by the radial
distributions of radial velocity angularly averaged in the
eruption-region at 201~ms (see Fig.~\ref{fig.exp3.50}). It is found
that a radial velocity is mostly larger in model~$\eta_{14}$. This
seems a key factor to know why the explosion 
energy is larger in model $\eta_{14}$. 

\begin{figure}
\epsscale{1}
\plotone{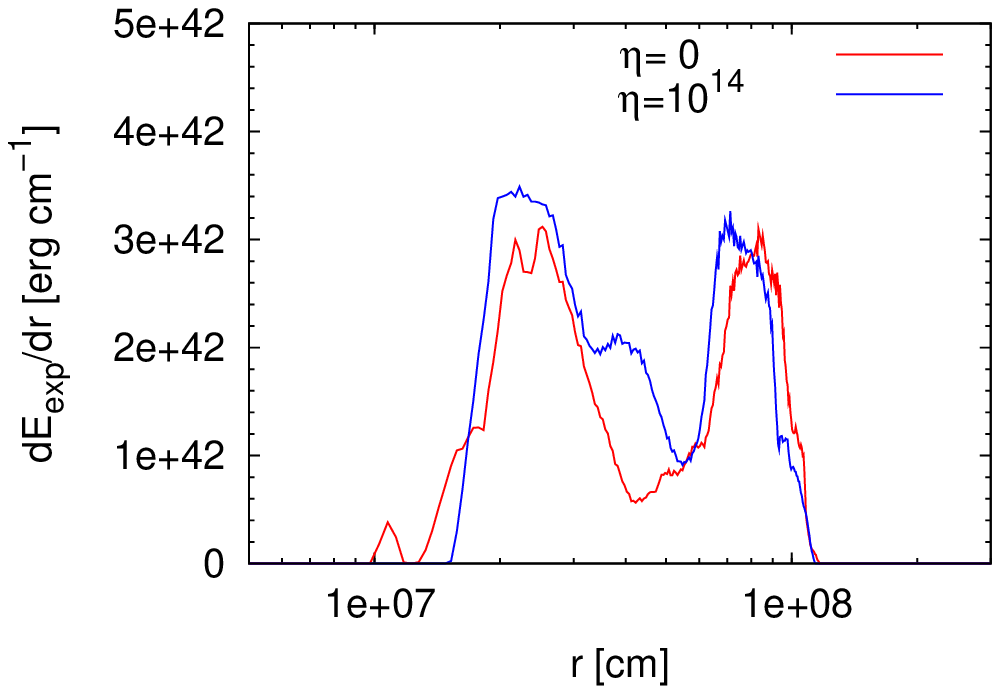}
  \caption{Radial distributions of $dE_{\textrm{exp}}/dr$ at
    $t=201$~ms in model B\textit{ss}-$\mho$-$\eta_{-\infty}$ and
    B\textit{ss}-$\mho$-$\eta_{14}$.}  
 \label{fig.exp2.50}
\end{figure}

\begin{figure}
\epsscale{1}
\plotone{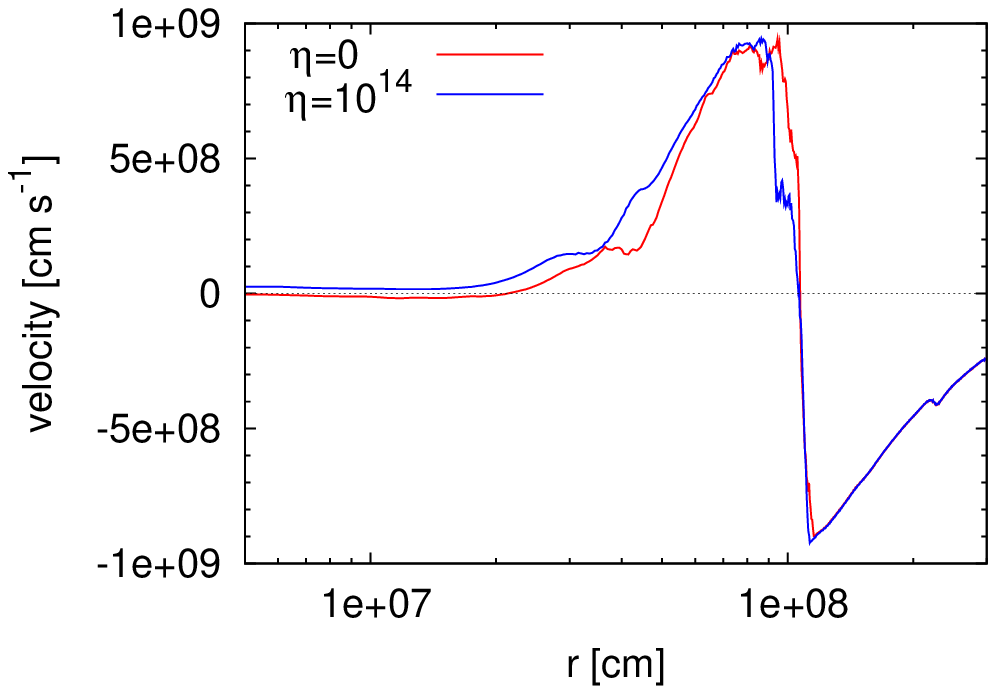}
  \caption{Radial distributions of radial velocities
    angularly averaged 
    in the eruption-region, at $t=201$~ms in model
    B\textit{ss}-$\mho$-$\eta_{-\infty}$ and
    B\textit{ss}-$\mho$-$\eta_{14}$.} 
 \label{fig.exp3.50}
\end{figure}

Then a question is why a radial velocity is larger in model
$\eta_{14}$, despite comparable strengths of radial accelerations.  
In Fig.~\ref{fig.vradz.50}, the velocity distributions in $\varpi$-$z$
plane at $t=201$~ms are depicted for the two models with a little
zooming towards the center. For model $\eta_{-\infty}$, it is
observed that matters flow from the infall-region into the
eruption-region around $r\lesssim 200$~km, and there damps a matter
ejection. In model $\eta_{14}$, on the other hand, a
matter flow into the eruption region is inhibited by a
"positive velocity island" observed around 150~km~$\lesssim
r\lesssim$~350~km in the infall-region, and a coherent
outflow of matter in the eruption-region is well kept even for
$r\lesssim 200$~km. In this way, the appearance of the positive
velocity island seems to be related to the difference in the velocity
distribution observed between the two models.  

\begin{figure*}
\epsscale{1}
\plottwo{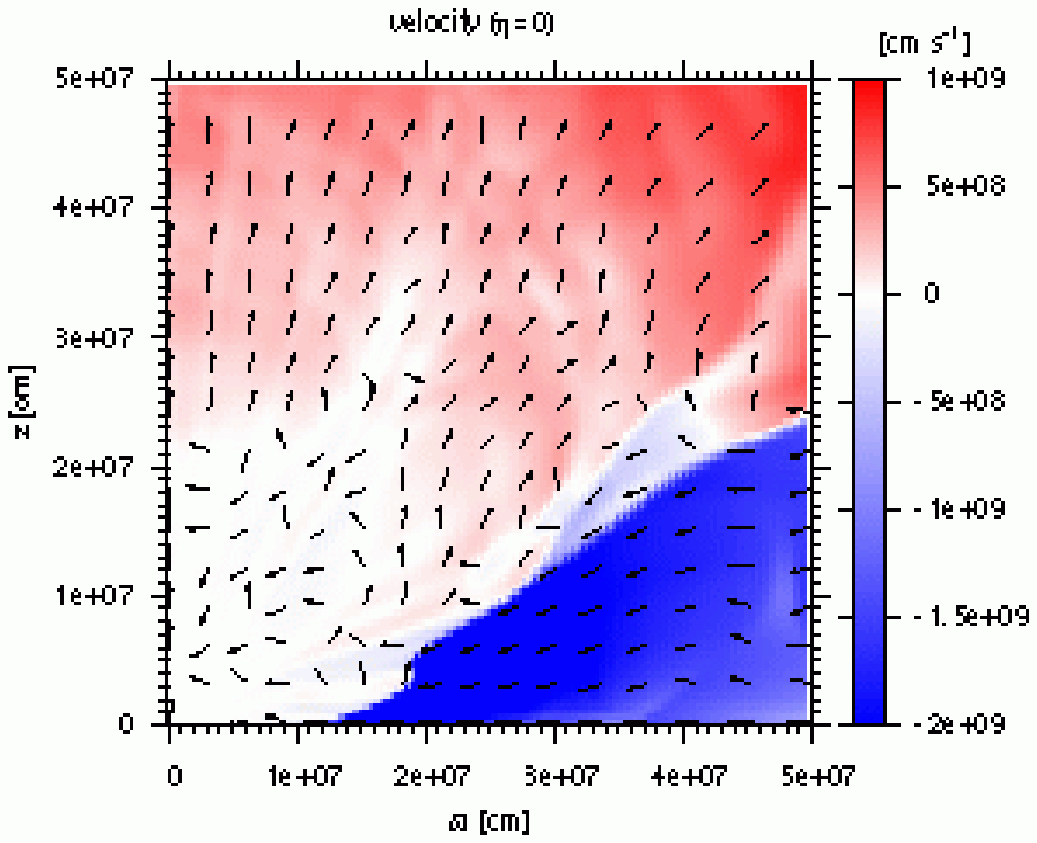}{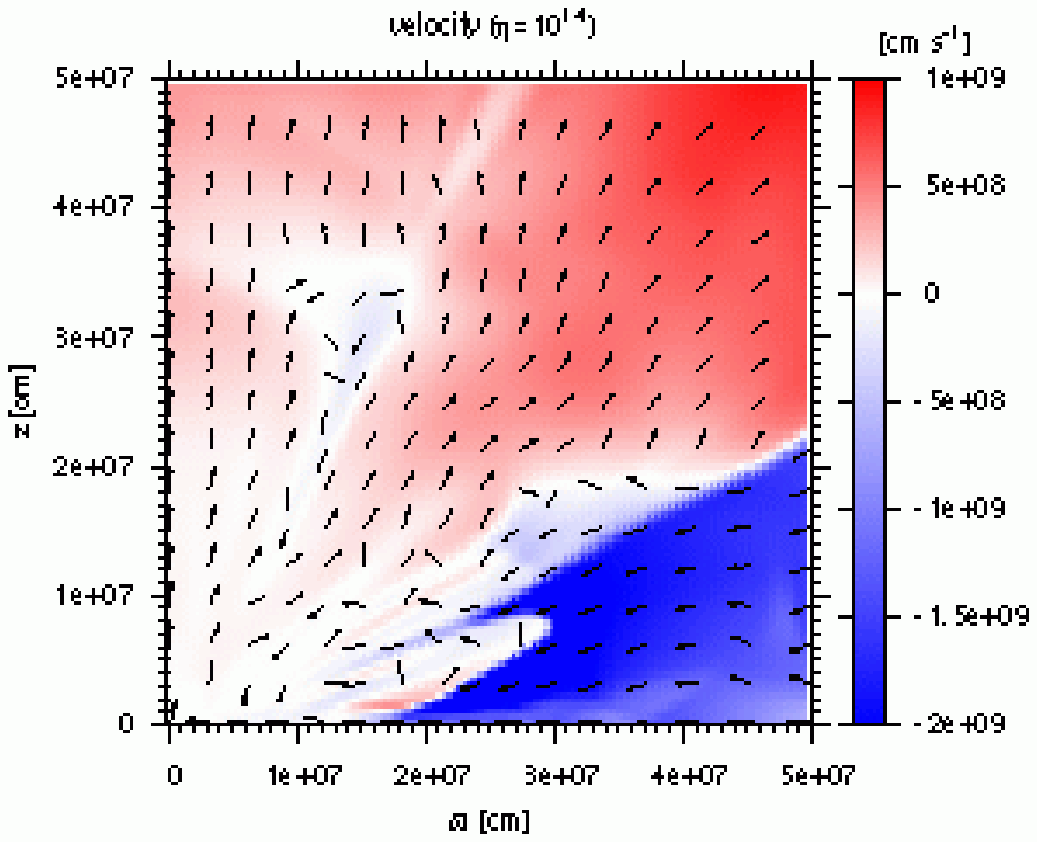}
  \caption{Distributions of radial velocity magnitude (color) and
  velocity directions (vectors) at $t=201$~ms for models
  B\textit{ss}-$\mho$-$\eta_{-\infty}$ (left) and
  B\textit{ss}-$\mho$-$\eta_{14}$ (right).} 
 \label{fig.vradz.50}
\end{figure*}

We found that the positive velocity island begins to emerge
around $170$~ms (15~ms after bounce) in model
$\eta_{14}$. Fig.~\ref{fig.vr_2.50} shows the radial distributions of
radial velocities, angularly averaged in the infall-region at 175~ms,
for the two models. It is observed in the radial range of
60~km$\lesssim r \lesssim$180~km that a infall velocity is slower in
model $\eta_{14}$, which seems due to 
the positive velocity island. In Fig.~\ref{fig.ar_2.50}, the radial
distributions of radial accelerations, angularly averaged in the
infall-region and time averaged during $t=170$-$175$~ms, is plotted. As
expected, a radial acceleration in the infall-region is larger in
model $\eta_{14}$ around a similar radial range to the above,
90~km$\lesssim r \lesssim$180~km. 
We found that a larger pressure and magnetic acceleration are
responsible for this acceleration superiority (see right panel of
Fig.~\ref{fig.ar_2.50}). We speculate that a larger
pressure acceleration in model 
$\eta_{14}$ will not be caused by the Joule heating, since the heating
timescale in the above range is too long, $\sim$1-10~s, to produce an
extra pressure. It is more likely that the accumulation of falling
matter stemmed by magnetic acceleration 
makes a pressure in the positive velocity island larger. It
seems that a magnetic acceleration is the primary factor for the
formation of the positive velocity island.

\begin{figure*}
\epsscale{0.5}
\plotone{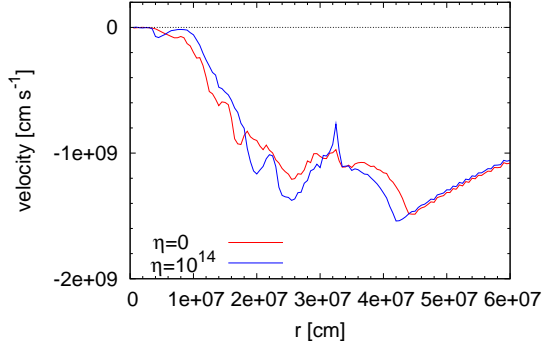}
  \caption{Radial distributions of average velocities in the
    infall-region at $t=175$~ms in models
    B\textit{ss}-$\mho$-$\eta_{-\infty}$  and 
    B\textit{ss}-$\mho$-$\eta_{14}$.} 
 \label{fig.vr_2.50}
\end{figure*}

\begin{figure*}
\epsscale{1}
\plottwo{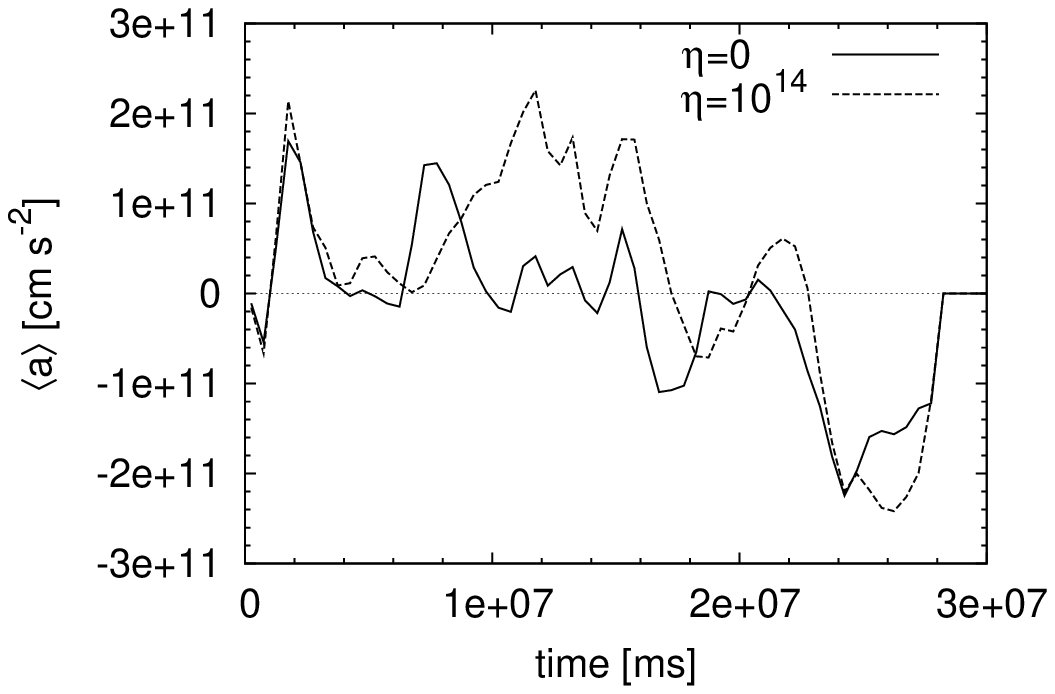}{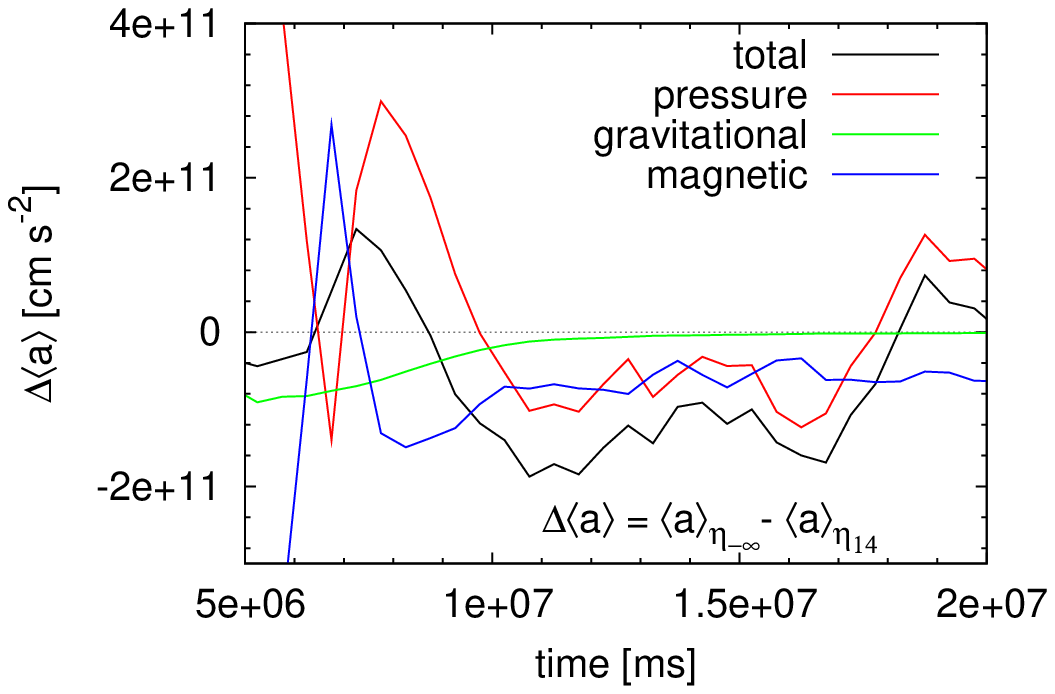}
  \caption{\textit{Left}: Radial distributions of radial
    acceleration, $\langle a \rangle$, angularly averaged in the
    infall-region and 
    time averaged during $t=170-175$~ms, for
    B\textit{ss}-$\mho$-$\eta_{-\infty}$ and
    B\textit{ss}-$\mho$-$\eta_{14}$. \textit{Right}: The radial
    distributions of differences in the averaged radial accelerations,
    $\Delta \langle a \rangle \equiv \langle a
    \rangle_{\eta_{-\infty}}-\langle a \rangle_{\eta_{14}}$,
    subtracting that of model $\eta_{14}$ from that of model
    $\eta_{-\infty}$.} 
 \label{fig.ar_2.50}
\end{figure*}

In the absence of toroidal magnetic field, 
a radial magnetic acceleration is written as 
\begin{equation}
a_{\textrm{B}}=\frac{B_\theta}{\rho}\left(-\frac{\partial
    B_\theta}{\partial r} + \frac{1}{r}\frac{\partial B_r}{\partial
    \theta} - \frac{B_\theta}{r}\right).\label{eq.fb.result}
\end{equation}
The radial distributions of magnetic field in the infall-regions at
175~ms plotted in the left panel of Fig.~\ref{fig.mag2.50} indicates
that $B_r$ is far 
grater than $B_\theta$ in the above radial range of 90-180~km,
and thus the second term in r.h.s. of Eq.~(\ref{eq.fb.result}), a part
of magnetic tension, is dominant. The left panel of
Fig.~\ref{fig.mag2.50} also shows 
that, in the above radial range, $B_\theta$ is larger in model
$\eta_{14}$ than in model $\eta_{-\infty}$, while $B_r$ is smaller in
model $\eta_{14}$. 
We found that the superiority in $B_\theta$ overwhelms the inferiority
in $B_r$ , i.e. $B_r(r)B_\theta(r)$ is averagely larger in model
$\eta_{14}$ in the above radial range. To put it simply, a larger
$B_\theta$ in in model $\eta_{14}$ is crucial to the formation of the
positive velocity island.

The radial distribution of $B_\theta$ in model $\eta_{14}$ shown in
the left panel of Fig.~\ref{fig.mag2.50} compared with that of
model $\eta_{-\infty}$ invokes a magnetic field diffusion. 
Indeed, it is found that a magnetic Reynolds number is smaller
than unity almost everywhere for $r\lesssim 85$~km in the
infall-region (see the right panel of Fig.~\ref{fig.rme.50}), viz. a
diffusion effectively advects a magnetic field outwards against matter
infall. The right panel of Fig.~\ref{fig.mag2.50} 
shows that $B_\theta$  in model $\eta_{-\infty}$ is highly confined
into the central region of 30~km radius, while the
$B_\theta$-distribution in model $\eta_{14}$ is rather gentle. This
implies that in the latter model, a resistivity effectively diffuses
$B_\theta$ out of the central region. In model 
$\eta_{-\infty}$, the gradient of $B_\theta$ suddenly becomes shallow
around $r=30$~km. Then with a resistivity, the incoming diffusion flux
there, $\eta\nabla B_\theta$, would be larger than the outgoing one,
which would result in a 
increase of $B_\theta$ outside the radius of 30~km. In this way, a
magnetic field diffused out from a deep inside of the core suppress a
matter infall and responsible for a larger explosion energy in model
$\eta_{14}$.

\begin{figure*}
\epsscale{1}
\plottwo{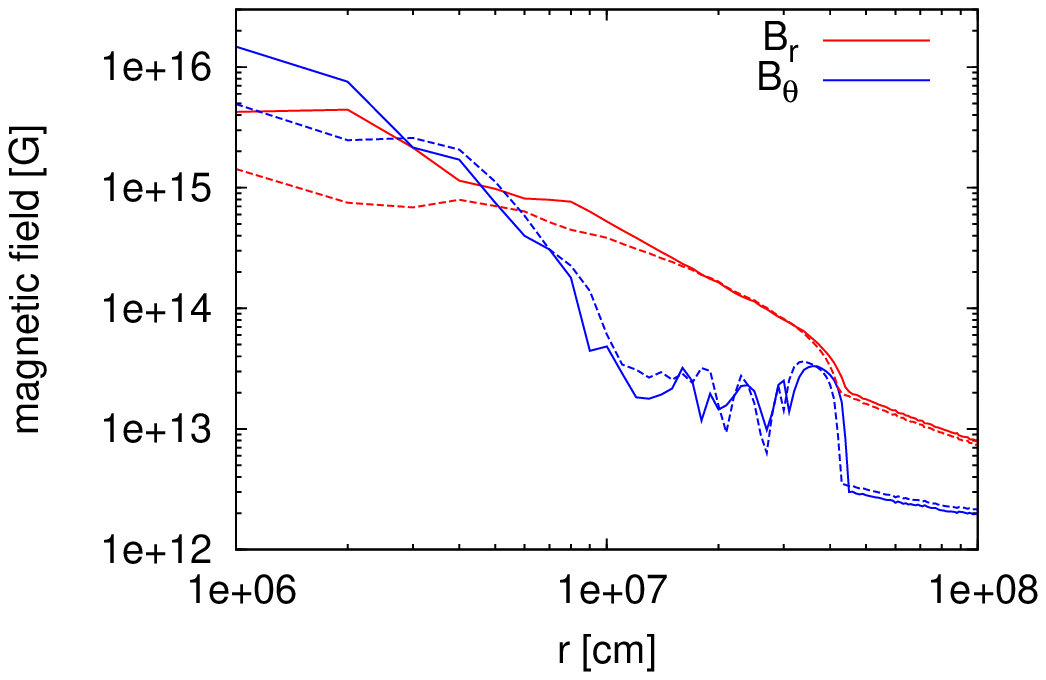}{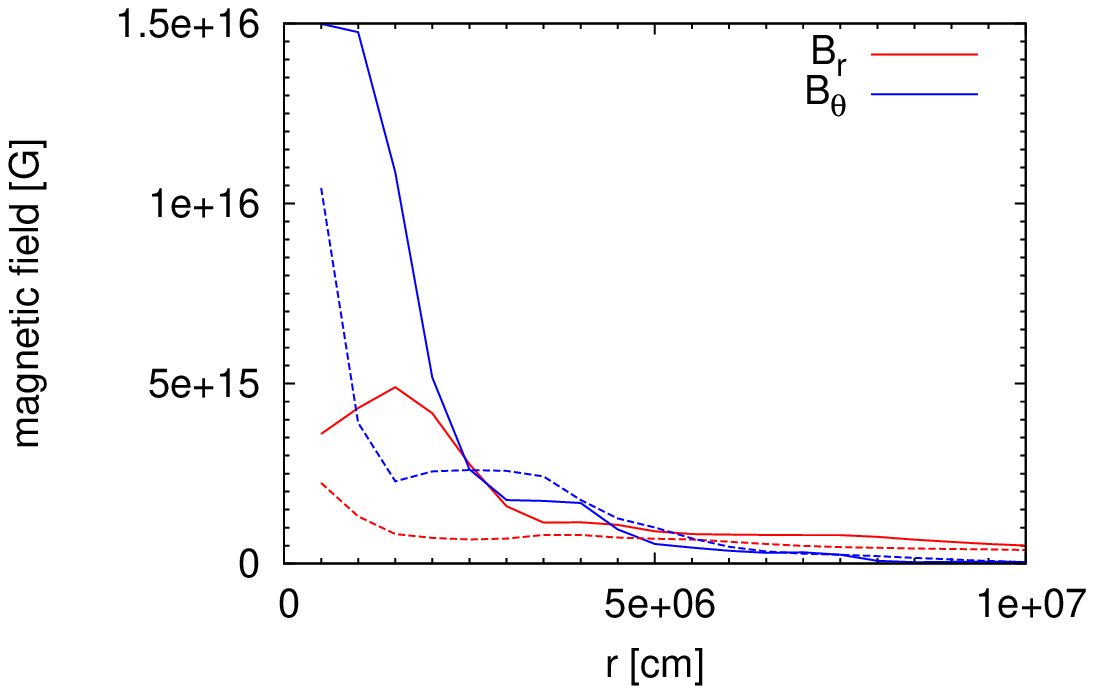}
  \caption{\textit{Left}: The radial distributions of magnetic field
    strengths in the infall-regions at $t=175$~ms in model
    B\textit{ss}-$\mho$-$\eta_{-\infty}$ (solid lines) and
    B\textit{ss}-$\mho$-$\eta_{14}$ (dashed lines). \textit{Right}:
    Same as the left panel but in linear scale, zooming towards
    the center.}   
 \label{fig.mag2.50}
\end{figure*}

A similar mechanism also works in model $\eta_{13}$. As shown in the
left panel of Fig.~\ref{fig.rme.50}, diffusion of magnetic field
in the infall-region also occurs well in this model. However, due to
a smaller resistivity, diffusion sites are limited to a smaller
volume compared with model $\eta_{14}$. Nonetheless, the explosion
energy in model $\eta_{13}$ is comparable to that of model $\eta_{14}$
at the end of the simulations (see Fig.~\ref{fig.exp1.50})

\begin{figure*}
\epsscale{1}
\plottwo{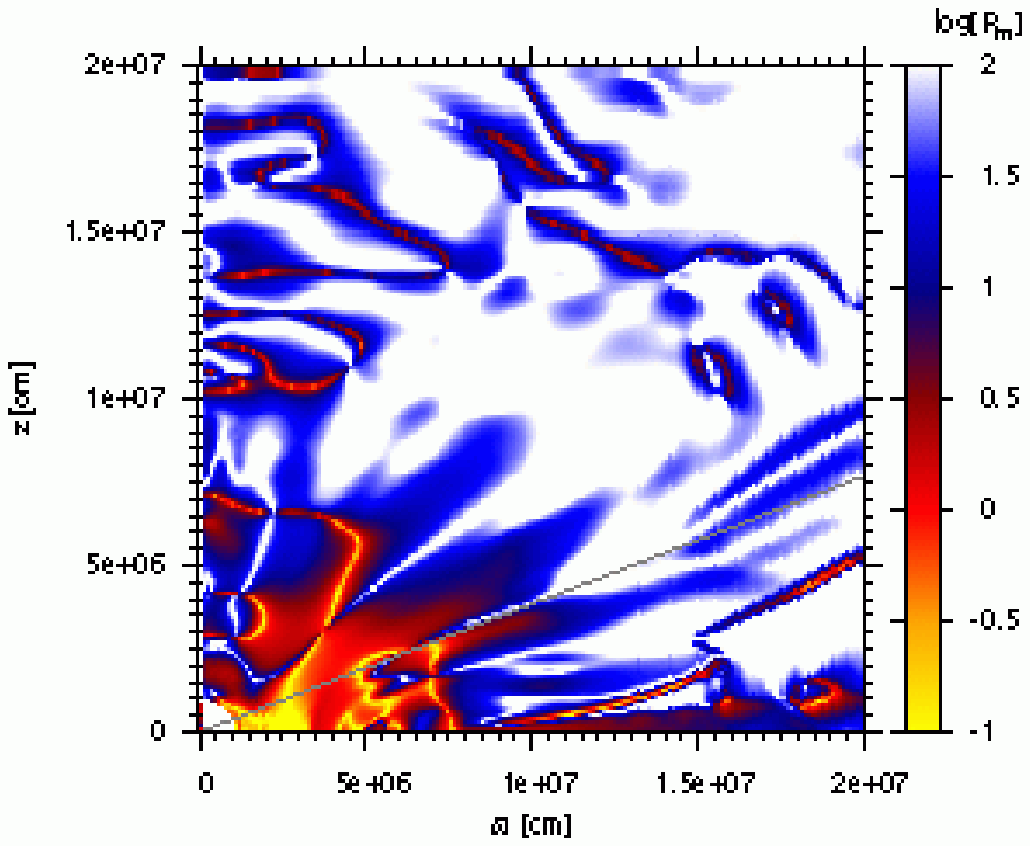}{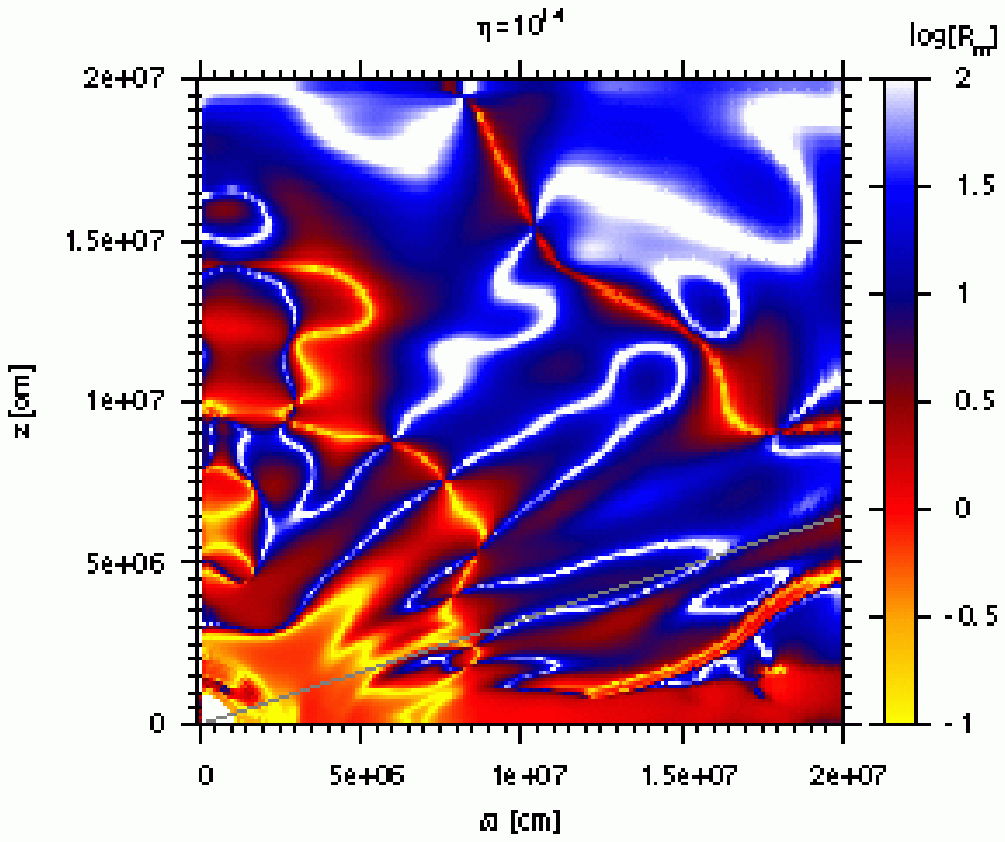}
  \caption{Distributions of magnetic Reynolds number
    in logarithmic scale at $t=175$~ms in model
    B\textit{ss}-$\mho$-$\eta_{13}$ (left) and 
    B\textit{ss}-$\mho$-$\eta_{14}$ (right). The volume
    below the gray line corresponds to the infall region.}
 \label{fig.rme.50}
\end{figure*}

\subsubsection{Aspect Ratio}
The aspect ratios in model-series B\textit{ss}-$\mho$ are also larger
than unity, i.e. the shape of ejecta is prolate, but are not as large
as in the rotating cases (see Fig.~\ref{fig.aspect.50}). Since
the initial magnetic field assumed in the present model-series is very
strong, it affects the dynamics even before bounce. Due to the dipole-like
configuration, a magnetic field hampers a matter infall especially
well in the lateral direction. This causes a weak bounce around the
equator, and thus a prolate matter ejection. 

The difference from the
former model-series is that the aspect ratios do not grow drastically
after bounce (see Fig.~\ref{fig.aspect.50}). We expect that this is
caused by a roughly flat angular distribution of magnetic energy per
unit mass in the eruption-region as indicated in
Fig.~\ref{fig.ethm.50} for model~B\textit{ss}-$\mho$-$\eta_{-\infty}$
and B\textit{ss}-$\mho$-$\eta_{14}$, reminding that 
a polarly concentrated magnetic field distribution is essential for a
large aspect ratio in the rotating cases.

We found that the change from the rather polarly concentrated
distribution of magnetic field at the beginning to a roughly flat
angular distribution occurs during the collapsing phase, i.e. before
bounce. Fig.~\ref{fig.induc.50} shows the angular distributions of
$\dot{\mathcal{E}_r}$, radially averaged over $50<r<1000$~km at 122~ms
(33~ms before bounce) in model
B\textit{ss}-$\mho$-$\eta_{-\infty}$. It is shown that a total
$\dot{\mathcal{E}_r}$ is larger around the equator, and a compression
of $B_r$ in the $\theta$-direction plays an important role.
Since $B_\theta$ is a priori strong around the equator due to the
assumed magnetic field configuration, an infall matter there is a
little deflected into non-radial direction by a Maxwell stress of
$B_rB_\theta/4\pi$, viz. $\theta$-component of velocity is 
generated. In this way, $B_r$ is compressed into the
$\theta$-direction, which leads to an increase of $B_r$ preferentially
around the equator and rather flat angular distribution of magnetic field.

A notable point here is that a rather flat distribution of magnetic
field, which causes a mild value of the aspect ratio, is established
before bounce when a resistivity does not work well because of a large
scale height of magnetic field. This is why the aspect ratio for a
different resistivity results in a similar value. 

\begin{figure}
\epsscale{1}
\plotone{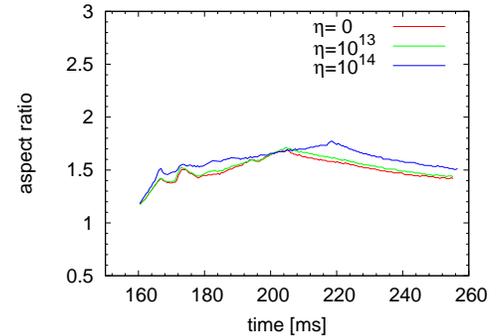}
\caption{Evolutions of aspect ratios 
  in model series B\textit{ss}-$\mho$.}
\label{fig.aspect.50}
\end{figure}

\begin{figure}
\epsscale{1}
\plotone{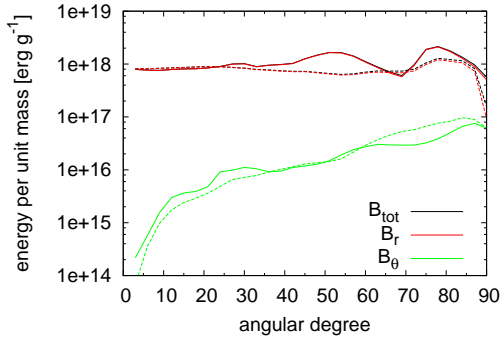}
\caption{Angular distributions of magnetic energies per unit mass
  averaged over $50$~km$<r<0.9\times r_{\textrm{sh}}$ at
  $t=180$~ms (25~ms after bounce). The solid and dashed lines are for
  models B\textit{ss}-$\mho$-$\eta_{-\infty}$ and
  B\textit{ss}-$\mho$-$\eta_{14}$, respectively.}
\label{fig.ethm.50}
\end{figure}

\begin{figure}
\epsscale{1}
\plotone{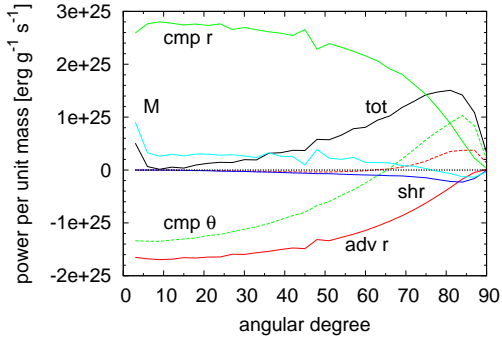}
\caption{Angular distributions of $\dot{\mathcal{E}_r}$, radially
  averaged over $50<r<1000$~km
  (see Eq.~(\ref{eq.amp2})) at 122~ms (33~ms before bounce) in model
  B\textit{ss}-$\mho$-$\eta_{-\infty}$. The black line shows total
  $\dot{\mathcal{E}_r}$, while the red-solid,
  red-dashed, green-solid, green-dashed, blue, and cyan lines
  represent the contribution to $\dot{\mathcal{E}_r}$ from the radial
  advection, angular advection, radial compression, angular
  compression, shear, and change in mass, respectively.  }
\label{fig.induc.50}
\end{figure}

\subsection{Numerical Resistivity}\label{sec.numdif}
In the preceding sections, we have seen how a resistivity impacts on
the dynamics of a strongly magnetized supernova. However, we should
remember that the numerical results are not only affected by a
physical resistivity but may also be influenced by a numerical
resistivity. Here, we estimate a possible impact of a numerical
resistivity on the dynamics.

First, we directly compared the strengths of a physical resistivity
and a numerical resistivity. We consider dividing a numerical flux into
two terms, the advection term and numerical diffusion term, where the
former may account for an advection of a physical flux, while the
latter for a numerical diffusion. For example, in the equation of
numerical fluxes  $\mbox{{\boldmath$F$}}^x$ in
Eqs.~(\ref{eq.kt.numfluxf}), the first and second term in r.h.s. are
the advection term and numerical diffusion term,
respectively. Accordingly, the ideal part of a numerical flux vector
for the induction equations $\mbox{{\boldmath$\bar{F}$}}$, which are
defined by a combinations of 6th - 8th components of numerical flux
vectors  $\mbox{{\boldmath$F$}}^{x}$ an $\mbox{{\boldmath$F$}}^{y}$ (see
Eqs.~(\ref{eq.kt.fbar})), are 
also divided into the advection term and numerical diffusion
term. Then, the total numerical flux vector for the
induction equations can be written as
$\mbox{{\boldmath$\bar{E}$}}=\mbox{{\boldmath$\bar{F}$}}_{\textrm{adv}}
+\mbox{{\boldmath$\bar{F}$}}_{\textrm{nd}}+\mbox{{\boldmath$\bar{G}$}}$, 
where subscripts "adv'' and "nd" stand for advection and numerical
diffusion, respectively. Recall that a term
$\mbox{{\boldmath$\bar{G}$}}$ arises due to a physical resistivity.

Now, to compare the strengths of a physical resistivity and a
numerical resistivity we evaluate
$\zeta\equiv|\mbox{{\boldmath$\bar{G}$}}|
/|\mbox{{\boldmath$\bar{F}$}}_{\textrm{nd}}|$, in which a physical
resistivity is dominant for $\zeta\gg 1$ while a numerical resistivity
is dominant for $\zeta\ll 1$. The distribution of $\zeta$ for model
B\textit{s}-$\Omega$-$\eta_{13}$ at 164~ms is shown in
Fig.~\ref{fig.numdiff}. This implies that a 
physical resistivity dominates over a numerical resistivity in a
considerable volume inside the shock surface, but in some locations a
numerical resistivity is not negligible. 

Next, we examined how much a numerical resistivity affects the
dynamics. This is done by evaluating $R_{\textrm{m,num}}\equiv
|\mbox{{\boldmath$\bar{F}$}}_{\textrm{adv}}|
/|\mbox{{\boldmath$\bar{F}$}}_{\textrm{nd}}|$, which may corresponds to
a magnetic Reynolds number for a numerical resistivity. Even if a
numerical resistivity dominates over a physical resistivity, it does
not make substantial impacts on the dynamics provided
$R_{\textrm{m,num}}\gg 1$. Fig.~\ref{fig.Rdiff.5513} shows the
distribution of $R_{\textrm{m,num}}$ for model
B\textit{s}-$\Omega$-$\eta_{13}$ at 164~ms. It is found that
$R_{\textrm{m,num}}$ is much larger than unity almost everywhere except
around the shock surface, where a steep variation of a magnetic
field exists. Meanwhile, we have seen in
\S~\ref{sec.diff} that a magnetic Reynolds number of physical
resistivity in model B\textit{s}-$\Omega$-$\eta_{13}$ is $\sim 30$
around the equator (see the left panel of Fig.~\ref{fig.rme.55}).
These facts imply that a numerical resistivity is mostly too small to
influences the dynamics, while a physical resistivity affects the
dynamics, albeit a little. Thus, it seems that a physical resistivity
``effectively'' dominates over a numerical resistivity in this model.

From the above discussion, it is likely that a comparison between
$R_{\textrm{m}}$ and $R_{\textrm{m,num}}$ is more meaningful than
that between a physical resistivity and numerical resistivity
themselves in order to assess an influence of a numerical resistivity.
Then, we also compare $R_{\textrm{m}}$ and $R_{\textrm{m,num}}$ for model
B\textit{m}-$\Omega$-$\eta_{13}$ and
B\textit{ss}-$\mho$-$\eta_{13}$.
In model B\textit{m}-$\Omega$-$\eta_{13}$, $R_{\textrm{m,num}}$ is
much larger than unity almost everywhere except around the shock
surface, while a relatively small $R_{\textrm{m}}\sim 1$-10 is
found around the equator (see Fig.~\ref{fig.RRdiff.4513}). Hence, a
physical resistivity also 
effectively dominates over a numerical resistivity in this model.
In model B\textit{ss}-$\mho$-$\eta_{13}$, a mildly small
$R_{\textrm{m,num}}\sim 10$ appears not only around the shock surface
but also other locations (see the left panel of
Fig.~\ref{fig.RRdiff.5013}). This means that a numerical resistivity 
somewhat affects the dynamics in this model. However, the right panel
of Fig.~\ref{fig.RRdiff.5013} shows that a physical resistivity is
more influential than a numerical resistivity especially at small
radii. Recalling that the essential role of a resistivity in this 
model is to diffuse a magnetic field from a small radius to
a larger radius, a physical resistivity seems to play primary role to
yield a different result for B\textit{ss}-$\mho$-$\eta_{13}$ from the
ideal model. For confirmation, we carried out the above analysis in
the models with $\eta=10^{14}$~cm$^2$s$^{-1}$, and found that the
``effective'' dominance of a physical resistivity over a numerical
resistivity is more pronounced than the counterpart models with
$\eta=10^{13}$~cm$^2$s$^{-1}$.

\begin{figure}
\epsscale{1}
\plotone{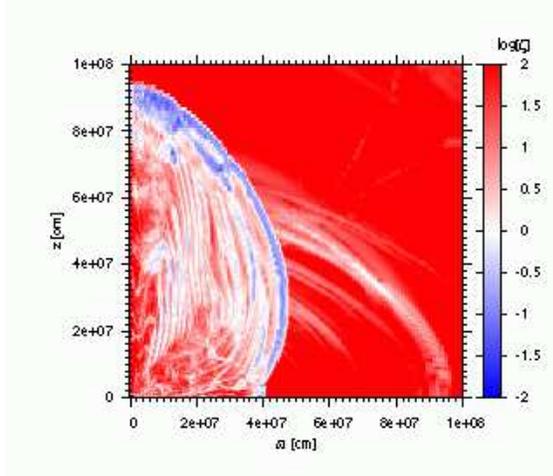}
\caption{Distribution of $\zeta\equiv|\mbox{{\boldmath$\bar{G}$}}
/\mbox{{\boldmath$\bar{F}$}}_{\textrm{nd}}|$ in logarithmic scale for
model B\textit{s}-$\Omega$-$\eta_{13}$ at 164~ms.}
\label{fig.numdiff}
\end{figure}

\begin{figure}
\epsscale{1}
\plotone{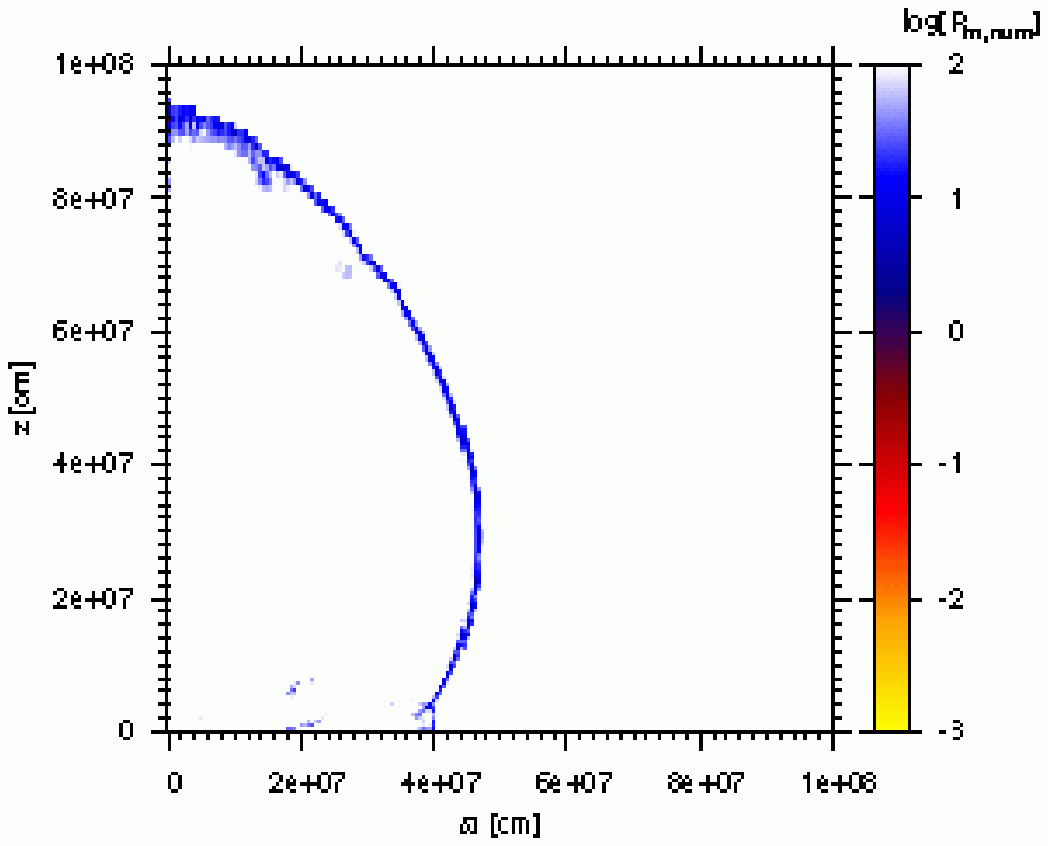}
\caption{Distribution of $R_{\textrm{m,num}}\equiv
|\mbox{{\boldmath$\bar{F}$}}_{\textrm{adv}}|
/|\mbox{{\boldmath$\bar{F}$}}_{\textrm{nd}}|$ in logarithmic scale for
model B\textit{s}-$\Omega$-$\eta_{13}$ at 164~ms.}
\label{fig.Rdiff.5513}
\end{figure}

\begin{figure*}
\epsscale{1}
\plottwo{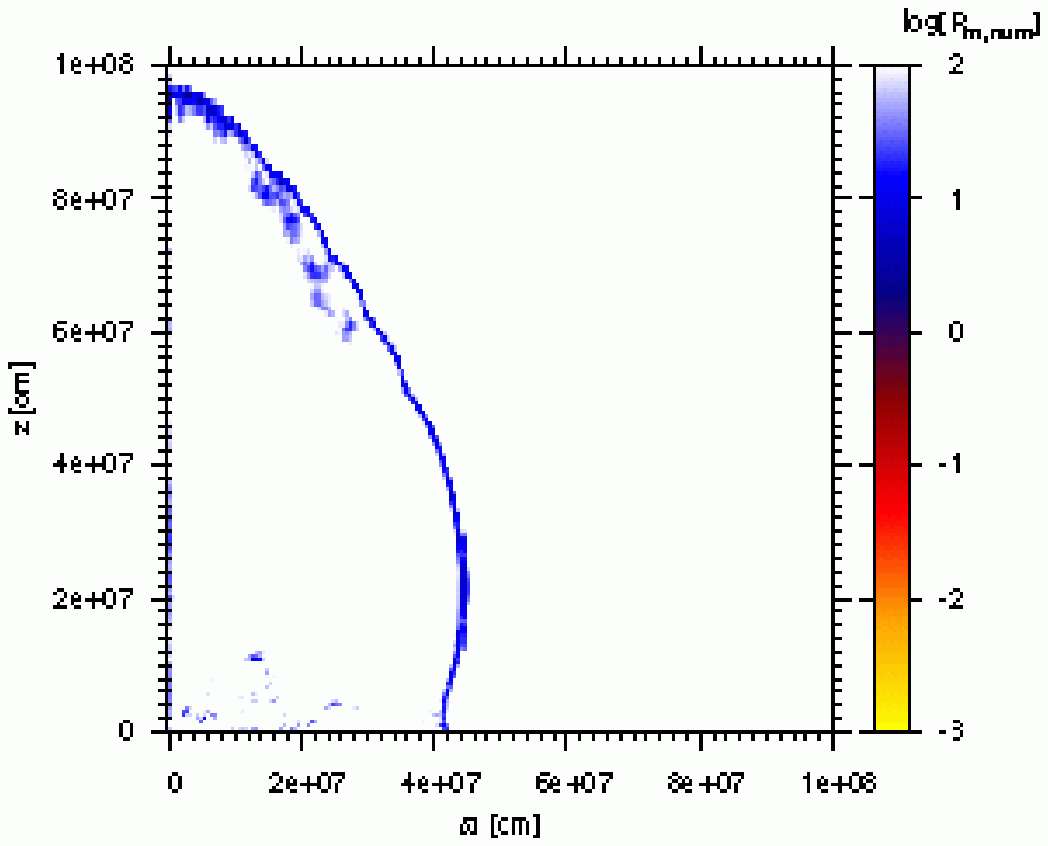}{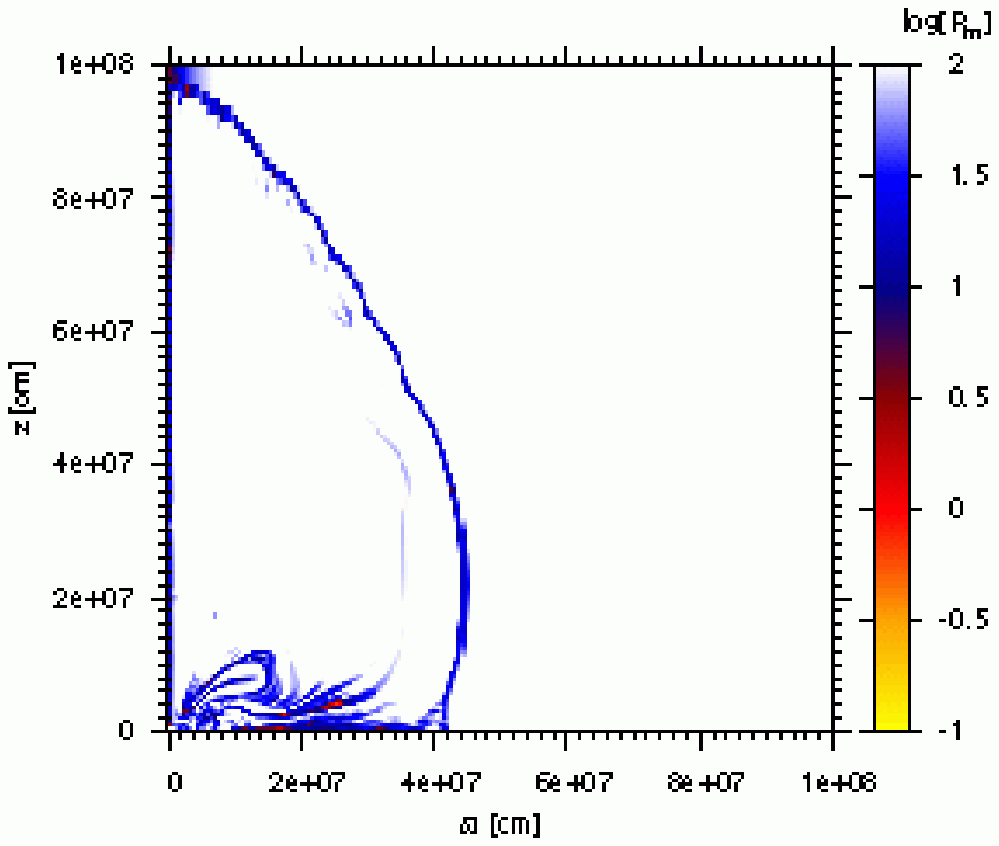}
\caption{Distribution of $R_{\textrm{m,num}}$ (\textit{left}) and
  $R_{\textrm{m}}$ (\textit{right}) in logarithmic scale for
model B\textit{m}-$\Omega$-$\eta_{13}$ at 182~ms.} 
\label{fig.RRdiff.4513}
\end{figure*}

\begin{figure*}
\epsscale{1}
\plottwo{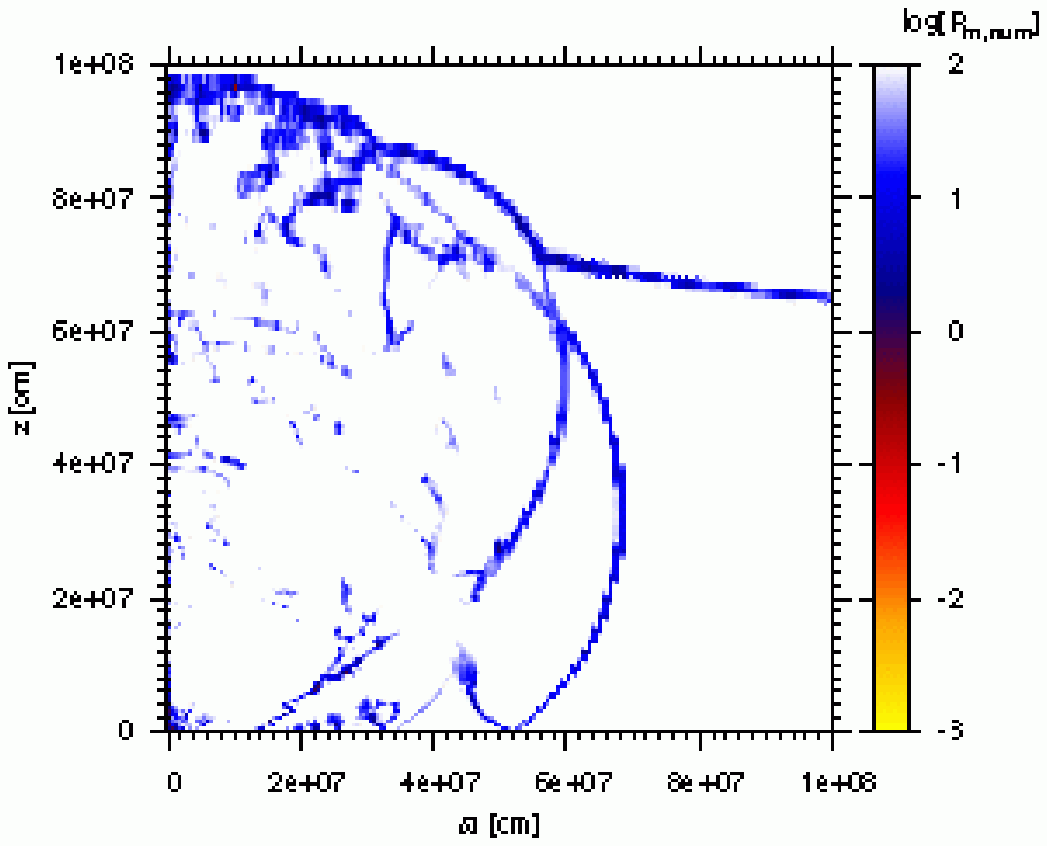}{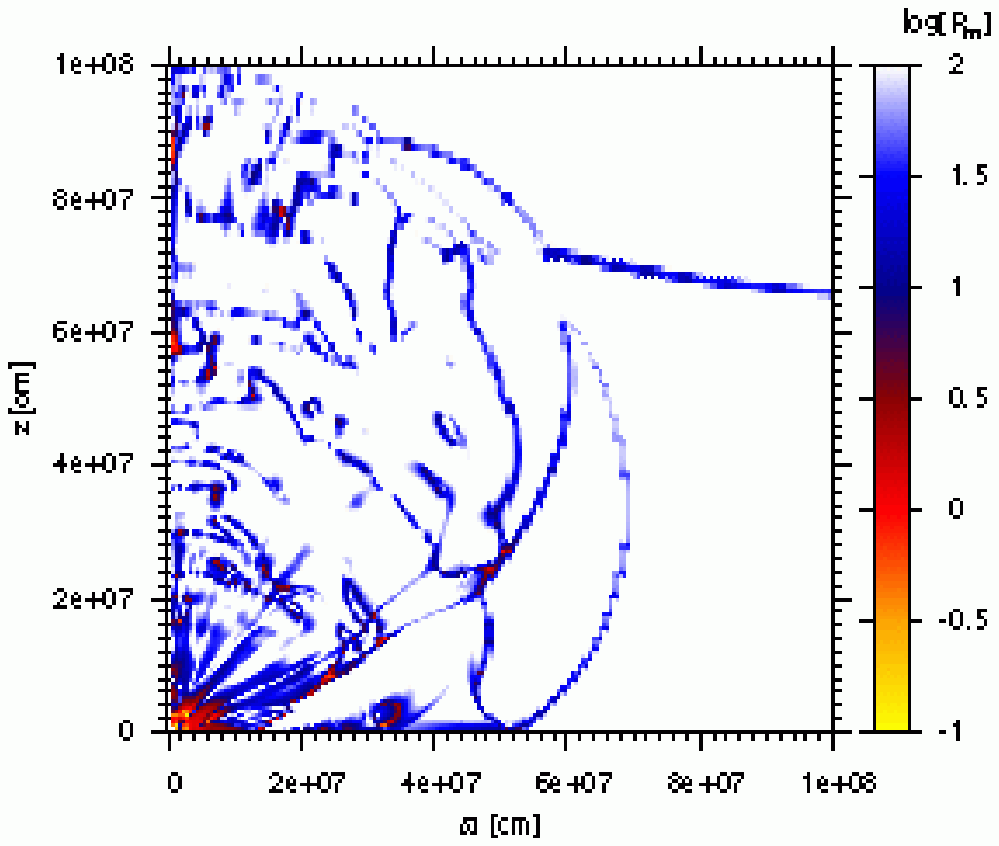}
\caption{Distribution of $R_{\textrm{m,num}}$ (\textit{left}) and
  $R_{\textrm{m}}$ (\textit{right}) in logarithmic scale for
model B\textit{ss}-$\mho$-$\eta_{13}$ at 195~ms.} 
\label{fig.RRdiff.5013}
\end{figure*}

\section{Discussion and Conclusion}\label{sec.conc}
We have done MHD simulations of CCSNe to understand roles of
a turbulent resistivity in the dynamics. As a result, we found that a
resistivity possibly has a great impact on the explosion energy,
a magnetic field amplification, and the aspect
ratio. How and how much a resistivity affects on these
depend on the initial strengths of magnetic field and rotation
together with the strength of a resistivity.

In the rotating cases (model-series B\textit{s}-$\Omega$ and
B\textit{m}-$\Omega$), a resistivity makes the explosion energy 
small. This is mainly ascribed to a small magnetic and
centrifugal acceleration due to an inefficiency in magnetic field
amplification and angular momentum transfer, respectively.
Meanwhile, in the case of very strong magnetic field and no rotation
(model-series B\textit{ss}-$\mho$), a resistivity enhances the
explosion energy. In the ideal
model, an inflow of negative radial momenta from the 
infall-region into the eruption-region inhibits a powerful mass
eruption. With a resistivity, a magnetic field in the infall-region 
diffuses outwards from a deep inside of the core to counteract the
negative momentum invasion.

The ideal models involving rotation shows a
polarly concentrated distribution of magnetic energy per unit
mass. Since an initial magnetic energy per unit mass is somewhat
larger around the pole than the equator, an amplification
preferentially occurs there, which makes the contrast stronger. We
found that the main mechanisms of amplification around the pole is an
outward advection of a magnetic energy from a small radius and the
winding of poloidal magnetic field-lines by differential rotation.
In a resistive 
model, an amplification occurs qualitatively similar way to the
corresponding ideal model, but proceeds less efficiently. As a result,
the angular distribution of magnetic field becomes less concentrated
towards the pole. Although we found signs of a MRI growth in these
models, it seems not to play substantial role in amplifying a magnetic
field. The MRI will at best help to keep the strength of a magnetic
field. We should note here that in our 2D-axisymmetric
simulations, the MRI may not be followed correctly, since
non-axisymmetric modes are also important in the MRI. A comparison
between a 2D-axisymmetric and 3D simulations in the behavior of the MRI
will be made and presented elsewhere in the near future.

Every model involving a rotation shows a large aspect ratio, $>2.5$,
at the end of the simulation, where a model with a larger
resistivity results in a smaller aspect ratio. The impact of
resistivity on the aspect ratio is more standout in the moderate magnetic
field case, model-series B\textit{m}-$\Omega$. The aspect ratios
obtained in the rotation models
are expected to grow even larger later on, since a radial velocity is
still faster around the pole than around the equator at the end of the
simulations. Further long time computations are necessary to 
predict the aspect ratio after the shock surface breaks out the stellar
surface. In a model without rotation, on the other hand, the
aspect ratio keeps mild value, $\sim 1.5$, through each simulation. No
significant difference is found among models with a different
resistivity. It is found that the aspect ratios attained in the present
simulations are related to the angular distribution of magnetic energy
per unit mass. That is, the more a magnetic energy is concentrated
towards a small $\theta$, the larger the aspect ratio of ejected matters.

In literature, we have found eight CCSNe in that both the upper and
lower limit of the aspect ratio is measured; SN1987a, SN1993J,
SN1997X, SN1998S, SN2002ap, SN2005bf, SN2007gr, and SN2010jl. The
observed aspect ratios are 
at most $\approx 3$; e.g. $\approx2-3$ for SN1987A  \citep{pap89},
$\approx 3$ for 1997X
and $1.2-2.5$ for SN1998S \citep{wan01}. It seems that aspect ratios
obtained in our rotational models are too large compared with these
observations, reminding that ours will grow larger later on.
However, we note that the above sample does not necessarily contain
supernovae that leave a magnetar. Applying SGR/AXP birth rate of
$\sim 0.1-0.2$ per century \citep{lea07} and Galactic CCSN rate of
$\sim 0.8-3.0$ per century \citep{die06}, the rate of magnetar
production among all CCSNe is $\sim 3-25$~\%. 

A constant resistivity that we assume in our computations may
not be natural. A resistivity arising from turbulent convective motions
may appear only around convectively unstable regions and may take a
different value at a different position. Computations 
implementing such a non-constant resistivity should be done in
future works. Effects of viscosity, which is expected to arise
together with resistivity in a turbulence, also have to be
investigated. Although the ideal way to study effects of turbulent
resistivity and viscosity is a direct numerical simulation of
turbulence itself, it is currently quite unfeasible. That kind of
simulation may demand a computational facility dozens orders of
magnitude more powerful than those of the present day.

Although we carried out 2D-MHD simulations with the assumptions of
axisymmetry and a purely poloidal magnetic field at the beginning, it
is well known that a purely poloidal and purely toroidal
magnetic fields are both unstable against non-axisymmetric
perturbations. According to \citet{bra09}, the ratio of the
poloidal magnetic to total magnetic energy, $E_p/E$, should be 
less than 0.8 for the stability. In each of our rotational model, we find
that $E_p/E<0.8$ is always satisfied after bounce. Although the
stability criterion is not fulfilled before bounce, the instability
will not grow substantially, since it takes at most one Alfv\'en
timescale until bounce. Thus, we expect that the results obtained in
our rotating models will not change drastically even if we carry out
3D-MHD simulations. In fact, 3D core-collapse simulations with strong 
magnetic field and rapid rotation done by \citet{kur10} shows
qualitatively similar results to those of our 2D simulations.
However, a purely poloidal magnetic field at the beginning itself may
be unnatural. Also, in our non-rotating models, the above stability
criterion is always not satisfied. The toroidal component of magnetic
field should be added to the initial condition to deal with more
realistic situations. It may be necessary in the future work to
investigate how such an additional component affects the results
presented here.

In the present work, we studied the role of a resistivity in the limited
parameter range of magnetic field and rotation. As we found that a
resistivity affects the dynamics differently for a different strength
of magnetic field and rotation, it may be interesting to carry out a
systematic study with a wide range of the parameters in the future.

\acknowledgments
H.S. would like to gratefully thank Kenta Kiuch, Hiroki Nagakura, and
Nobutoshi Yasutake to give him helpful advice on developing the numerical
MHD code. H.S. is also grateful to Ken'ichiro Nakazato and Kosuke
Sumiyoshi to provide Shen's EOS table and to instruct him in handling
that. This work is supported by Grant-in-Aid for Research
Activity Start-up of JSPS (21840050).

\appendix

Here we present numerical results of various HD and
MHD test problems calculated with \textit{Yamazakura}. Among
them, the numerical solution of 1D Riemann problem, Sedov's self-similar
point source explosion, Yahil's self-similar collapse, linear wave
propagation problem, magnetic field diffusion
problem, can be compared with the analytic solutions. The others are
multi-dimensional problem, and no reference solution exits. 
One can find numerical results of such problems in literature, and may check
whether our results match up with those. In each test calculation, the 
ideal gas EOS, $p=(\gamma-1)e$, is used. The distribution of numerical cells 
are uniform unless otherwise stated.

\subsection{Riemann Problem in 1D}\label{sec.app.rie1}
An 1D-Riemann problem, which is also referred as a shock tube
problem or Sod problem, is carried out to
check that 1D-HD 
equations are properly solved. The initial set-up is as follows:
\begin{eqnarray}
\left(
\begin{array}{c}
\rho\\
v   \\
p   
\end{array}
\right)
=
\left(
\begin{array}{c}
1.0\\
0.0\\
1.0
\end{array}
\right)
\hspace{1pc}\textrm{for}\hspace{1pc}
x<0.5,
&&
\left(
\begin{array}{c}
\rho\\
v   \\
p   
\end{array}
\right)
=
\left(
\begin{array}{c}
0.125\\
0.0\\
0.1 
\end{array}
\right)
\hspace{1pc}\textrm{for}\hspace{1pc}
x>0.5
\end{eqnarray}
with the computational domain $x\in[0,1]$.
The adiabatic index is set $\gamma=1.4$. Numerical results with
800~cells are shown in 
Fig.~\ref{fig.test.riemann}, in which a good agreement with the reference
solution is found.

\begin{figure}
\epsscale{0.8}
\plotone{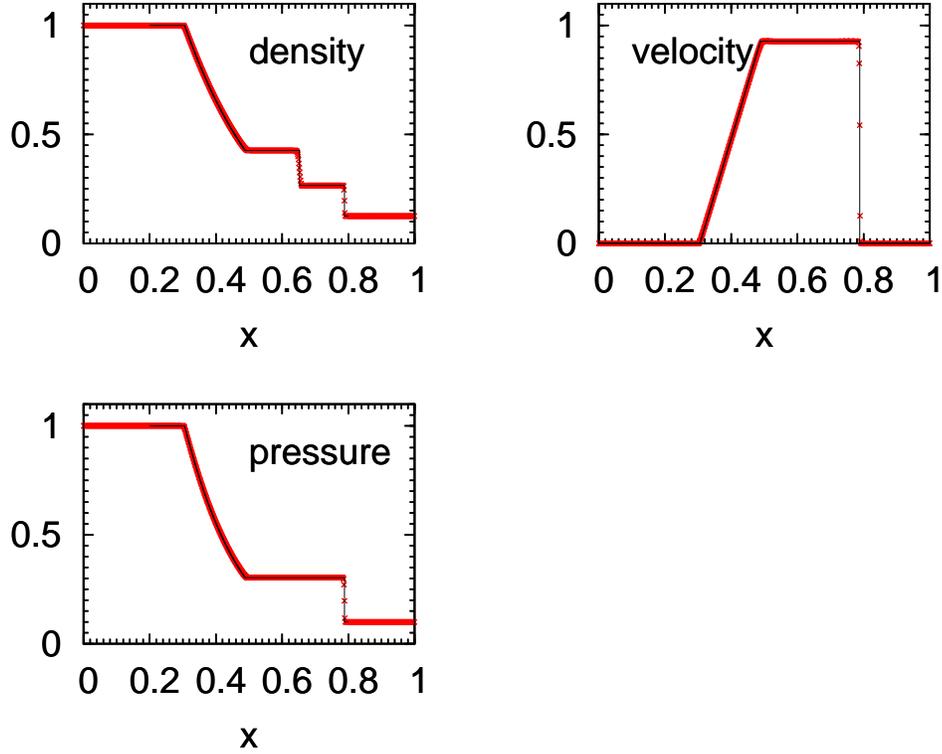}
  \caption{Numerical results (red crosses) of the 1D-Riemann problem
    on the reference solutions (black lines) at $t=0.164$}  
 \label{fig.test.riemann}
\end{figure}

\subsection{Riemann Problem in 2D}\label{sec.app.rie2}
In order to test a multi-dimensional HD computation, a Riemann
problem in 2D is carried out. A square
numerical domain $(x,y)\in[0,1]\times[0,1]$ is divided into the four
parts, and constant hydrodynamical values are set in each part. The
initial condition is 
\begin{eqnarray}
\left(
\begin{array}{c}
\rho\\
p\\
v_x \\
v_y 
\end{array}
\right)
&=&
\left(
\begin{array}{c}
1.0\\
0.8\\
0.0\\
0.0
\end{array}
\right)
\hspace{1pc}\textrm{for}\hspace{1pc}
x<0.5\hspace{1pc}\textrm{and}\hspace{1pc}y<0.5,\nonumber\\
\left(
\begin{array}{c}
\rho\\
p\\
v_x \\
v_y 
\end{array}
\right)
&=&
\left(
\begin{array}{c}
1.0\\
1.0\\
0.7276\\
0.0
\end{array}
\right)
\hspace{1pc}\textrm{for}\hspace{1pc}
x<0.5\hspace{1pc}\textrm{and}\hspace{1pc}y>0.5,\nonumber\\
\left(
\begin{array}{c}
\rho\\
p\\
v_x \\
v_y 
\end{array}
\right)
&=&
\left(
\begin{array}{c}
1.0\\
1.0\\
0.0\\
0.7276
\end{array}
\right)
\hspace{1pc}\textrm{for}\hspace{1pc}
x>0.5\hspace{1pc}\textrm{and}\hspace{1pc}y<0.5,\nonumber\\
\left(
\begin{array}{c}
\rho\\
p\\
v_x \\
v_y 
\end{array}
\right)
&=&
\left(
\begin{array}{c}
0.4\\
0.5313\\
0.0\\
0.0
\end{array}
\right)
\hspace{1pc}\textrm{for}\hspace{1pc}
x>0.5\hspace{1pc}\textrm{and}\hspace{1pc}y>0.5,\nonumber\\
\end{eqnarray}
which is symmetric across a diagonal line $y=x$. An adiabatic index of
$\gamma=1.4$ 
is used here. These setups are same with \textit{Case 12} of \citet{lis03}.
The calculation is done with $400 \times 400$ numerical cells. A
numerical result is shown in Fig.~\ref{fig.test.riemann2d}, which is
good agreement with that of \citet{lis03}. 

\begin{figure}
\epsscale{0.5}
\plotone{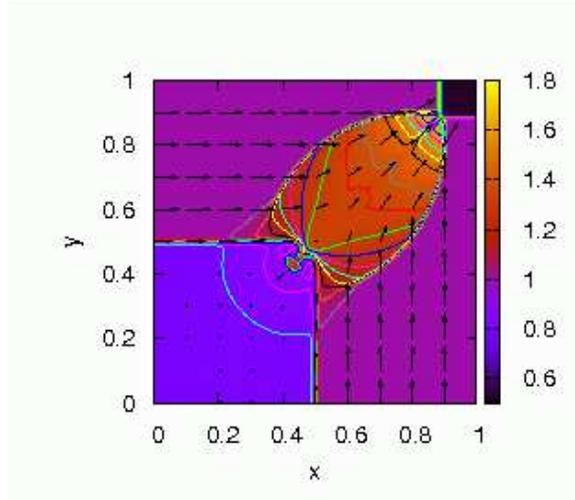}
  \caption{Numerical result of the 2D-Riemann problem at $t=0.25$ 
    (density color map with contours and velocity field vectors). The
    computation is done with $400\times 400$ cells.}
 \label{fig.test.riemann2d}
\end{figure}

\begin{figure}
\epsscale{0.8}
\plotone{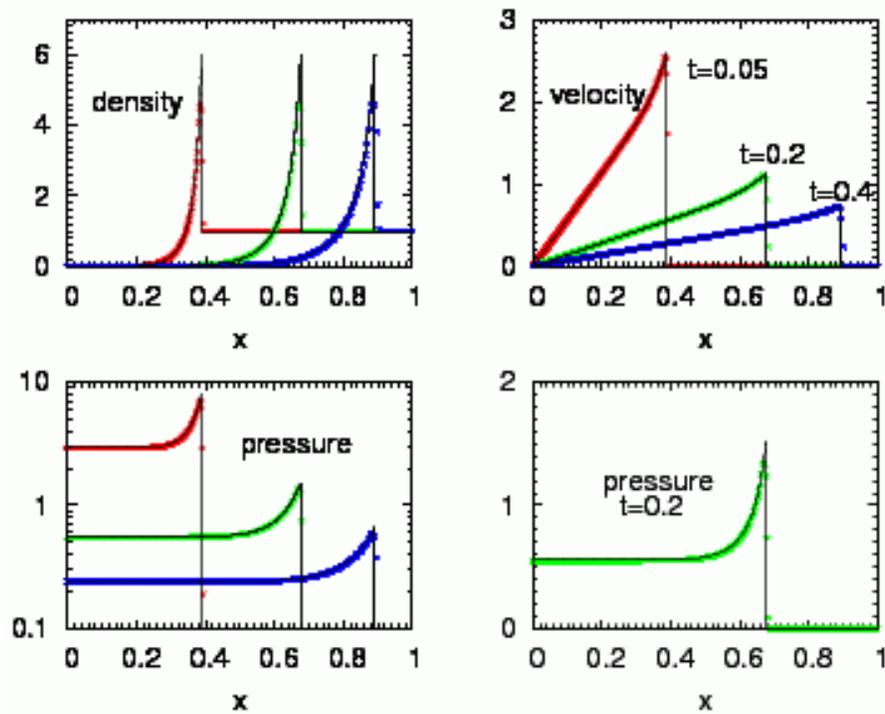}
  \caption{Numerical results of a point source spherical explosion
    for density (left-top), velocity (right-top), and pressure
    (left-bottom) distribution. The right-bottom panel is shown to
    see, with linear scale, a sharpness of the numerical solution for
    pressure. In each figure, the red, green, and blue 
    crosses represent solutions at $t=0.05$, $0.2$, and $0.4$,
    respectively.}  
 \label{fig.test.sedov}
\end{figure}

\subsection{Point Source Spherical Explosion}~\label{sec.app.sedov}
Another HD test calculation we have done is a spherical
point source explosion, the solution of that known as the Sedov's
solution. An 1D spherically symmetric computation is done with polar
coordinate. We use non-uniform 
numerical cells with the spatial resolution of the $i$-th cell being
$\Delta r_i = \Delta r_{\textrm{min}} \alpha^{i-1} $, where $\Delta
r_{\textrm{min}}=10^{-4}$, $\alpha = 1.0056$, and the maximum of $i$ is 720.
This cell distribution is similar to that of the MHD core-collapse
simulations presented in the present work, where the same cell
distributions are set along the $\varpi$ and $z$ directions
separately. Hence, we may be able to speculate here how much a
propagating shock wave diffuses in our MHD core-collapse simulations.
We put a static sphere of radius $r=1.0$ with the uniform density of
$\rho_0=1.0$. A point source explosion is initiated by injecting the
internal energy of $E_0=3.0$ at the central sphere of radius
$r=0.01$. Numerical results are shown in Fig~\ref{fig.test.sedov}. It
is found that a sharpness of density peak in the numerical solution is
kept during propagation, despite that the spatial resolution is lower
for a larger radius. This implies that, in our computations, a
propagating shock wave is not damped crucially due to a numerical
diffusion.

\subsection{Self-Similar Collapse}\label{sec.app.yahil}
In order to check gravity is correctly dealt in \textit{Yamazakura},
we tested a self-similar collapse, the solution of that known as 
Yahil's solution \citep{yah83}. Both an 1D spherically symmetirc
calculation with polar coordinate and a 2D axially symmetric calculation
with cylindrical coordinate are done. In describing Yahil's solution,
a radius, density, and velocity are written in dimensionless form:
\begin{eqnarray}
R&=&\kappa^{-1/2} G^{(\gamma-1)/2}r(t_0-t)^{\gamma-2} \nonumber \\
D(R)&=&\rho G (t_0-t)^{2}\\ 
V(R)&=&v \kappa^{-1/2} G^{(\gamma-1)/2} (t_0-t)^{\gamma-1},   \nonumber
\end{eqnarray}
where $G$, $\gamma$, $\kappa$, and $t_0$ are respectively
the gravitational constant, adiabatic index, polytrope coefficient, and
the time when the central density diverge. We put $\gamma = 1.3$,
$\kappa=1.2435\times 10^{15} /2^{4/3}$~cgs, and $t_0$ =
0.1~s. The collapse of a sphere with 4000~km radius is followed. In
the cylindrical coordinate calculations, two different spatial
resolutions are tested, where the
resolutions of the inner-most cells are $\{\Delta\varpi_{\rm{min}},\Delta
z_{\rm{min}}\}=2\times 10^{4}$~and~$4\times 10^{4}$~cm. The number of
numerical cells are  $N_{\varpi}\times N_z=720\times 720$ in the both
calculations. In the polar coordinate calculation, only
one spatial resolution, $\Delta r_{\rm{min}} =4\times 10^{4}$~cm, with
$N_{r}=720$ are taken. 

The distributions of density and velocity when the central density
reaches $2.3 \times 
10^{14}$~g~cm$^{-3}$ are presented in Fig.~\ref{fig.test.yahil}, while
errors at the inner most numerical cells are shown in
Table~2. Note that, in our MHD core-collapse simulations, a
bounce occurs when the  
central density reaches $\sim 2 \times 10^{14}$~g~cm$^{-3}$. Although
the spherical collapse of the 15~$M_\odot$ progenitor used in the
present work is not 
very well described by Yahil's solution\footnote{We found that when the
  central density is in the range of $\sim
  10^{12}-10^{13}$~g~cm$^{-3}$ in a spherical collapse of the
  progenitor, the distributions of a density and 
  velocity is roughly described by Yahil's solution. Out of this
  range, a deviation from Yahil solution is large.},
the results obtained here may indicate how much error our collapse
simulations involve.  
From the left panel of Fig.~\ref{fig.test.yahil} one can see that, a
deviation from the analytic solution is large near the center in each
calculation\footnote{The deviations from the analytic solution around
  the outer boundary are a numerical artifact due to a boundary
  condition.}.
In the cylindrical coordinate calculation with
$\{\Delta\varpi_{\rm{min}},\Delta z_{\rm{min}}\}=4\times 10^{4}$~cm,
in which the grid construction is same with our standard MHD collapse
simulations, the
errors both in $D$ and $V$ are $\sim 10$~\% (see the second row of
Table~2). Although these errors seem too large to
pursue a proper numerical simulation, as discussed in \S~\ref{sec.res},
varying the spatial resolution in a MHD collapse simulation does not change
results very much. This implies that the errors around the center are
not fatal to the extent of the discussions developed in this paper.

As seen in Fig.~\ref{fig.test.yahil} and in
Table~2, the cylindrical coordinate calculations have
larger errors than the polar coordinate calculation does, which may
be due to the structure of numerical cells. However, we found that
numerically evaluated electric currents behave better in cylindrical
coordinate than in polar coordinate, and thus adopt the former in the
present MHD simulations.

\begin{table}
\label{tab.test.yahil}
\begin{center}
\caption{Errors in numerical solution at the inner-most cell.}
\begin{tabular}{cccc}
\tableline\tableline
Coordinate & $\{\Delta r_{\rm{min}}, \Delta\varpi_{\rm{min}},\Delta
z_{\rm{min}}\}$\tablenotemark{a} [cm]& 
D-error\tablenotemark{b} [\%]& V-error\tablenotemark{c} [\%]\\ 
\tableline
Cylindrical & $2\times 10^{4}$ & 7.28 & 22.0 \\
Cylindrical & $4\times 10^{4}$ & 13.1 & 22.9 \\
Polar       & $4\times 10^{4}$ & 4.18 & 1.22 \\
\tableline
\end{tabular}
\tablenotetext{a}{Spatial resolution of the inner-most cell}
\tablenotetext{b}{Error of D at the inner-most cell}
\tablenotetext{c}{Error of V at the inner-most cell}
\end{center}
\end{table}

\begin{figure}
\epsscale{1}
\plottwo{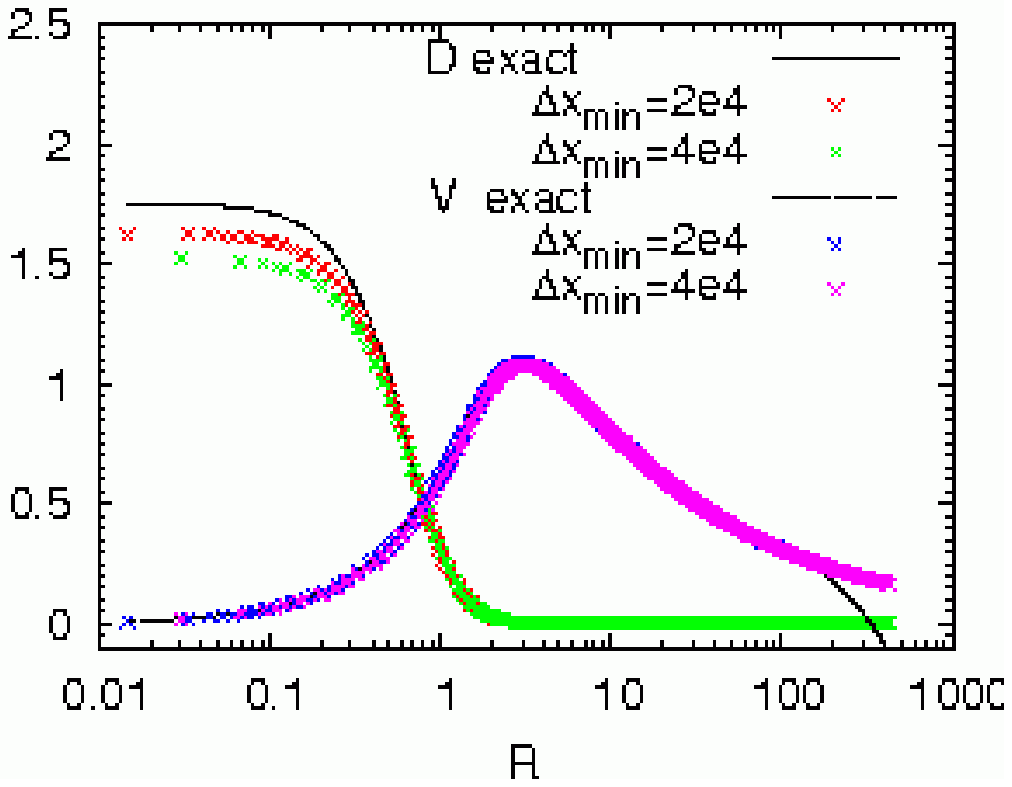}{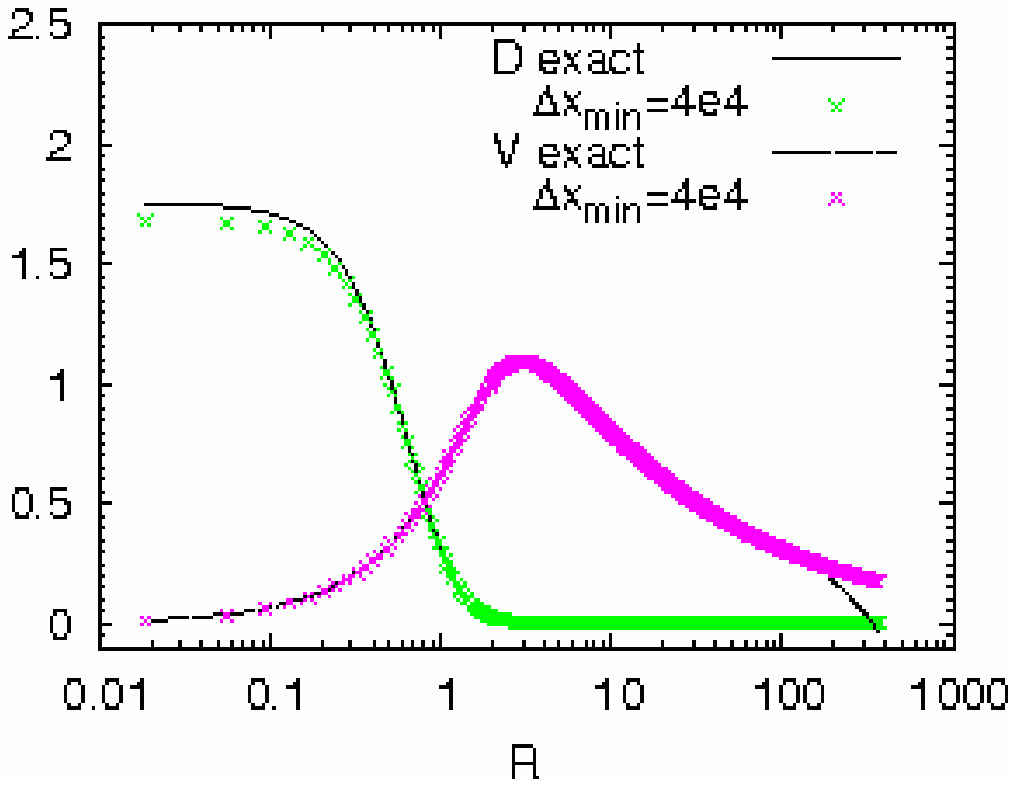}
  \caption{Results for the calculations of the Yahil's self-similar
    collapse with cylindrical coordinate (left) and polar coordinate
    (right). The solid lines and dotted lines are, respectively, for
    the analytic solution of D and V, while the colored crosses represent
    the numerical ones. For the cylindrical coordinate
    calculations, the numerical solutions along the pole ($\varpi$=0) are
    shown.}  
 \label{fig.test.yahil}
\end{figure}

\subsection{MHD Riemann Problem}\label{sec.app.brie}
As a basic test for the ideal-MHD part of our code, we carried out a
1D-MHD Riemann problem that are 
known as Brio-Wu Problem \citep{bri88}. The initial condition is set
as follows; 
\begin{eqnarray}
\left(
\begin{array}{c}
\rho\\
p   \\
B_x/\sqrt{4 \pi}\\   
B_y/\sqrt{4 \pi}   
\end{array}
\right)
=
\left(
\begin{array}{c}
1.0\\
1.0\\
0.75\\
1.0
\end{array}
\right)
\hspace{1pc}\textrm{for}\hspace{1pc}
x<0.5,
&&
\left(
\begin{array}{c}
\rho\\
p   \\
B_x/\sqrt{4 \pi}\\   
B_y/\sqrt{4 \pi}      
\end{array}
\right)
=
\left(
\begin{array}{c}
0.125\\
0.1\\
0.75\\
-1.0 
\end{array}
\right)
\hspace{1pc}\textrm{for}\hspace{1pc}
x>0.5\nonumber\\
\end{eqnarray}
where the computational domain is $x\in[0,1]$. The initial
velocities are set to be zero. The adiabatic index here is 2.0.
The results of the calculation at $t=0.1$ done with 800~numerical
cells are given in Fig.~\ref{fig.test.bw} together with those obtained
by a famous numerical code ZEUS-2D \citep{sto92}. Although a little
differences can be found, the two results well matches.

\begin{figure}
\epsscale{1}
\plotone{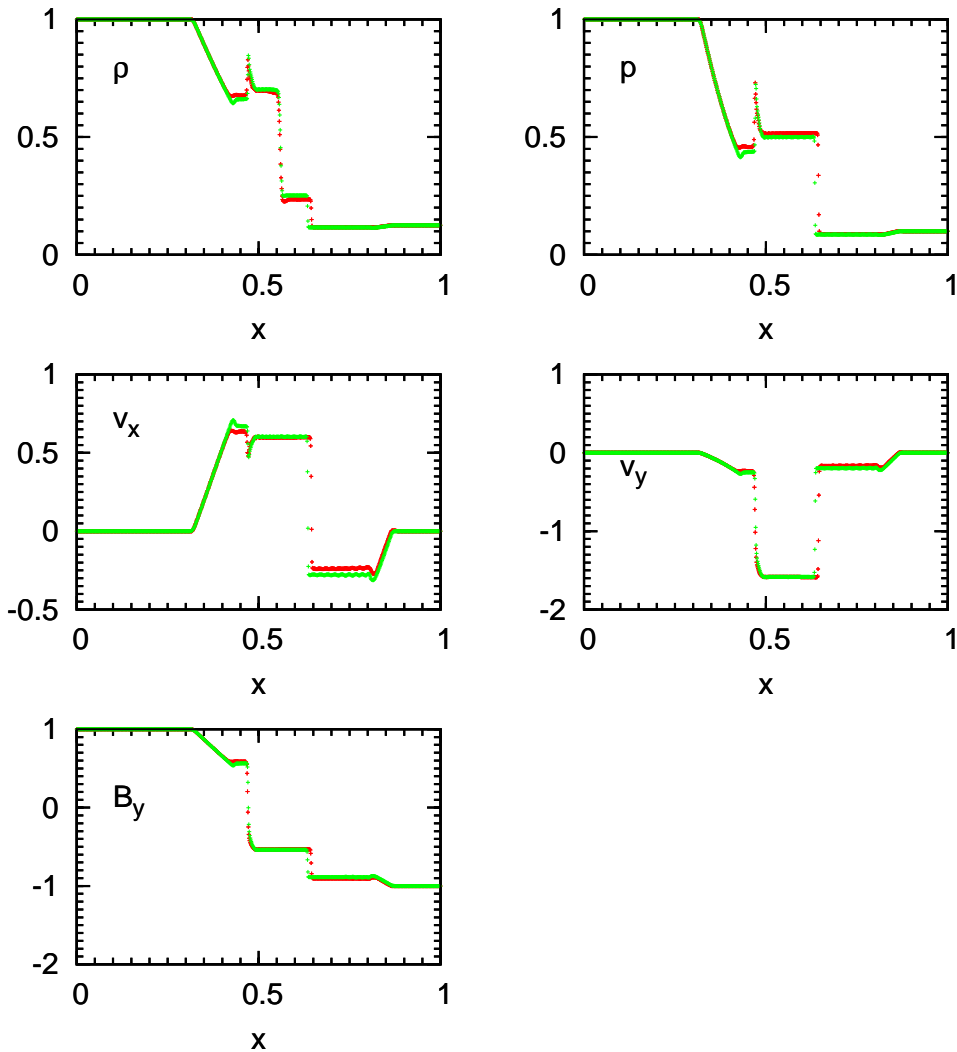}
  \caption{Results of the MHD-Riemann problem at $t=0.1$. The red and 
    Green crosses represent the solutions obtained by
    \textit{Yamazakura} and ZEUS-2D, respectively. The both
    calculations are done with 800~numerical cells.}
 \label{fig.test.bw}
\end{figure}

\subsection{Linear Wave Propagation}\label{sec.app.lw}
In \S~\ref{sec.eq}, we mentioned that the KT scheme adopted in our
numerical code is third order in time and second order in space. Here,
we will see whether \textit{Yamazakura} solve the MHD equations fairly with
these degree of accuracy, by carrying out a linear wave propagation
problem in 1D Cartesian coordinate. A right-going slow magnetosonic
wave is adopted as a linear wave. 

According to \citet{gar05}, we put the initial state of the conserved
variables as 
\begin{eqnarray}
\mbox{\boldmath$u$}_{in}=\mbox{\boldmath$u_0$} + \epsilon
\mbox{\boldmath$R$} \cos(2\pi x), 
\end{eqnarray}
where $\mbox{\boldmath$u_0$}$ is the background state,
$\epsilon=10^{-6}$ is the wave amplitude, and $\mbox{\boldmath$R$}$ is
the right eigenvector for the right-going slow magnetosonic wave in
conserved variables given by  
\begin{eqnarray}
\mbox{\boldmath$R$}=\frac{1}{6\sqrt{5}}\left(
\begin{array}{c}
12\\
6\\
8\sqrt{2}\\
4\\
9\\
0\\
-4\sqrt{2}\\
-2   
\end{array}
\right).
\end{eqnarray}
The background state is set as $\rho_0=1$, $p_0=1$, $v_{0x}=v_{0y}=v_{0z}=0$,
$B_{0x}=1$, $B_{0y}=\sqrt{2}$, and $B_{0z}=1/2$, with which the slow
magnetosonic speed is $c_s=1/2$. The adiabatic index is set as
$\gamma=5/3$. The computational domain is $x\in[0,1]$, and the
periodic boundary condition is adopted at each boundary. Each
calculation is run until $t=4$ when the wave has propagated the
distance of two wave length. 

In order to check the degree of accuracy in space, we first
tested two calculation serieses with uniform cell construction, each of
which contains four calculations with different grid numbers. In the
first series, series A, we fix $\Delta t$ by controlling the
Courant-Friedrichs-Lewy (CFL) number while varying $\Delta x$. The
numbers of cells N and CFL numbers $\nu$ for the four calculations are
$(N,\nu)=(8,0.49/64), (32,0.49/16), (128,0.49/4)$, and $(512,0.49)$.
In the second series, series B, we fix the CFL number at $\nu=0.49$
and tested the same set of the cell numbers as before. Note that in
series B, both $\Delta x$ and $\Delta t$ are halved by doubling the
cell number. 
The distributions of $v_x$ at $t=4$ for series B are plotted in
Fig.~\ref{fig.test.lw}, which shows that the numerical solution well
matches the analytic one for $N\ge 128$. The error from the analytic
solution is evaluated by the L1 error norm,
\begin{eqnarray}
|\delta \mbox{\boldmath$u$}| = \sqrt{\sum_m(\delta u^m)^2}, 
\end{eqnarray}
where
\begin{eqnarray}
\delta u^m=\frac{1}{N}\sum_i|u^m_i-(u^m_{in})_i|.
\end{eqnarray}
The left panel of Fig.~\ref{fig.test.lwerr} plots the
variation of L1 error norm with respect to $N$ for series A and
B. The red plots (series A) show that the scheme is approximately
second order in space. 
We found that the results from series B (blue plots) are almost
identical to those in series A, despite that $\Delta t$ is larger in
series B for a fixed $N$, viz. a fixed $\Delta x$, except for $N=512$.
This indicate that an error due to $\Delta t$ is dominated over by an
error due to $\Delta x$ for $\nu<0.5$ (the CFL condition for the KT
scheme), and brings difficulty in properly checking the 
degree of accuracy in time. Nonetheless, we can still state that the
scheme is at least approximately second order in time, by the fact
that the L1 error norms in series B, varying both $\Delta x$ and
$\Delta t$ at a same rate, result in $\sim N^{-2}$ (see the blue plots
in the left panel of Fig.~\ref{fig.test.lwerr}). 

Next, we evaluate the degrees of accuracy of our code for cases of
non-uniform cell construction. A structure of numerical cells we set
here is such that the $(2n-1)$-th cell has the size
$L/N\times(1-\alpha)$ while the $2n$-th cell has the size
$L/N\times(1+\alpha)$, where $L$ 
is the size of the whole numerical domain and $\alpha<1$ is the parameter to
represent the degree of non-uniformity. We tested three cases
$\alpha = 10^{-2}, 10^{-1}$, and $5\times 10^{-1}$. In each case, four
calculations fixing $\nu=0.49$ with $N=8, 32, 128$, and $512$ are
done. The results are plotted in the right panel of
Fig.~\ref{fig.test.lwerr}. It is shown that the second order in space
and at least second order in time are marginally kept even for 200~\%
size increase and 33~\% size decrease from the neighboring cell
($\alpha=5\times10^{-1}$). Note that a numerical cell size in our
MHD-collapse simulations increases outwards by $0.56~\%$.

\begin{figure}
\epsscale{0.5}
\plotone{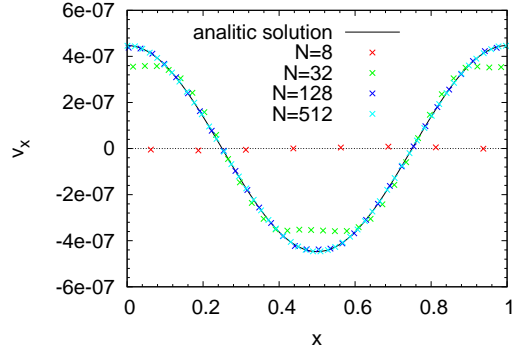}
  \caption{Numerical results of the linear slow magnetosonic wave
    propagation for series A for the distribution of $v_x$ at
    $t=4$. The plots for the calculations with $N=128$ and 512 are
    thinned for the purpose of clear illustration.}
 \label{fig.test.lw}
\end{figure}
\begin{figure}
\epsscale{1}
\plottwo{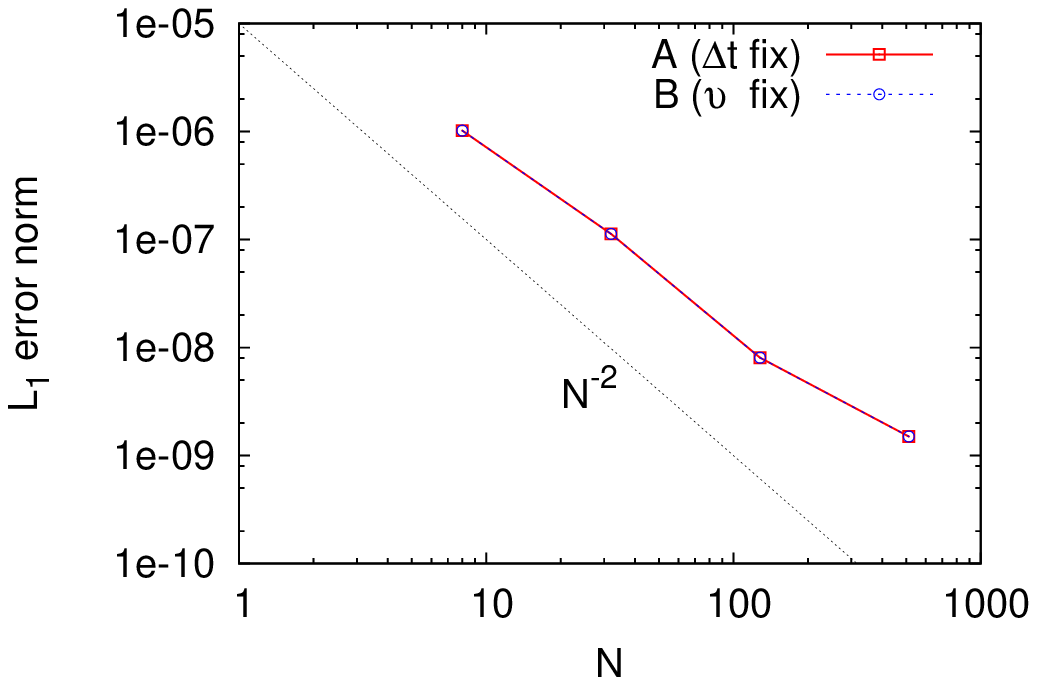}{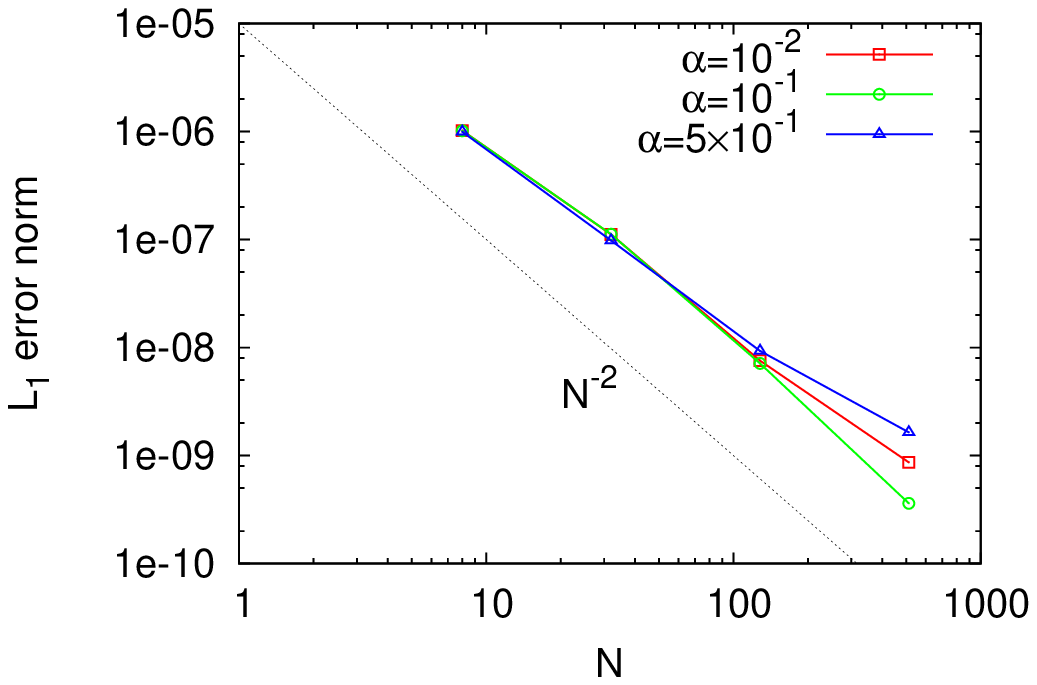}
  \caption{Distributions of L1 error norm with respect to the number of
    numerical cells for the linear slow magnetosonic wave
    propagation problem with uniform cell spacing (\textit{left}) and
    non-uniform cell spacing (\textit{right}).}
 \label{fig.test.lwerr}
\end{figure}

\subsection{Rotor Problem}\label{sec.app.rotor}
In order to check that the multi-dimensional ideal MHD equations are well
solved with our code, we perform two test calculations. One is a
so-called rotor problem to be presented here, 
and the other is a point source explosion with magnetic field that
will be shown in the next section.

The rotor problem is suggested by \citet{bal99}, which simulates the
propagations of torsional Alfv\'en waves.
The system initially consists of a rapidly rotating cylinder of
dense fluid surrounded by lighter gas at rest, which is threaded by
uniform a magnetic field parallel to the $x$-axis. With the computational
domain of $(x,y)\in[-0.5,0.5]\times[-0.5,0.5]$, the initial condition is  
\begin{eqnarray}
\rho&=&1+9f(r),\nonumber\\
p&=&1,\nonumber\\
v_x&=&-2f(r)y/r,\hspace{1pc}v_y=-2f(r)x/r,\hspace{1pc}v_z=0,\nonumber\\
B_x&=&5,\hspace{1pc}B_y=B_z=0,\\
\textrm{where}\nonumber\\
f(r)&=&
\begin{cases}
1 & \textrm{if}\hspace{1pc}r<0.1,\\
\frac{200}{3}(0.115-r) & \textrm{if}\hspace{1pc}0.1<r<0.115,\\
0 & \textrm{if}\hspace{1pc}r>0.115,
\end{cases}\nonumber\\
r&=&(x^2+y^2)^{1/2}.\nonumber
\end{eqnarray}
The adiabatic index is $\gamma=1.4$. The calculation is done with $400
\times 400$ numerical cells. Results of the calculation are displayed in 
Fig.~\ref{fig.test.rotor}, which show good agreement with those found
in \citet{zie04}.

\begin{figure}
\epsscale{1}
\plottwo{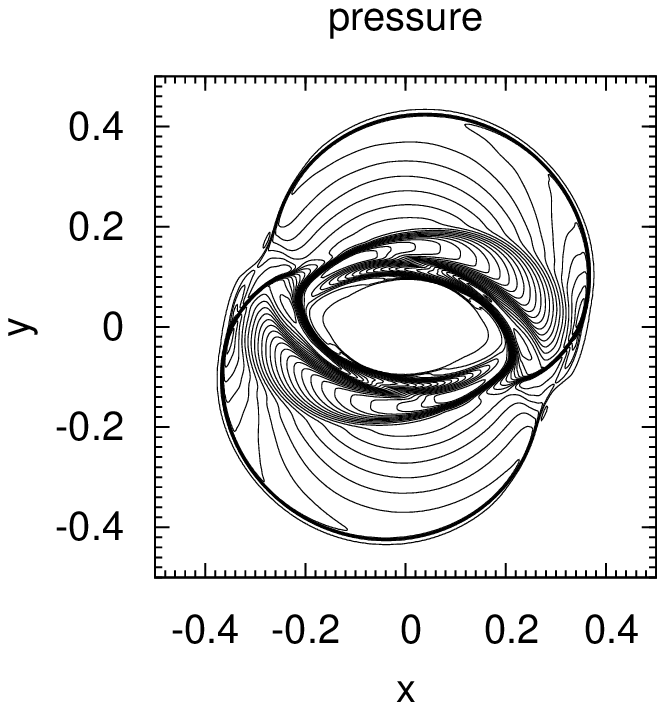}{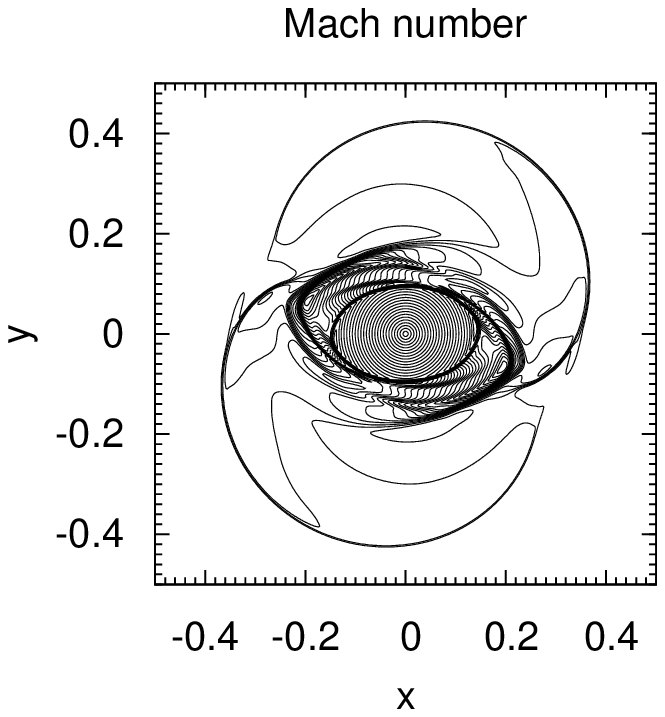}
  \caption{Numerical results of the rotor problem. Left and 
    Right panels respectively show pressure and Mach number
    contours at $t=0.15$}
 \label{fig.test.rotor}
\end{figure}

\begin{figure}
\epsscale{0.5}
\plotone{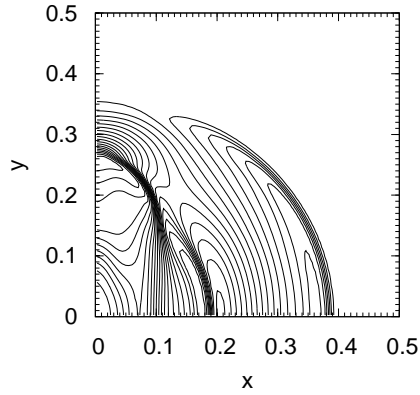}
  \caption{Magnetic pressure contours at $t=2.5\times 10^{-3}$ in the
    calculation of a point source explosion with magnetic field.} 
 \label{fig.test.bsedov}
\end{figure}

\subsection{Point Source 3D-Explosion with Magnetic
  Field}\label{sec.app.bsedov} 
A calculation presented here differs from the spherical explosion
test described above in that fluid is magnetized.  
The initial condition is given by
\begin{eqnarray}
\rho&=&1,\nonumber\\
p&=&
\begin{cases}
10^4 & \textrm{if}\hspace{1pc}x^2+y^2+z^2<r^2,\\
1    & \textrm{otherwise.}
\end{cases}\nonumber\\
v_x&=&v_y=v_z=0,\nonumber\\
B_z/\sqrt{4\pi}&=&100,\hspace{1pc}B_x=B_y=0.
\end{eqnarray}
The adiabatic index is chosen as $\gamma=5/3$.
We do a calculation in cylindrical coordinate with the numerical
domain of $(\varpi,z)\in[0,0.5]\times[0,0.5]$, taking the
symmetric axis in the magnetic field direction.
The number of numerical cells is $N_\varpi\times N_z = 96 \times
96$. A result of the calculation is given in
Fig.~\ref{fig.test.bsedov}. This also shows good agreement with that
of \citet{zie04} done with Cartesian coordinate.

\subsection{Magnetic Field Diffusion}\label{sec.app.dif}
In order to see that \textit{Yamazakura} properly handles the
resistive terms in the induction equation, we test a
magnetic field diffusion problem in 1D. Here, we solve not the full
set of the MHD equations but the magnetic diffusion equation with a
constant resistivity, 
\begin{eqnarray}
\frac{\partial B}{\partial t}&=&\eta\frac{\partial^2 B}{\partial x^2},
\end{eqnarray}by freezing fluid in the induction equation.
Setting the initial condition with an anti-parallel magnetic field,
$B=-B_0$ for $x<0$ and $B=B_0$ for $x>0$, the exact solution at time
$t$ is written as
\begin{eqnarray}
B(x,t)&=&-\frac{2B_0}{\sqrt{\pi}}\int^{x/\sqrt{4\eta t}}_0 e^{-u^2}du
\hspace{1pc}\textrm{for}\hspace{1pc}x<0,\nonumber\\
B(x,t)&=&\frac{2B_0}{\sqrt{\pi}}\int^{x/\sqrt{4\eta t}}_0 e^{-u^2}du
\hspace{1pc}\textrm{for}\hspace{1pc}x>0.\label{eq.test.diffuse}
\end{eqnarray}
We set $B_0=100$ and $\eta=10$. The calculation is done with
800~numerical cells for $x\in[-5,5]$. The  
Results are displayed in Fig.~\ref{fig.test.diff}, in which one
can find good agreement between the numerical and exact
solutions. 

\begin{figure}
\epsscale{0.5}
\plotone{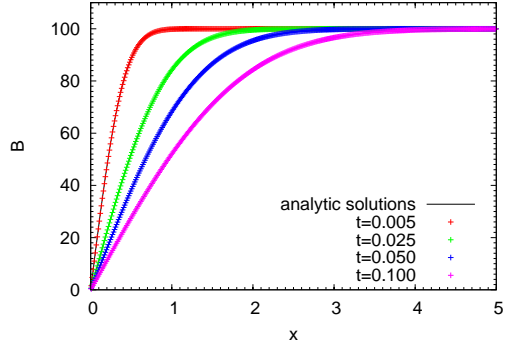}
  \caption{Results of the magnetic field diffusion
    calculation. The numerical solutions are shown 
    by the crosses while the analytic solutions are by the solid
    lines.} 
 \label{fig.test.diff}
\end{figure}

\subsection{Spontaneous Fast Reconnection}\label{sec.app.rec}
We finally check that \textit{Yamazakura} properly deals with
the multi-dimensional 
resistive-MHD equations, by the simulation of a spontaneous
fast reconnection firstly done by \citet{uga99}.
The initial current system is constructed as follows:
\begin{eqnarray}
B_x(y)/\sqrt{4\pi}&=&
\begin{cases}
\sin(\pi y/2)        & \textrm{for}\hspace{1pc}y<1,\\
1                    & \textrm{for}\hspace{1pc}1<y<3.6,\\
\cos[(y-3.6)\pi/1.2] & \textrm{for}\hspace{1pc}3.6<y<4.2,\\
0                    & \textrm{for}\hspace{1pc}4.2<y,\\
-B_x(-y)/\sqrt{4\pi}  & \textrm{for}\hspace{1pc}y<0,\\
\end{cases}\nonumber\\
p&=&(1+\beta_0-B_x^2/4\pi)/2,\nonumber\\
\rho&=&2p/(1+\beta_0),
\end{eqnarray}
where $\beta_0=0.15$.
A velocity and the $x,y$-component of magnetic field are initially
zero. The adiabatic index is taken as $\gamma=5/3$. A reconnection of
magnetic field-lines is initiated by an anomalous resistivity;
\begin{eqnarray}\label{eq.test.ugai2}
\eta(\mbox{\boldmath$r$},t)&=&
\begin{cases}
k_R[V_d(\mbox{\boldmath$r$},t)-V_c] & \textrm{for}\hspace{1pc}V_d>V_c,\\
0                                   & \textrm{for}\hspace{1pc}V_d<V_c,\\
\end{cases}\\
\end{eqnarray}
where
\begin{eqnarray}
V_d(\mbox{\boldmath$r$},t)&=&
\left|\mbox{\boldmath$J$}(\mbox{\boldmath$r$},t)/\rho(\mbox{\boldmath$r$},t)\right|,
\nonumber\\
V_c&=&V_{c,0}\left[\frac{2p}{\rho(1+\beta_0)}\right]^{\alpha}.
\end{eqnarray}
The parameters are taken as $k_R=0.03$, $V_{c,0}=4$, and $\alpha=0.5$. 
For details of the resistivity model, see \citet{uga99}.
The anomalous resistivity model (\ref{eq.test.ugai2}) is assumed for
$t>4$. During $0<t<4$, a localized resistivity bellow is imposed to
disturb the initial 1D-configuration:
\begin{eqnarray}
\eta(\mbox{\boldmath$r$})&=&\eta_d\exp[-(x/1.1)^2-(y/1.1)^2],
\end{eqnarray}
where $\eta_d=0.02$. The computation is done with
$800\times800$~numerical cells covering the domain of
$(x,y)\in[-20,20]\times[-6,6]$.

Some important results of the computation are shown in
Fig.~\ref{fig.test.ugai1} and \ref{fig.test.ugai2}. The global
magnetic-field distributions are displayed in
Fig.~\ref{fig.test.ugai1} for t=18 and 30. We found that they are 
similar to those obtained by \citet{uga99}, and also those of
\citet{fen06}. The left panel of
Fig.~\ref{fig.test.ugai2} represents profiles of $v_x$ and $B_y$ along
the x-axis at $t=24$ and $t=30$, while the right panel shows the
evolution of the resistivity $\eta$ and electric field $E$ at the
origin, $v_y$ at $(x,y)=(0,0.9)$, and a magnetic flux
$\Phi$ evaluated by
\begin{eqnarray}
\Phi=\int_{y\ge0}B_x(x=0,y)dy.\nonumber
\end{eqnarray}
For these results, we also find a rough agreement with the
above two works, although there are some insignificant differences. 

With all these results presented in the appendix, we expect that
the present MHD collapse simulations done by \textit{Yamazakura}
offer proper results.

\begin{figure}
\epsscale{0.8}
\plotone{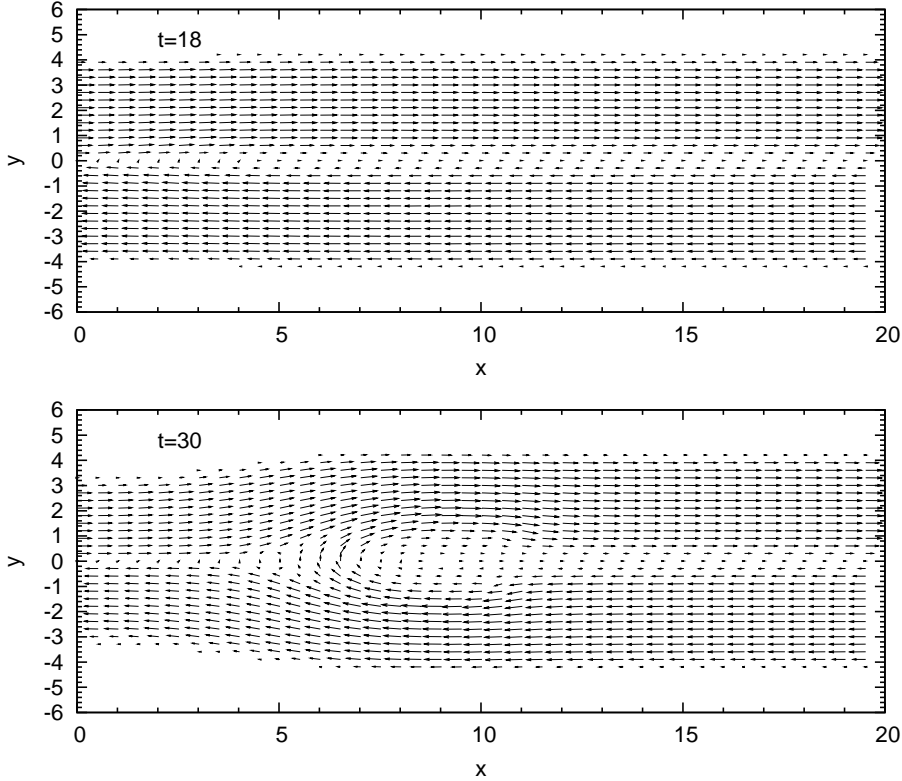}
  \caption{Magnetic field distributions at t=18 and 30 for the
    calculation of spontaneous fast reconnection.} 
 \label{fig.test.ugai1}
\end{figure}

\begin{figure}
\epsscale{1}
\plottwo{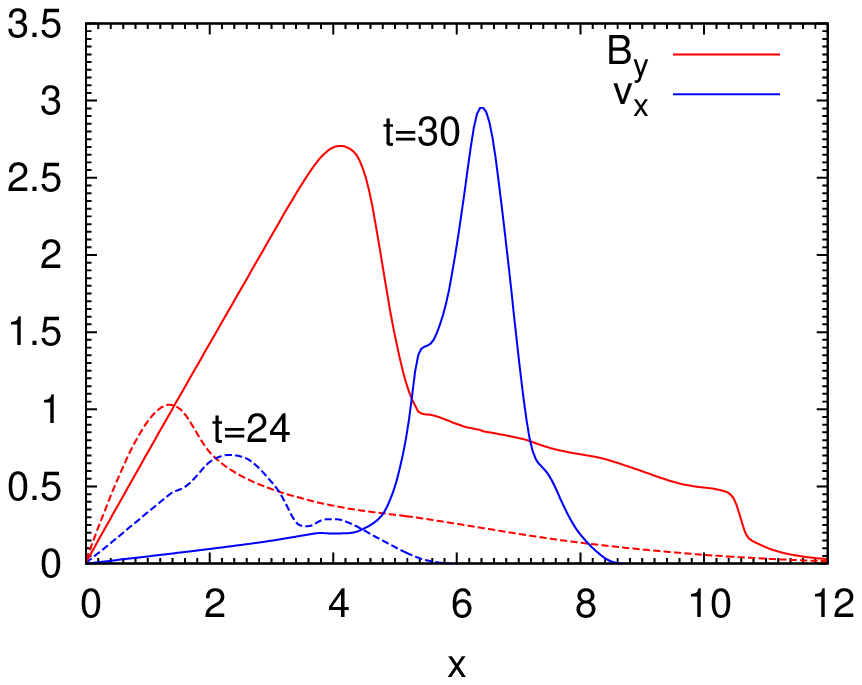}{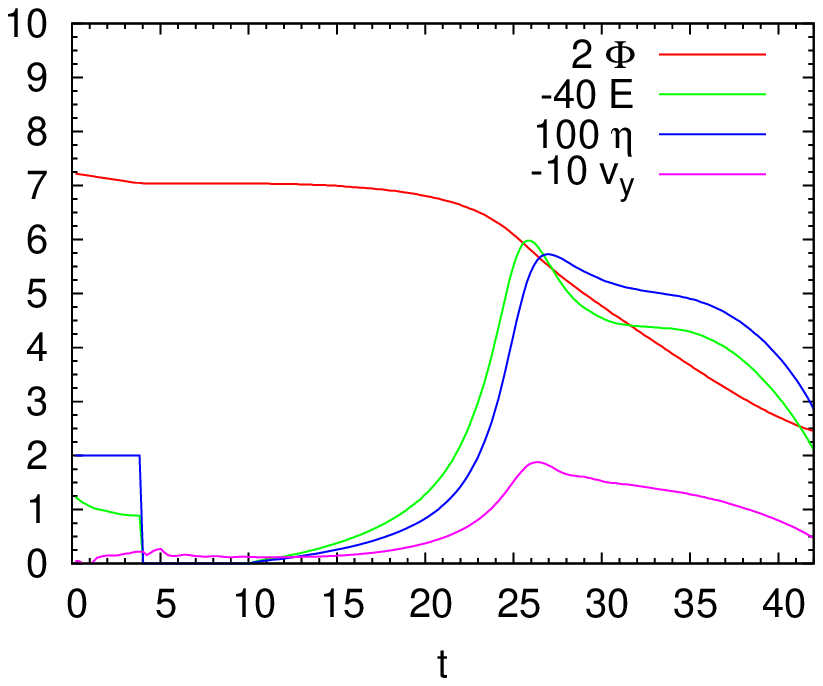}
  \caption{Results for the calculation of spontaneous
    fast reconnection. \textit{Left}:
    Profiles of $v_x$ and $B_y$ along the x-axis 
    at $t=24$ and $t=30$. \textit{Right}: Temporal variations of the
    magnetic flux $\Phi$, the resistivity $\eta$ and electric field
    $E$ at the origin, and $v_y$ at $(x,y)=(0,0.9)$.} 
 \label{fig.test.ugai2}
\end{figure}

\end{document}